\documentclass[phdthesis,12pt]{wuthesis}     %% LaTeX2e document.
\usepackage{ellipsis}
\usepackage[round]{natbib}
\usepackage{setspace}
\usepackage{amsmath}
\usepackage{textcomp}
\usepackage{indent first}
\usepackage{amssymb}
\usepackage{graphicx}
\usepackage{rotating}
\DeclareGraphicsExtensions{.pdf,.jpeg,.png}
\usepackage{subcaption}
\usepackage{xr}
\usepackage{lscape}
\usepackage{xcolor}
\usepackage{tocloft}

\definecolor{blue}{rgb}{0.0, 0.44, 1.0}

\renewcommand{\thesisauthor}{Janie Katherine Hoormann}
\renewcommand{\thesisauthorlastname}{Hoormann}

\renewcommand{\thesisdepartment}{Department of Physics}

\renewcommand{\thesismonth}{August}
\renewcommand{\thesisyear}{2016}

\renewcommand{\thesistitle}{Using Simulations of Black Holes to Study General Relativity and the Properties of Inner Accretion Flow}
\renewcommand{\thesisshorttitle}{Using Black Holes to Probe Gravity}

\renewcommand{\thesissupervisor}{Professor Henric Krawczynski}
\renewcommand{\thesiscommittee}{Henric Krawczynski, Chair \\
	James Buckley \\
	Martin Israel \\
	Jon Miller \\
	Ryan Ogliore}

\renewcommand{\thesisdedication}{DEDICATION GOES HERE}

\renewcommand\cftfigfont{\normalsize}

\begin{document}

\frontmatter
%%
%  This is all that frontmatter stuff
%
%  This way I can 'not' include it easily

% NOTE: do not put any text in the thesistitlepage, thesiscopyrightpage,
% or thesisdedicationpage sections.  If you want to use these pages, then you
% should remove the notes below (e.g., by uncommenting the \iffalse
% and \fi lines) and change the appropriate fields in thesis-main.tex.
% This will ensure that the copyright and dedication lines are positioned
% and formatted correctly.  Additionally, remove the
% thesisacknowledgmentpostscript and listoftablespostscript sections, since
% these are used to add explanatory notes which shouldn't be there in normal
% theses.

\begin{thesistitlepage}               %% Generate the title page.
%\iffalse
\end{thesistitlepage}

\begin{thesiscopyrightpage}                 %% Generate the copyright page.
%\iffalse
\end{thesiscopyrightpage}

\begin{singlespace}
\setcounter{page}{2}
\renewcommand{\contentsname}{Table of Contents}
\tableofcontents

\renewcommand{\listoftablespostscript}{
}
\newpage
\addcontentsline{toc}{chapter}{List of Figures}
\listoffigures
\newpage
\addcontentsline{toc}{chapter}{List of Tables}
\listoftables
\end{singlespace}

%\iffalse
\renewcommand{\thesisacknowledgmentpostscript}{
}
%\fi

\begin{thesisacknowledgments}

I could not have arrived at this point without the help and support from a vast number of people.  First, I need to thank Henric Krawczynski who, through advising me these past several years, has taught me how to be a physicist. Under his direction I not only learned about black holes, general relativity, and how to study them, but also how to share my work with others through both writing papers and developing presentations.  I am also grateful for guidance from Fabian Kislat, Paul Dowkontt, Anna Zajczyk, and Matthias Beilicke and in particular for their help and patience in teaching me how to do experimental work.  I also couldn't have asked for better colleagues/office-mates/friends than my fellow grad students, Banafsheh Behestipour and Rashied Amini.  It would be remiss of me to not thank those I have shared an office/floor with, including Nathan Walsh, Andrew West, Brian Rauch, Ryan Murphy, Avery Archer, and Kelly Lave who, in addition to those mentioned above, helped make this a great place to come and work every day.  I am thankful for the support from all of the administrative staff, in particular Sarah Akin and Julia Hamilton, who made sure my time as a graduate student ran smoothly.

I would also like to thank Jim Buckley and Ram Cowsik for serving on my faculty mentoring committee, as well as Marty Israel, Ryan Ogilore, and Jon Miller for joining them to participate on my defense committee.

I would like to thank the McDonnell Center for the Space Sciences at Washington University in St. Louis and NASA grant \# NNX14AD19G for funding this work.

Last but most certainly not least, I must thank my family who has supported and encouraged me not only during my graduate career but throughout my entire life.  In particular I would like to thank my mom, who is responsible for any correctly placed commas found in this and any of my other written works.

\end{thesisacknowledgments}

%\begin{thesisdedicationpage}                %% Generate the dedication page.
%\iffalse
%\fi
%\end{thesisdedicationpage}

\begin{thesisabstract}
While Albert Einstein's theory of General Relativity (GR) has been tested extensively in our solar system, it is just beginning to be tested in the strong gravitational fields that surround black holes.  As a way to study the behavior of gravity in these extreme environments, I have used and added to a ray-tracing code that simulates the X-ray emission from the accretion disks surrounding black holes.  In particular, the observational channels which can be simulated include the thermal and reflected spectra, polarization, and reverberation signatures.  These calculations can be performed assuming GR as well as four alternative spacetimes.  These results can be used to see if it is possible to determine if observations can test the No-Hair theorem of GR which states that stationary, astrophysical black holes are only described by their mass and spin.  Although it proves difficult to distinguish between theories of gravity, it is possible to exclude a large portion of the possible deviations from GR using observations of rapidly spinning stellar mass black holes such as Cygnus X-1.  The ray-tracing simulations can furthermore be used to study the inner regions of black hole accretion flows.  I examined the dependence of X-ray reverberation observations on the ionization of the disk photosphere. My results show that X-ray reverberation and X-ray polarization provides a powerful tool to constrain the geometry of accretion disks which are too small to be imaged directly.  The second part of my thesis describes the work on the balloon-borne X-Calibur hard X-ray polarimetry mission and on the space-borne PolSTAR polarimeter concept.

\end{thesisabstract}

%\chapter{Preface}

%%
%% For List of Abbreviations, Glossary or Nomenclature also
%% use \chapter, but put some kind of list environment inside.

%%% Local Variables: 
%%% mode: latex
%%% TeX-master: "thesis-main"
%%% End: 

\mainmatter
\doublespacing
\setlength{\parindent}{1cm}
\chapter{Introduction}
\label{Introduction}		

\section{Motivation}

In 1915, Albert Einstein proposed his now famous theory of general relativity (GR) which describes gravity as a result of the curvature of spacetime.  Now, having just celebrated its 100th birthday, GR remains our best thoery to describe gravitational effects and the expansion of the universe.  Karl Schwarzschild found the first black hole (BH) solution of GR describing a non-rotating mass (with the solution for a rotating mass being later found by Roy Kerr).  This solution had the peculiar feature that once something came within a particular radius surrounding the mass ($r_s = 2GM/c^2$ where $G$ is gravitational constant, $M$ is the mass of the BH, and $c$ is the speed of light) it was impossible for it to escape the gravitational pull and emerge again.  The surface at $r = r_s$ surrounds the BH and became known as the event horizon. The first BH to be discovered was Cygnus X-1, the black hole binary (BHB) system containing a stellar mass BH and companion star \citep{Tananbaum1972}.  While stellar mass BHs can have masses up to tens of solar masses, it was predicted and later confirmed that in the center of our galaxy is a supermassive BH (SMBH), Sagittarius $A^\ast$, with a mass on the order of $10^6 M_{\odot}$. It has since been seen that finding a SMBH in the center of galaxies is a common occurrence, with many of them being even more massive than Sgr $A^\ast$ (see \citet{Raine2005}).  Since their discovery significant progress has been made in understanding these extreme systems but there is still much that needs to be done.  This thesis will present the research I performed to study the inner regions of accretion flow onto BHs and to use these systems to study GR.

\section{General Relativity}
	
	\subsection{Introduction to General Relativity}

An important aspect of theories of gravity is the equivalence principle.  This begins in Newtonian gravity with the weak equivalence principle, which states that the internal structure and composition of a test body does not affect its trajectory when it is freely falling.  Here a test body is described as an object that is neither affected by tidal gravitational fields nor electromagnetic forces.  Einstein extended this by not only stating that the weak equivalence principle is valid but also that local non-gravitational experiments do not depend on the location and the velocity of the freely-falling reference frame in which the experiment is performed, leading to what is termed the Einstein equivalence principle (EEP).  The strong equivalence principle extends the EEP by stating that freely falling objects are not affected by their structure and composition even if they are self-gravitating.  In order to have a theory of gravity which meets the requirements of the EEP the metric describing the theory must satisfy the following criteria - the metric is symmetric (such that $g_{ab} = g_{ba}$), freely falling test bodies have trajectories along geodesics, and special relativity holds in local freely falling reference frames (see \cite{Will2006} and references therein).

The line element, the invariant quantity describing the geometry of spacetime given the metric $g_{\mu\nu}$, is defined to be 
	
 	\begin{equation}  \label{eq:lineElement}
 		ds^2 = g_{\mu \nu} dx^\mu dx^\nu.
 	\end{equation}
In flat spacetime this is given by the Minkowski metric where 	

 	\begin{equation}
 	g_{\mu \nu} = \begin{pmatrix}
		-1 & 0 & 0 & 0 \\
		0 & 1 & 0 & 0 \\
		0 & 0 & 1 & 0 \\
	 \vspace{12pt}	0 & 0 & 0 & 1    
    \end{pmatrix}
    \end{equation}  
It is worth noting that we adopt the convention that the time component has a negative sign while the spatial components are positive.  This convention will be used for all metrics presented throughout the rest of this thesis.  In order to solve for the metric describing spacetime around BHs it is necessary to start with Einstein's equation.  This is given by
    
    \begin{equation}
  G_{\mu\nu}=  R_{\mu\nu} - \frac{1}{2}Rg_{\mu\nu}  = \frac{8 \pi G}{c^4}T_{\mu\nu}
    \end{equation}   
where $R_{\mu\nu}$ is the Ricci tensor, $R$ is the Ricci scaler (both of which describe the curvature of space),  $T_{\mu\nu}$  is the stress energy tensor which describes the density, energy, and momentum in a given spacetime, $G$ is Newton's gravitational constant, and $c$ is the speed of light.  In a vacuum this reduces to
   \begin{equation} \label{eq:einstein}
   G_{\mu\nu} = 0.
   \end{equation}  
There are only a few analytic solutions to the Einstein equation, i.e. BH solutions and gravitational wave solutions.  In this thesis I am interested in the former. For a non-rotating BH Equation \ref{eq:einstein} can be solved to find the following solution for the metric referred to as the Schwarzschild solution \citep{Schwarzschild1916}. 
   	 \begin{equation}
 	g_{\mu \nu} = \begin{pmatrix}
		 \frac{2M}{r}-1 & 0 & 0 & 0 \\
		0 &  \left(1-\frac{2M}{r}\right)^{-1} & 0 & 0 \\
		0 & 0 & r^2  & 0 \\
       0 & 0 & 0 & r^2 \sin ^2 \theta    
    \end{pmatrix}
    \end{equation}   
    Similarly, the metric for a rotating BH with angular momentum $J = aMc$, where $a$ is a dimensionless number characterizing the spin of the BH, is again derived by solving Equation \ref{eq:einstein} resulting in the well-known Kerr metric \citep{Kerr1963} where

	 	\begin{equation}
 	g_{\mu \nu} = \begin{pmatrix}
		 \frac{2Mr}{\Sigma}-1 & 0 & 0 & -\frac{2aMr\sin^2\theta}{\Sigma} \\
		0 &  \frac{\Sigma}{\Delta} & 0 & 0 \\
		0 & 0 & \Sigma  & 0 \\
       -\frac{2aMr\sin^2\theta}{\Sigma} & 0 & 0 & \sin ^2 \theta \left( r^2+a^2+\frac{2a^2Mr\sin^2\theta}{\Sigma}\right)    
    \end{pmatrix}
    \end{equation}	
	with
	\begin{equation}
	\Sigma \equiv r^2 + a^2 \cos^2 \theta\ \\
	\end{equation} 
	\begin{equation}
	\Delta \equiv r^2 -2Mr +a^2.
	\end{equation}	
The angle $\theta$ is measured with respect to the rotation axis of the BH with the disk lying in the plane $\theta = Pi/2$. Trajectories of particles in these spacetimes are calculated by solving the geodesic equation which will be discussed in detail in Chapter 2. The calculations performed in this thesis use the common notation of general relativity which sets $G = c = \hbar = 1$. 
	
\subsection{Testing General Relativity} \label{Introduction:testingGR} 
	
Included in GR are a number of new phenomena having observational consequences that allow for the opportunity to test Einstein's theory. Early tests were performed using the gravitational fields found in our solar system and confirmed the predictions of GR.  Examples of these tests include, but are not limited to, the advance of the perihelion of Mercury, light bending, the time delay of light, and the E\"otv\"os experiment, all of which will be reviewed below (see e.g. \citet{DInverno1992} for a review of solar system tests of GR).  Previously, astronomers observed that the perihelion of Mercury, the point of nearest approach to the sun, was advancing but not at a rate consistent with the predictions of classical theories.  The difference observed was 43.1 $\pm$ 0.5 arcseconds per century, which was significantly larger than the expected observational error.  Upon the introduction of GR, they were able to redo the calculations taking into account this new gravitational theory and found that the effects of GR they were previously neglecting would account for an additional 43 arcseconds per century, finally bringing theory and observation into agreement \citep{Clemence1947}.  GR also predicts that light will bend when passing near a massive object.  For example, light from stars which passes near the sun will get bent due to its gravitational field, changing the apparent location of the stars.  In 1919, Eddington used observations of star locations taken during a total eclipse and compared them to the locations of the same stars when the sun was no longer obstructing their view to first test this phenomenon \citep{Dyson1920}.  Along with the bending of light, GR also predicted that the curvature in spacetime would cause a delay in the travel time of light.  To test this prediction, radar signals were sent to Venus right before it would pass behind the sun.  The time it would take for these signals to be reflected and observed on Earth was calculated and compared with both the classical and GR theories.  There was a measured delay with respect to the classical theories of around 200 $\mu$s which was within 5\% of the delay predicted by GR \citep{Shapiro1971}.  The E\"otv\"os experiment was designed to test the equivalence principle.  In order to do this, two objects were connected by a rod and suspended by a fine wire.  Each object was made of a different composition and if the equivalence principle did not hold each object would experience a different gravitational acceleration which would produce a torque.  E\"otv\"os found the limit on the difference in acceleration between the two objects to be less than $5 \times 10^{-9}$ and later experiments improved upon this measurement \citep{Nordtvedt1971}.  

Tests of GR moved out of the solar system upon the discovery of the Hulse-Taylor pulsar.  This system, PSR 1913+16, is composed of two rapidly rotating neutron stars and the orbit was found to decay at a rate of 1.01 $\pm$ 0.01 times the GR prediction \citep{Taylor1989}.  GR predicted that this decay was a result from the production of gravitational waves which would carry off some of the energy, which would result in this system requiring a smaller orbit (see \citet{Will2006} and references therein). While the Hulse-Taylor pulsar provided the first indirect detection of gravitational waves, it wasn't until this year that the first direct detection of gravitational waves was announced.  In February, 2016, it was announced that LIGO, the Laser Interferometer Gravitational-Wave Observatory, made its first detection of a gravitational wave.  This event, GW150914, was found to correspond to the merger of two black holes with masses of $36^{+ 5}_{-4} M_{\odot}$ and $29 \pm 4 M_{\odot}$ resulting in a final black hole of mass $62 \pm 4 M_{\odot}$ with $3 \pm 0.5 M_{\odot} c^2$ being radiated away in the form of gravitational waves with a second event detected later that year \citep{Abbott2016,Abbott2016b}.

\subsection{General Relativity in the Strong Gravity Regime}
Section \ref{Introduction:testingGR} details the significant successes made in testing the predictions of GR since its introduction.   However, a majority of these experiments are only testing the predictions of GR in the weaker gravitational fields found in our solar system.  Even tests performed with the Hulse-Taylor pulsar are still unable to probe the extreme gravitational fields predicted by GR such as those present around BHs. 
The strength of a gravitational field is quantified through the potential 
\begin{equation}
\epsilon \equiv \frac{GM}{rc^2}
\end{equation}
where the strong gravity regime corresponds to a larger value for $\epsilon$.  At the event horizon of a Schwarzschild BH $\epsilon = 1$ as opposed to $\epsilon \sim 10^{-11}$ around the moon and $\epsilon \sim 5 \times 10^{-6}$ for the Hulse-Taylor pulsar. While the first direct detection by LIGO of gravitational waves from a binary BH merger made progress in this regard, there is still much work to be done to fully understand the behavior of gravity in the strong gravity regime (see \citet{Psaltis2008} for a review of the subject).  To this end, the bulk of the work presented in this thesis makes use of the No-Hair theorem of GR.  This theorem states that a BH can only be defined by up to three parameters: mass ($M$), spin ($a$), and in some cases charge ($Q$) \citep{MTW}.  In order to test this important theorem, metrics have been introduced which contain additional parameters that violate the No-Hair theorem of GR.  These metrics can then be used to study how observations of BHs can be used to constrain potential deviations from GR (see \citet{Johannsen2016} and references therein). The work presented here examines four non-GR metrics \citep{Johannsen2011a,Aliev2005,Glampedakis2006,Pani2011} to determine the effect deviations from GR has on combined X-ray spectral, timing, and polarization signatures \citep{Hoormann2016}.

\section{Black Hole Models}			
	\subsection{Thin Disk Model} \label{Introduction:thinDisk}
		
	In 1973, Shakura and Sunyaev introduced a Newtonian model to describe the emission from the gas forming the accretion disk which surrounds BHs.  These calculations were then extended by \citet{Novikov1973} to include the effects from GR and became the standard model (referred to as the NT model) to describe the thermal emission from BH accretion disks.  In the NT model, the accretion disk is described as being geometrically thin and optically thick resulting from an axisymmetric, radiatively efficient accretion flow.  The local radiative flux can then be calculated at each radius for a given black hole of mass, spin, and mass accretion rate, $\dot{M}$, resulting in a thermal, black-body like spectrum.  However, there are many conditions that must be met in order for the NT model to hold.  These conditions are described in \citet{McClintock2011} and references therein and are summarized in the following.  First, it is important to note that this model can only be used to describe BHs that exhibit a strong thermal component. The next assumption is that the inner edge of the accretion disk is truncated at the location of the radius of the innermost stable circular orbit, $r_{ISCO}$.  This assumption is valid if there are only geodesic forces in the midplane of the disk but needs to be re-evaluated in the presence of magnetohydrodynamics (MHD).  Lastly, the NT model requires that the torque vanish at the $r_{ISCO}$.  It has been found that this assumption is valid for thin disks where $H/R$ $\ll$ 1 with $H$ corresponding to the height of the disk, but again is a condition that must be reconsidered in the presence of MHD \citep{Afshordi2003}.
	
\subsection{Corona Models} \label{sec:Corona}
	
\begin{figure}
        \centering
        \begin{subfigure}[b]{0.8\textwidth}
   				 \includegraphics[width=\textwidth]{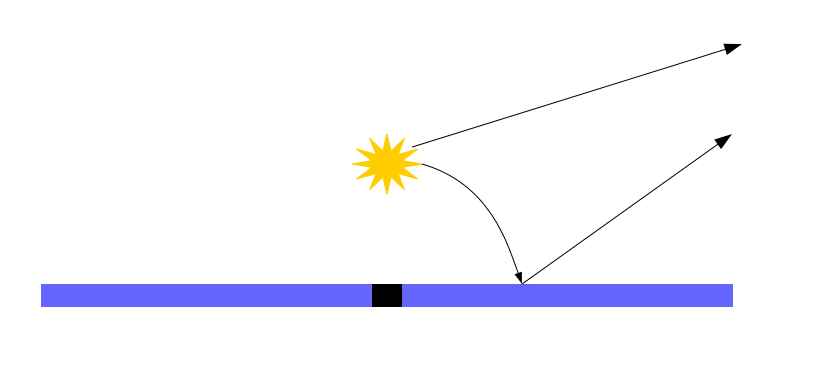}
                \caption{\label{fig:CoronaDiagrama} Lamp-post corona}
        \end{subfigure} 
        	\quad
        \begin{subfigure}[b]{0.8\textwidth}
   				 \includegraphics[width=\textwidth]{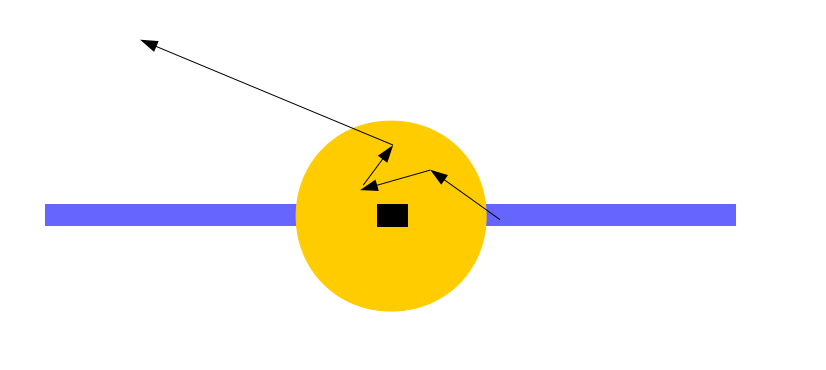}
                \caption{\label{fig:CoronaDiagramb} Spherical corona}
        \end{subfigure}
        \quad
        \begin{subfigure}[b]{0.8\textwidth}
   				 \includegraphics[width=\textwidth]{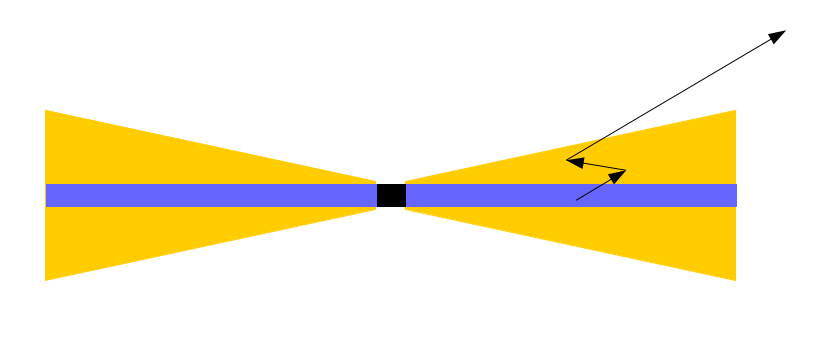}
                \caption{\label{fig:CoronaDiagramc} Wedge corona}
        \end{subfigure}
        \caption{\label{fig:CoronaDiagram} Schematic diagrams of different corona models where the black region represents the BH, the blue represents the accretion disk, and the yellow region is the corona. Arrows represent sample photon trajectories.}
\end{figure}
	
The NT thin disk model provides a good description for some scenarios but it does not explain all of our observations.  For example, the NT model characterizes the thermal emission from the disk but does not explain the observed power-law component seen in BH spectra.  This requires the existence of a corona of hot electrons in which the thermal photons from the disk can comptonize \citep{Thorne1974, Katz1976}.  It is unclear how exactly this corona is formed and therefore what its resulting geometry is \citep{Reynolds2003}.  The simplest and most commonly implemented model is the lamp-post corona where photons are emitted from a point source on the rotation axis above the black hole and irradiates the accretion disk as shown in Figure \ref{fig:CoronaDiagrama}.  While this is a simple model it does produce results which fit the observations (see e.g. \citet{Dovciak2004,Dovciak2012, Matt1991, Wilkins2012}) including recent X-ray reverberation results \citep{Cackett2014, Uttley2014}.  A lamp-post corona might result from hot gas accumulating in the lower density region above the BH or from internal shocks at the base of a jet.   Competing models have been introduced which describe extended corona.  In the spherical corona (Figure \ref{fig:CoronaDiagramb}) the accretion disk is truncated at some radius larger than $r_{ISCO}$ and within that point the BH is surrounded by a sphere of hot gas \citep{Schnittman2010}.  This model can be physically motivated by the existence of Advection Dominated Accretion Flow (ADAF) in which the region within the truncation radius is filled with hot gas because instead of being radiated away the energy released from viscous dissipation remains in the disk.  In this case the corona will have temperatures of 100-300 keV with optical depths ranging from 0.1-1 \citep{McClintock2006}. Another model for the corona is the wedge geometry (Figure \ref{fig:CoronaDiagramc}) where the accretion disk is sandwiched by a corona with a similar temperature and optical depth as in the spherical scenario \citep{Schnittman2010}.  Observations of the active galactic nuclei (AGN) Markarian 335 indicate corona geometry can evolve over time from being a large corona similar to the spherical geometry to being a compact, vertically collimated corona above the BH itself \citep{Wilkins2015}. There is also evidence to suggest that most AGN corona have a temperature near the maximum value allowed by pair production \citep{Fabian2015}.

\subsection{General Relativistic Magnetohydrodynamic Simulations} \label{GRRMHD}

While the methods described above provide a way to model geometry of BH corona, they are still primarily phenomenological in nature.  In order to model the BH accretion disk from first principles, simulations were introduced which solve the full MHD equations present in the disk.  Great strides have been made over the past several years with these simulations, which include the full GR calculations and the magneto-rotational instability which is necessary to explain the existence of the plasma viscosity required for accretion.  These became known as GRMHD - General Relativistic Magnetohydrodynamic - simulations and those which also track the emission from the disk are referred to as GRRMHD - General Relativistic Radiative Magnetohydrodynamic- simulations.  For example, the code HARM was introduced to calculate the gas density, velocity, and viscous heating rate for a disk which is kept thin using an artificial cooling prescription \citep{Gammie2003,Noble2006,Noble2011}.  In order to model supercritial accretion rates in a geometrically thick disk, another code, Koral, was then introduced  \citep{Sadowski2013,Sadowski2014}.  As a way to model the emission from both Koral and HARM, the post processor HEROIC (the Hybrid Evaluator for Radiative Objects Including Comptonization) was created to provided a multi-dimensional way to calculate the emission from the accretion disks described by these models  in both optically thin and thick environments while including the comptonization of the photons \citep{Narayan2016}. Code has also been developed to do similar calculations for the relativistic radiation transport in the presence of an optically thick disk and thin corona while including the resulting polarization of the emission \citep{Schnittman2013a, Schnittman2013b}. GRMHD simulations have shown that the NT model is valid for systems with luminosities less than 30 \% of the Eddington limit \citep{Kulkarni2011,Penna2012}.
		
\section{Observational Signatures}
	\subsection{Spectral Observations}
	
		\begin{figure}
			\begin{center}
        		\vspace{0pt}
   			 \includegraphics[width=.8\textwidth]{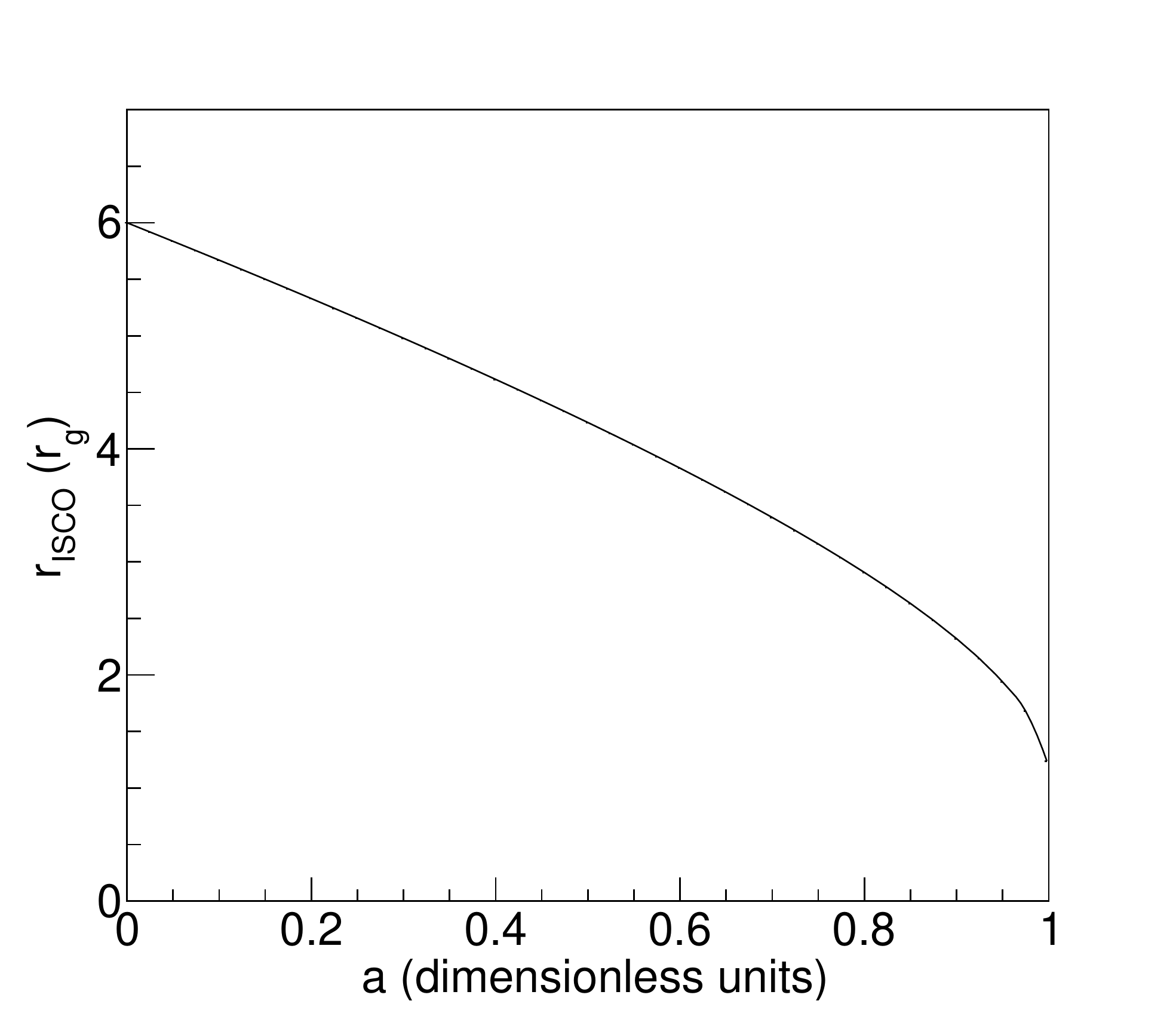}
   			\vspace{10pt}
        		\caption{\label{fig:riscova} $r_{ISCO}$ as a function of spin ($a$).}
        		\end{center}
		\end{figure}
	
\begin{figure}
			\begin{center}
        		\vspace{0pt}
   			 \includegraphics[width=.9\textwidth]{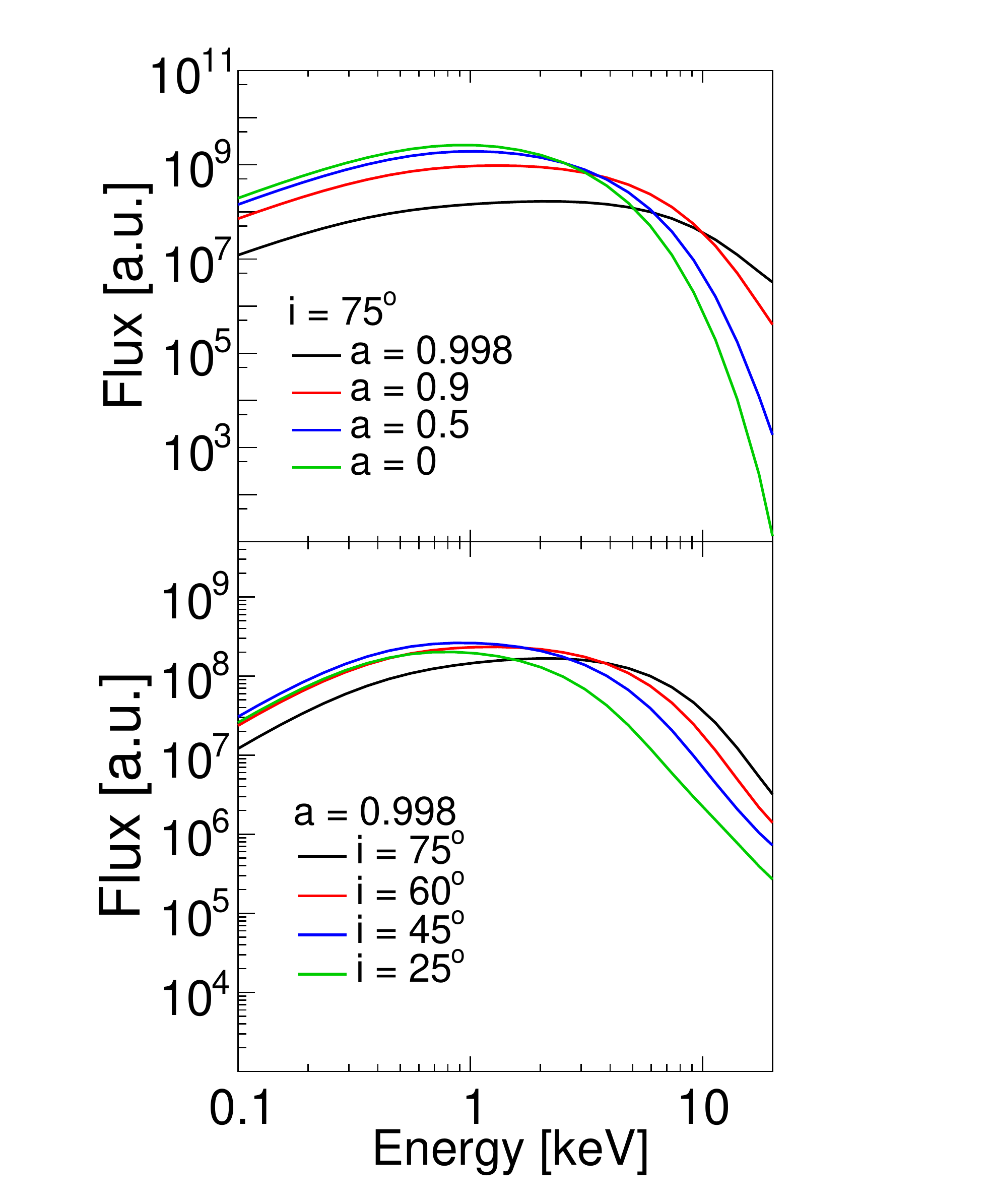}
   			\vspace{10pt}
        		\caption{\label{fig:ContinuumFlux} Thermal disk emission illustrating the effects on the flux of the thermal emission as a function of energy while varying both spin (top panel) and inclination (bottom panel). The flux is given in arbitrary units having been calculated assuming $G = c = \hbar = 1$ with distances measured in terms of the gravitational radius $r_g$. }
        		\end{center}
		\end{figure}

	Spectral observations of BHs reveal three main components thought to originate close to the BH: the thermal emission, broadened Fe-K$\alpha$ line, and the Compton hump.  The thermal emission component is seen in X-ray observations of stellar mass BHs.  The temperature of the accretion disk scales like $ T \propto M^{-1/4}$ and leads to blackbody emission in the X-ray band (BHBs) and in the UV (SMBHs).  This "blue bump emission" is difficult to observe owing to the presence of a large number of emission lines contaminating the signal and will not be discussed further in this thesis (see e.g. \citet{Gilfanov2014}).  The thermal emission from accretion disks surrounding BHBs is modeled using the NT prescription described in Section \ref{Introduction:thinDisk}. Since the $r_{ISCO}$ is thought to correspond to the inner edge of the disk, none of the thermal emission will emanate from within this point.  This results in a thermal continuum flux which depends on the $r_{ISCO}$.  Because the $r_{ISCO}$ is a monotonic function of spin, the observation of the continuum emission provides a way to measure spin in these systems.  The relationship between $r_{ISCO}$ (measured in units of gravitational radius $r_g = GM/c^2$) and spin, $a$, is illustrated in Figure \ref{fig:riscova}. This method of determining the spin of BHBs requires that the  $M$, distance to the source, $D$, and inclination of the observer, $i$, be known in advance and also requires that the emission be dominated by the thermal emission.  Because $M, i$ , and $D$ must be known in order to perform the continuum fitting method the uncertainties in these values dominate the uncertainty in the determined value for spin (see e.g. \citet{Gou2011,Steiner2011}).  It is also assumed that the accretion disk is described by the NT model (see e.g.\citep{McClintock2014} for a review of the topic).  The dependence of the continuum emission on spin and inclination can be seen in Figure \ref{fig:ContinuumFlux} where the flux vs. energy spectra are shown for various values of $a$ in the top panel and various $i$ values in the bottom panel. These results along with the other figures presented in this chapter were generated using our GR ray-tracing code which will be described in Chapter 2. As mentioned previously in this chapter, the NT model makes several assumptions which could potentially prove invalid when considering the presence of MHD.  To address this issue \citet{Kulkarni2011} examined the strength of the continuum fitting method using GRMHD simulations and concluded that while these simulations do not yield the exact same values for spin as those obtained with the NT model, they are similar enough that it is not currently the limiting factor when measuring BH spin using this method.

\begin{figure}
			\begin{center}
        		\vspace{0pt}
   			 \includegraphics[width=1\textwidth]{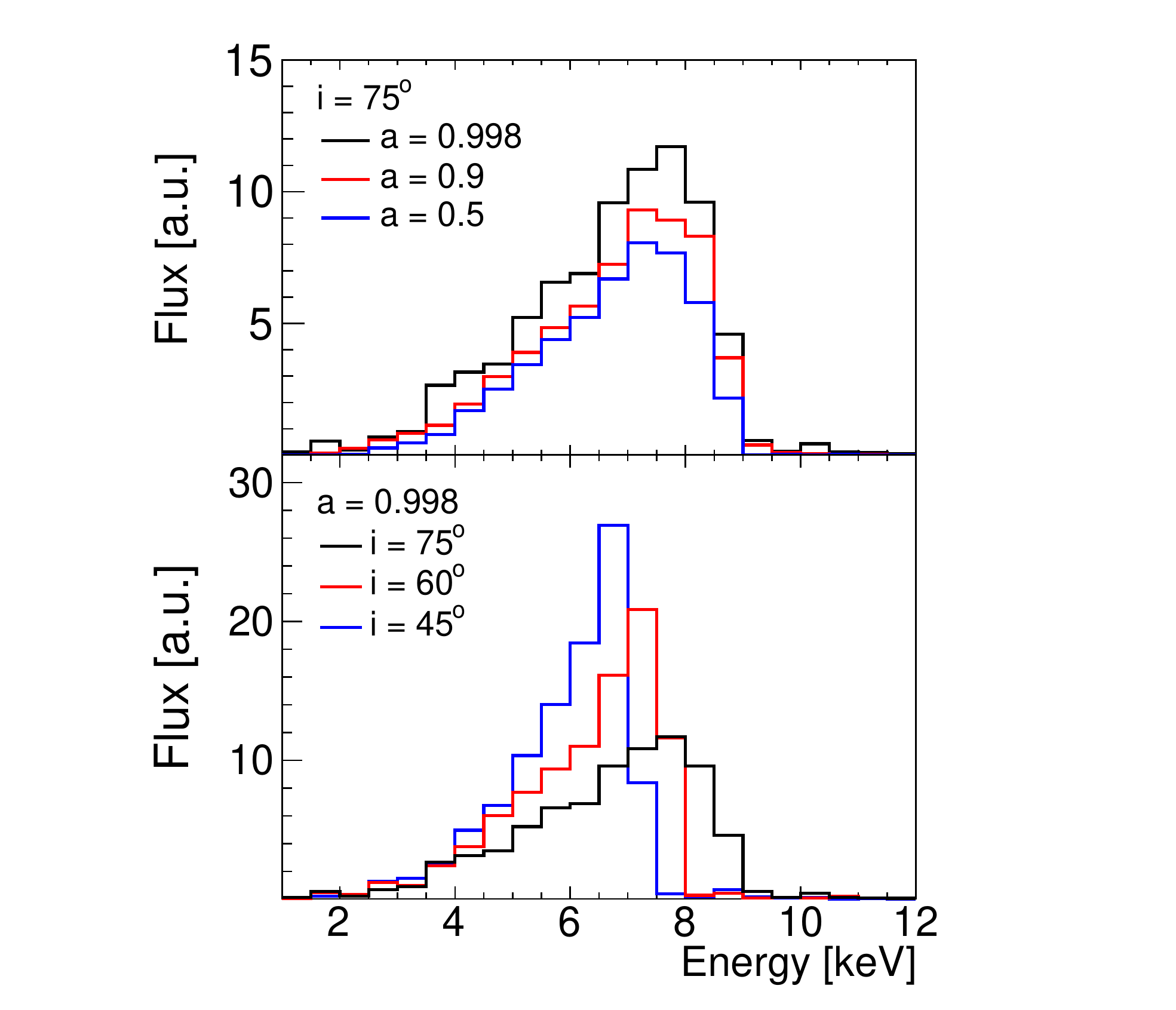}
   			\vspace{10pt}
        		\caption{\label{fig:IronFlux} Fe-K$\alpha$ line illustrating the change seen by varying both spin (top panel) and inclination (bottom panel) for a lamp-post at a height of $h =$7 $r_g$.}
        		\end{center}
		\end{figure}

The other main spectral emission component is the reflected power-law component.  This component is created due to the Comptonization of disk photons in the corona leading to an emission with an energy profile of $E^{-\Gamma}$ where $\Gamma$ is the spectral index.  This emission irradiates the disk causing some photons to then reflect off of the accretion disk.  Fluorescence gives rise to the prominent Fe-K$\alpha$ line and scattering produces the so-called Compton hump.  The iron line is the strongest line observed in the X-ray spectra of BHs ranging in energy from 6.40-6.97 keV.  This line is created because when a photon hits the accretion disk it can be reprocessed into a fluorescent line with the Fe-K$\alpha$ line emitted at 6.4 keV being the most common for typical Fe ionization states in BHBs and SMBHs.  While the line itself is narrow it is broadened due to gravitational redshift, Doppler effects, and relativistic beaming.  Instead of creating a fluorescent line it is also possible for the photon to be destroyed by Auger de-excitation or scatter of the disk. At higher energies, above around 20 keV, Compton recoil reduces the backscattered emission leading to the creation of the Compton hump (\citet{Fabian2000} for a review of the subject).  Because the shape of the Fe-K$\alpha$ line is so significantly affected by the gravitational effects it provides a way to measure the spin and inclination of both BHBs and SMBHs.  An example of this is shown in Figure \ref{fig:IronFlux}.  The top panel shows the shape of the Fe-K$\alpha$ line when varying the BH spin.  As $a$ increases the gravitational effects increase the broadening of the red wing of the line.  The blue wing of the line is more significantly affected by the inclination of the observer with respect to the BH as seen in the bottom panel.  It is worth noting that like with the continuum fitting method, using Fe-K$\alpha$ signatures to determine BH spin requires the assumptions that the disk is geometrically thin, optically thick, as well as being radiatively efficient down to the $r_{ISCO}$ (see e.g. \citet{Reynolds2014} and references therein). Early uses of this method include studying the Fe-K$\alpha$ observed in the SMBH MCG-6-30-15 \citep{Iwasawa1996, Wilms2001} and the BHB XTE J1650-500 \citep{Miller2002}.
		
\subsection{Timing} \label{Intro:timing}
\subsubsection{Quasi-Periodic Oscillations: BHBs}
In addition to spectral observations, timing signatures have proven to be a useful tool to further study BHs.  An example of these are QPOs, Quasi-Periodic Oscillations in the light curves of BHBs.  Low-frequency QPOs (LFQPOs) are high amplitude, high coherence signals which vary on time scales of less than one minute.  It is thought that these LFQPOs could be related to the flow of the accreting matter in the disk as the frequency and amplitudes are correlated with the parameters describing both the thermal and power-law spectral components while being relatively stable (see \citet{Remillard2006} and references therein). One possible explanation for these observations is the presence of Lense-Thirring precession, an effect of GR where the spacetime around a spinning object becomes twisted, resulting in a precession of the matter in the inner accretion disk \citep{Miller2005, Schnittman2006,Ingram2009}.  Combining LFQPO observations with high energy X-ray polarimetry observations could provide a way to verify this theory \citep{Ingram2015}.  High-Frequency QPOs (HFQPOs) have been observed in multiple sources.  It has been seen that multiple QPO frequencies are often observed in a single source which are in a 3:2 ratio. These HFQPOs are of particular interest because they originate near where the $r_{ISCO}$ is expected and they don't shift in frequency in spite of changes in luminosity.  Once a model is known which describes these oscillations it will be possible to use HFQPOs to determine the spin of a BH whose mass is known regardless of $i$ and $D$ \citep{Narayan2005,Remillard2006}.  One of the proposed explanations for these signatures is the existence of a hot spot in the accretion disk orbiting the BH \citep{Schnittman2004,Schnittman2005,Stella1998, Stella1999, Abramowicz2001, Abramowicz2003}.  The polarization signatures of the hot spot model for HFQPOs has also been studied in \citet{Beheshtipour2016}.

\begin{figure}
        \centering
        \begin{subfigure}[b]{0.6\textwidth}
   				 \includegraphics[width=\textwidth]{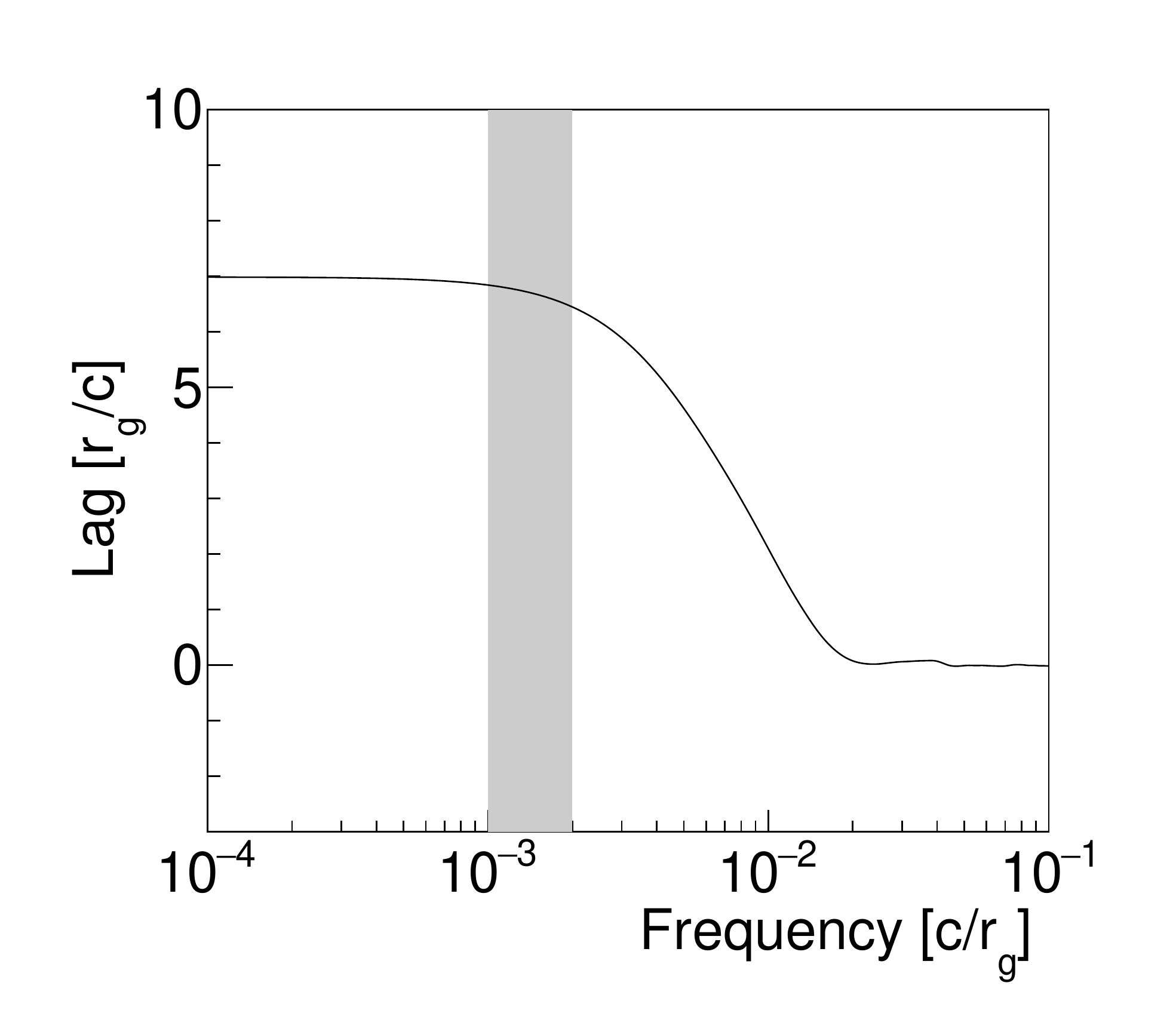}
                \caption{\label{fig:lagfreq} Lag-Frequency spectrum.}
        \end{subfigure} 
        	\quad
        \begin{subfigure}[b]{0.6\textwidth}
   				 \includegraphics[width=\textwidth]{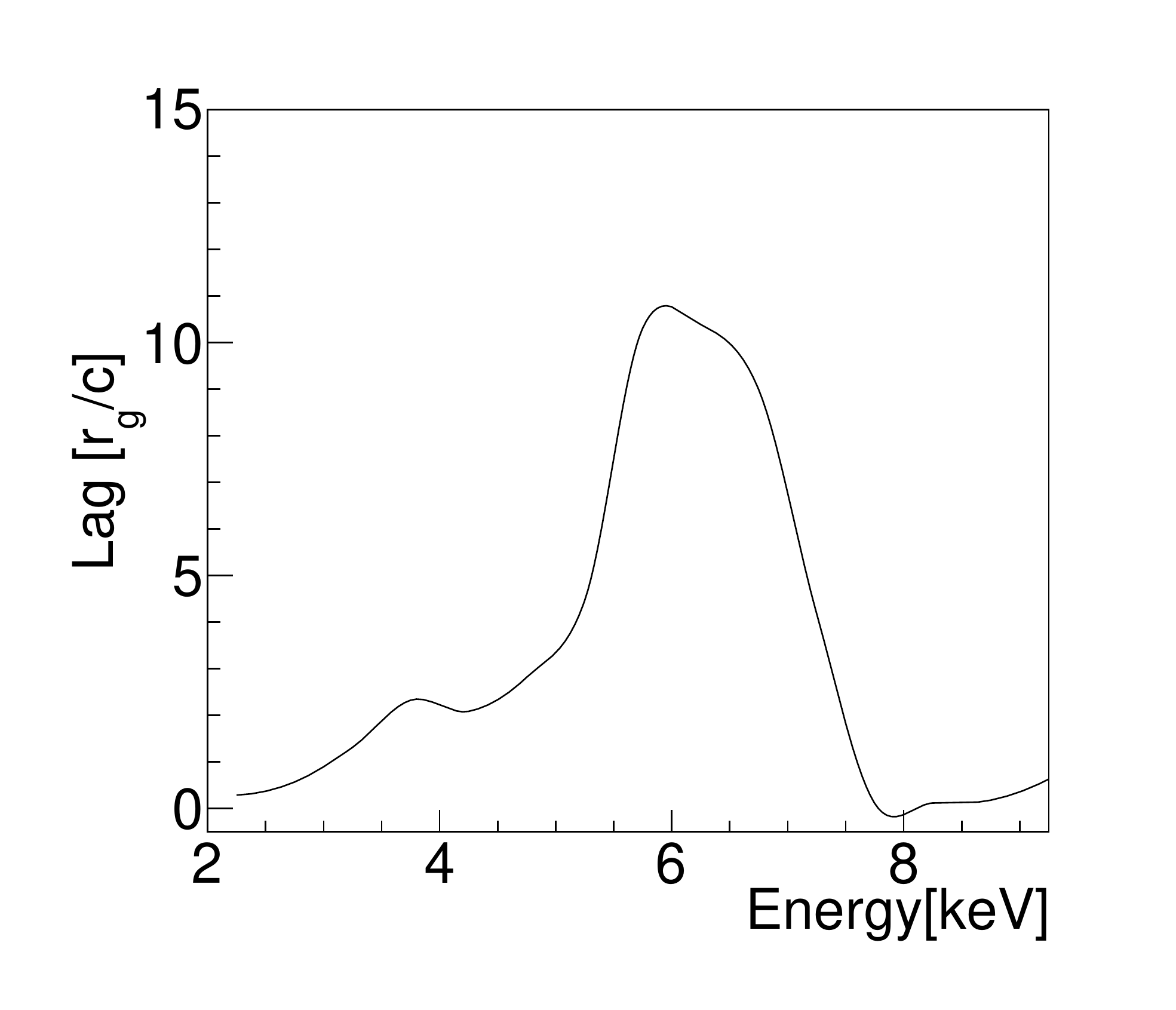}
                \caption{\label{fig:lagenergy} Lag-Energy spectrum calculated in the frequency range indicated by the grey band in Figure \ref{fig:lagfreq}}
        \end{subfigure}
        \caption{\label{fig:reverb} Iron K reverberation for a BH with $a=$0.998, $h =$ 5 $r_{g}$, $i =$ $45^{o}$.}        
\end{figure}

\subsubsection{Reverberation: SMBHs}		
In recent years X-ray reverberation has come into its own as a powerful tool to dissect the accretion flows of SMBH.  Originally, reverberation was used to observe changes in the UV continuum seen in the broad line region of unobscured AGN which was then used to study the geometry and mass of the AGN \citep{Blandford1982}.  In order to begin to probe the inner regions of accretion flow the lag between the soft excess and the harder continuum was studied.  However, this method proves to be difficult for multiple reasons including the large number of lines observed in the soft excess, the presence of absorption along with the tail of the disk's black body radiation, and the higher signal to noise ratio at lower energies.  Significant progress has been made to extend the use of reverberation to study the inner regions of accretion flow and the geometry of the corona through the discovery of the lag between the direct corona emission and the Fe-K line.  The Fe-K line has the benefit of being free from contamination from other line and emission components.  It has been found that Fe-K reverberation can be used to study different regions of the accretion disk being that the red wing of the line responded on shorter time scales from inner radii followed after a longer time by the rest frame of the line coming from larger radii.  The observations also indicate that the corona in the BHs producing these signals must be compact (see \citet{Uttley2014} for a review).   Using theoretical ray-tracing simulations to model the observed Fe-K reverberation, it has been shown that the lamp-post model fits the data very well and can be used to constrain the height of the lamp-post and the inclination of the BH \citep{Cackett2014,Wilkins2016}. While the methods used to calculate these signatures will be described in detail in later chapters, an example of the lag-frequency (Figure \ref{fig:lagfreq}) and lag-energy (Figure \ref{fig:lagenergy}) spectra  are shown comparing the direct emission to the Fe-K emission.  Even more recently, a lag between the direct emission and the emission forming the Compton hump has been observed \citep{Zoghbi2014, Kara2015}.  In particular, a larger delay has been seen at higher energies indicating interactions with outer regions of the disk, which also implies that the ionization of the accretion disk is not uniform.  A study showing how GR ray-tracing simulations can be used to model X-ray reverberation to study the structure and ionization of the accretion disk will be presented in Chapter 4.
		
\subsection{Polarization}

While both spectral and timing observations can provide valuable information about the structure of the accretion flow, X-ray polarization gives complimentary information about these regions around BHs which are too small to image directly.  The polarization of emission from an optically thick medium has been extensively studied by \citet{Chandrasekhar1960}.  Once emitted from the disk the photons can either scatter in a corona or  follow the spacetime curvature of the black hole, return to the disk and scatter off the disk with each scattering drastically changing the polarization of the photon.  In \citet{Li2009} it was shown that polarization can be used to determine the inclination of the inner disk.  Furthermore, polarization can be used to measure the BH spin \citep{Schnittman2009} to detect the warping of the disk due to Lense-Thirring precession \citep{Ingram2015}, and a misalignment between the inner disk and the orbital plane \citep{Krawczynski2015}.  Additionally, polarization can be used to constrain the geometry of the corona \citep{Schnittman2010,Dovciak2011}. 

\subsection{X-ray Missions} 

Numerous X-ray missions have been launched to observe the emission from BHs with several additional missions being planned.  Current missions which observe the X-ray spectral and timing signatures include \textit{Chandra}, \textit{XMM-Newton}, \textit{Swift}, and \textit{NuSTAR}.  \textit{Chandra} and \textit{XMM-Newton} are both low energy X-ray missions.  \textit{Chandra} is a NASA satellite launched in July 1999, operating from 0.1-10 keV \citep{Weisskopf1999} while \textit{XMM-Newton} was launched by the European Space Agency (ESA) in December 1999 \citep{Cottam2001}. \textit{Swift}, a mission designed to study gamma-ray bursts, was launched by NASA in 2004 as part of their Medium Explorer (MIDEX) program.  It was designed to perform spectroscopy from 180-600 nm and 0.3-150 keV \citep{Hurley2003}.  \textit{NuSTAR} launched in June 2012 as a NASA Small Explorer Mission (SMEX) which is a focusing high energy telescope working in the energy range from 3-79 keV \citep{Harrison2013}. ESA is currently working on a new satellite, \textit{Athena}, which will observe 0.3-12 keV X-rays with an estimated launch date of 2028 \citep{Barret2013}.  There are no space-borne X-ray polarimetry missions currently operational.  The last X-ray polarimetry satellite flown was \textit{OSO-8}, launched in June 1975 \citep{Weisskopf1976}.  However, there are currently three missions under consideration.  NASA is currently studying two low energy X-ray polarimeters as part of its SMEX call including the imaging x-ray polarimeter explorer, \textit{IXPE} \citep{Weisskopf2014}, and the polarimeter for relativistic astrophysical X-ray sources, \textit{PRAXyS} \citep{Jahoda2015}.  ESA is also considering the X-ray imaging polarimetry explorer (\textit{XIPE}) for an upcoming mission to detect low energy X-ray polarization \citep{Soffitta2013}.

\section{What Will Follow}
The rest of this thesis will detail the research I performed using ray-tracing simulations to further study the inner regions of BH accretion flow and to constrain deviation from GR. Chapter 2 will describe the GR ray-tracing code that I used and refined to model the thermal, timing, and polarization X-ray signatures of stellar and supermassive black holes.  This includes a description of the modeling of both the thermal disk emission and the power-law emission from a lamp-post corona. In Chapter 3 I will discuss the implementation of four non-GR metrics into the ray-tracing code and the resulting simulated X-ray observations.  Using these simulations it is possible to constrain a large portion of potential deviations from GR.  The results from my studies showing the dependence of X-ray reverberation signatures on the ionization of the accretion disk will be presented in Chapter 4.  As X-ray polarization promises to be a valuable tool to further understand these systems, I will then go on to discuss the work I did on the alignment of the balloon-borne hard X-ray polarimetry experiment X-Calibur in Chapter 5.  In Chapter 6 I will conclude by discussing how a satellite version of X-Calibur, \textit{PolSTAR}, could be used to study properties of BHs such as their spin, the orientation of the inner disk, and the geometry of the corona.
	
%%% Local Variables: 
%%% mode: latex
%%% TeX-master: "thesis-main"
%%% End: 

\chapter{General Relativistic Ray-Tracing Code}
\label{RayTracingCode}

\section{Ray-Tracing} 

\begin{figure}			
	\begin{center}
        		\vspace{0pt}
   			 \includegraphics[width=.8\textwidth]{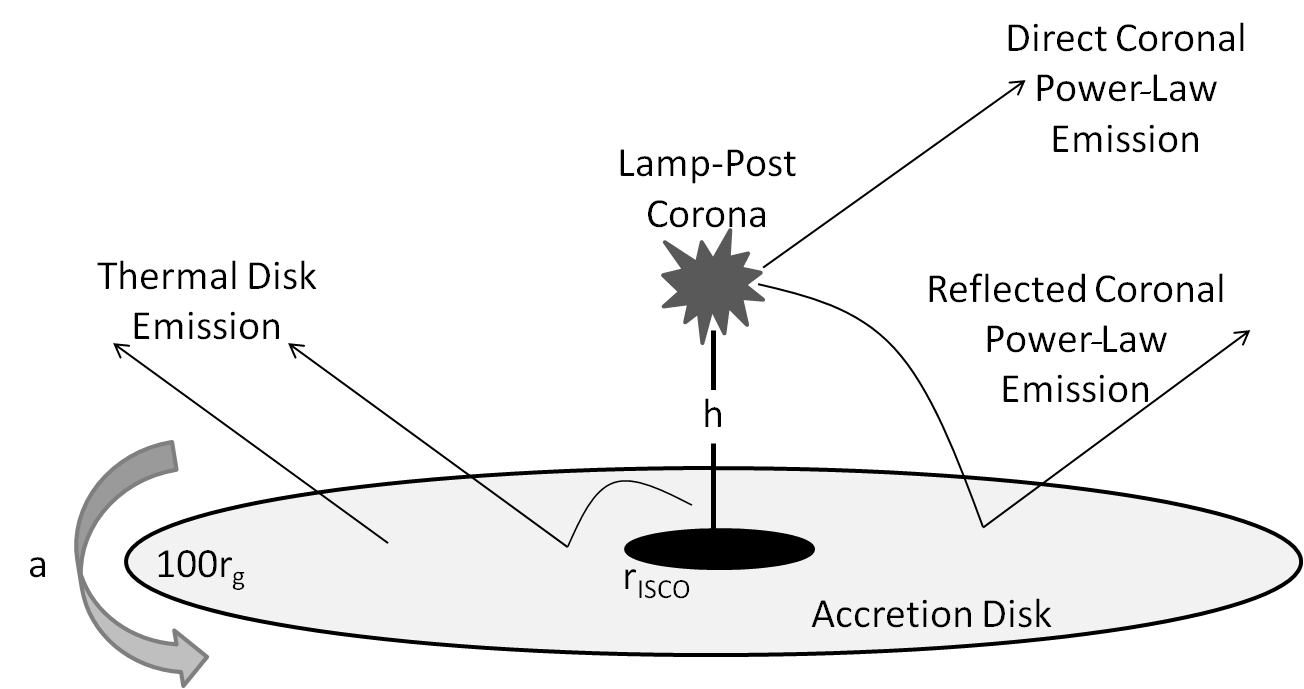}
   			\vspace{10pt}
        	\caption{\label{fig:Diagram} Diagram of the simulated emission. This diagram was taken from Figure 1 of \citet{Hoormann2016}}
    \end{center}
\end{figure}

	The results presented in this thesis are obtained through the use of a GR ray-tracing code.  This code was originally developed by \citet{Krawczynski2012} to model the X-ray thermal spectral and polarization signatures from mass accreting stellar mass BHs.  This code was further modified in \citet{Hoormann2016, Hoormann2016b} to simulate the X-ray power-law emission from stellar and supermassive BHs which leads to the broadened Fe-K$\alpha$ and the Compton hump, which can be used to study spectral, polarization, and reverberation observations.  Figure \ref{fig:Diagram} illustrates the emission that will be simulated with this code. In addition to performing these calculations in general relativity the simulations can also be performed assuming four alternative theories of gravity.  These simulation tools are also used to study quasi-periodic oscillations using the hotspot model \citep{Beheshtipour2016} and model extended corona geometries \citep{Beheshtipour2016b}.  This code is detailed below.
	
\subsection{Solving the Geodesic Equation}

\begin{figure}
    	\begin{center}
		\begin{subfigure}[b]{0.8\textwidth}
   			 \includegraphics[width=\textwidth]{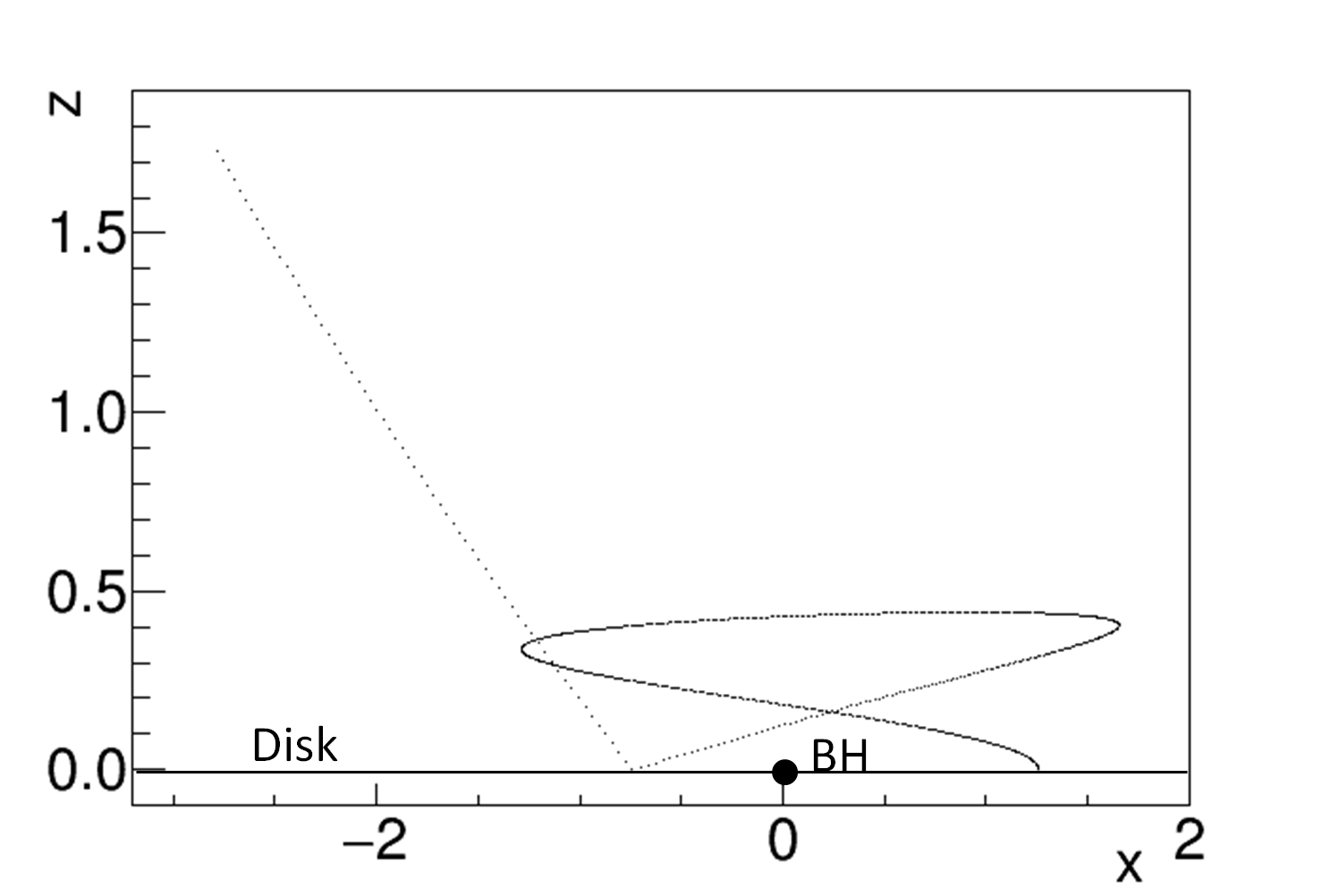}
   			\caption{\label{fig:trackThermal} Thermal disk emission from a rapidly rotating BH.}
   		\end{subfigure}
   		\quad
		\begin{subfigure}[b]{0.8\textwidth}
   			 \includegraphics[width=\textwidth]{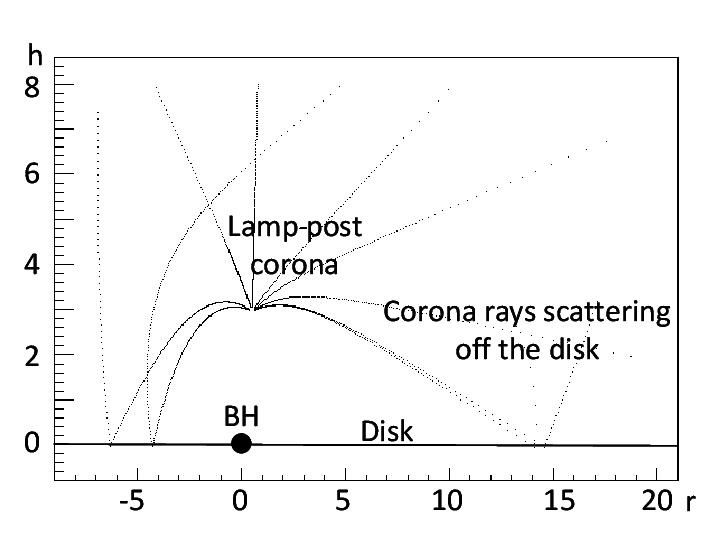}
   			\caption{\label{fig:trackPL} Power-law emission from a lamp-post at height of 3 $r_g$.}
   		\end{subfigure}
    \end{center}
    \caption{\label{fig:trackDiagram} Tracks of the thermal disk emission (Figure \ref{fig:trackThermal}) and power-law emission (Figure \ref{fig:trackPL}) calculated with the ray-tracing code.}
\end{figure}
	The X-ray photons are emitted from either the accretion disk or the corona and tracked until they reach the observer.  First, the mechanisms used to track the photons will be described and then it will be discussed how this can be applied to study the thermal and power-law spectral components.  The accretion disk is defined to range from the $r_{ISCO}$ out to 100 $r_g$ where
	\begin{equation}
		r_g = \frac{G M}{c^2}.
	\end{equation}
The photons are tracked until they reach the observer, located at 10,000 $r_g$, or until they reach the event horizon and it is assumed the photon falls into the BH.  Each photon is tracked by integrating the geodesic equation
	\begin{equation} \label{eq:geodesic}
	\frac{d^2 x^\mu}{d\lambda^{'2}} = -\Gamma^\mu_{\ \sigma \nu} \frac{dx^\sigma}{d \lambda^{'}} \frac{d x^\nu}{d \lambda^{'}}.
	\end{equation}
 which describes an external distance between two points in curved spacetime.  In the global coordinate frame (GC) $x^\mu = (t,r,\theta,\phi)$ and the basis vectors are $\textbf{e}_\mu \equiv \partial/\partial x^\mu$.  Here $\lambda^{'}$ is the affine parameter which is a multiple of the proper time and  $\Gamma^\mu_{\ \sigma \nu}$ are the Christoffel symbols \citep{MTW}. The geodesic equation is integrated using the fourth order Runga Kutta method (see e.g. \citet{Press2002,Psaltis2012}).   Because this  equation is written in terms of the Christoffel symbols it is straightforward to solve the geodesic equation assuming any metric allowing us to track photons in various spacetimes.  Once the metric is known the Christoffel symbols, which describe the gravitational force field, are calculated using the following relationship \citep{DInverno1992}
\begin{equation}
\Gamma^\alpha_{\beta \gamma} = \frac{1}{2} g^{\alpha \mu}(\partial_\beta g_{\mu \gamma} + \partial_\gamma g_{\mu \beta} - \partial_\mu g_{\beta \gamma}).
\end{equation}
Figure \ref{fig:trackDiagram} shows the results of these calculations with Figure \ref{fig:trackThermal} showing an example path of a thermal photon being emitted from the accretion disk and Figure \ref{fig:trackPL} showing the trajectories of power-law photons emitted from a lamp-post corona.

\subsection{Modeling Polarization}		
As the photon is tracked the polarization vector, $f^\mu$, is parallel transported along with it using the equation
	\begin{equation}
	\frac{df^\mu}{d \lambda^{'}} = -\Gamma^\mu_{\ \sigma \nu} f^\sigma \frac{dx^\nu}{d \lambda^{'}}.
	\end{equation}
The polarization of each photon is determined through the use of Stokes parameters which possess the property that the Stokes parameters can be summed for each photon to determine the polarization of the total emission.  This makes Stokes parameters particularly useful in determining polarization of light in both simulations and in polarization experiments (see e.g. \citet{Kislat2015}).  The Stokes parameters are a set of four values which describe the polarization of a photon including
\begin{itemize}
	\item I: intensity of the beam
	\item Q: polarization with respect to an orthogonal set of axis $x$,$y$
	\item U: polarization with respect to axis $x$',$y$' obtained by rotating $x$,$y$ by $45^{o}$
	\item V: circular polarization
\end{itemize}
These can be written in terms of the time averaged electric field of a quasi-monochromatic wave packet projected along two perpendicular axes $\hat{x}$ and $\hat{y}$ such that
	\begin{eqnarray}
		I & = & \langle E_x^2 + E_y^2 \rangle \\ \nonumber
		Q & = & \langle E_x^2 - E_y^2 \rangle \\ \nonumber
		U & = & \langle 2 E_x E_y \cos \delta \rangle \\ \nonumber
		V & = & \langle 2 E_x E_y \sin \delta \rangle \nonumber
	\end{eqnarray} 
where $\delta$ is the lag of $E_y$ behind $E_x$.  The polarization fraction is then defined as
	\begin{equation}
	p = \frac{\sqrt{Q^2 + U^2 + V^2}}{I}
	\end{equation}
where $p$ = 1 for a 100\% polarized wave.  The Stokes parameters $Q$ and $U$ can be written in terms of the polarization angle $\psi$:
	\begin{eqnarray}
		Q & = & I\, p \cos 2 \psi \\ \nonumber
		U & = & I\, p\sin 2 \psi \nonumber
	\end{eqnarray} 
which implies
	\begin{equation}
	\psi = \frac{1}{2}\tan ^{-1} \frac{U}{Q}.
	\end{equation}
Because our simulations and experiments are only concerned with linearly polarized light we assume $V$ = 0 throughout the following (see e.g. \citet{McMaster1954, McMaster1961} for a review of Stokes parameters).  In order to determine the polarization of the initial emission from the disk and the change in polarization after scattering, this work makes use of the calculations presented in \citet{Chandrasekhar1960} where the polarization direction is measured with respect to the projection of the spin axis of the BH in the sky.  Table XXIV of \citet{Chandrasekhar1960} gives the parameters describing the polarization of photons emitted from an optically thick atmosphere while Table XXV describes the change in polarization upon scattering.
		
\subsection{Frame Transformations}
	In order to scatter a photon off of the accretion disk the wave vector of the photon must be transformed into the local inertial frame of the plasma orbiting the BH.  This Plasma Frame (PF) is defined in terms of orthonormal basis vectors (indicated with hats) where $\textbf{e}_{\hat{t}}$ is chosen to be parallel to the four-velocity of the orbiting particles.  Given the vector \textbf{p}, $p_\mu p^\mu = m^2$ where $m=1$ making $p^\mu$ a unit vector.  The basis vectors (a "tetrad") for the PF are defined as 
	\begin{eqnarray} \label{eq:gc_pf}
	\textbf{e}_{\hat{t}} & \equiv & \textbf{p} = p^t \textbf{e}_t+p^\phi \textbf{e}_\phi \\ \nonumber
	\textbf{e}_{\hat{r}} & \equiv & \textbf{e}_r/\sqrt{g_{rr}} \\ \nonumber
	\textbf{e}_{\hat{\theta}} & \equiv & \textbf{e}_\theta/\sqrt{g_{\theta \theta}} \\ \nonumber
	\textbf{e}_{\hat{\phi}} & \equiv & \alpha \textbf{e}_t + \beta \textbf{e}_\phi \nonumber
	\end{eqnarray} 
where $\textbf{e}_{i}$ are the basis vectors in the GC system and $\alpha$ and $\beta$ are determined by the orthonormal nature of the vectors through finding the positive solutions to the equations $\textbf{e}_{\hat{\phi}} \cdot \textbf{e}_{\hat{\phi}} = 1$ and $\textbf{e}_{\hat{t}} \cdot \textbf{e}_{\hat{\phi}} = 0$. This leads to the following expressions,
\begin{equation}
\alpha = \mp\frac{g_{\phi\phi}p^\phi+g_{t\phi}p^t}{\sqrt{g_{\phi\phi}g_{tt}(p^\phi)^2-g_{t\phi}^{\ \ 2}p^t(2g_{t\phi}p^\phi+g_{tt}p^t)+g_{\phi\phi}(-g_{t\phi}^{\ \ 2}(p^\phi)^2+2g_{t\phi}g_{tt}p^\phi p^t+g_{tt}^{\ \ 2}(p^t)^2)}} 
\end{equation}
and
\begin{equation}
\beta = \pm \frac{g_{t\phi}p^\phi+g_{tt}p^t}{\sqrt{-(g_{t\phi}^{\ \ 2} - g_{\phi\phi}g_{tt})(g_{\phi\phi}(p^\phi)^2+p^t(2g_{t\phi}p^\phi+g_{tt}p^t))}}.
\end{equation}
The transformation matrices are defined such that 
\begin{equation}
	\textbf{e}_{\hat{\mu}} = e^\nu_{\ \hat{\mu}} \textbf{e}_{\nu}
\end{equation}	
and 
\begin{equation}
	\textbf{e}_{\nu} = \bar{e}^{\hat{\mu}}_{\ \nu} \textbf{e}_{\hat{\mu}}.
\end{equation}
The transformation matrix $\bar{e}^{\hat{\mu}}_{\ \nu}$ can be used to transform each component of the wave vector from the GC to the PF yielding the following relationship,
\begin{equation} \label{eq:wave_transform}
	k^{\hat{\mu}} = \bar{e}^{\hat{\mu}}_{\ \nu} k^{\nu}.
\end{equation}	
After the scattering has been performed the wave and polarization vectors are transformed from the PF back to the GC by inverting Equation \ref{eq:wave_transform} until the photon either scatters again or reaches the observer.  Once the photon reaches the observer whose momentum is given by $\textbf{p}_r = \textbf{e}_t/\sqrt{g_{tt}}$ the vectors are transformed into a new set of basis vectors (marked with a tilde) defining the Coordinate Stationary (CS) frame which are given by
\begin{eqnarray}
	\textbf{e}_{\tilde{t}} & \equiv & \textbf{p}_r =  \textbf{e}_t/\sqrt{g_{tt}}  \\ \nonumber
	\textbf{e}_{\tilde{r}} & \equiv & \textbf{e}_r/\sqrt{g_{rr}} \\ \nonumber
	\textbf{e}_{\tilde{\theta}} & \equiv & \textbf{e}_r/\sqrt{g_{\theta \theta}} \\ \nonumber
	\textbf{e}_{\tilde{\phi}} & \equiv & \gamma \textbf{e}_t + \delta \textbf{e}_\phi. \nonumber
\end{eqnarray}
The constants $\gamma$ and $\delta$ are calculated in the same way as those in Equation \ref{eq:gc_pf} leading to 
\begin{equation}
\gamma = 0
\end{equation}
and
\begin{equation}
\delta = 1/\sqrt{g_{\phi\phi}}
\end{equation}

\section{Thermal Model}

\subsection{Emission}
\begin{figure}
	\begin{center}
		\vspace{-25pt}
       \begin{subfigure}[b]{0.65\textwidth}
   			 \includegraphics[width=\textwidth]{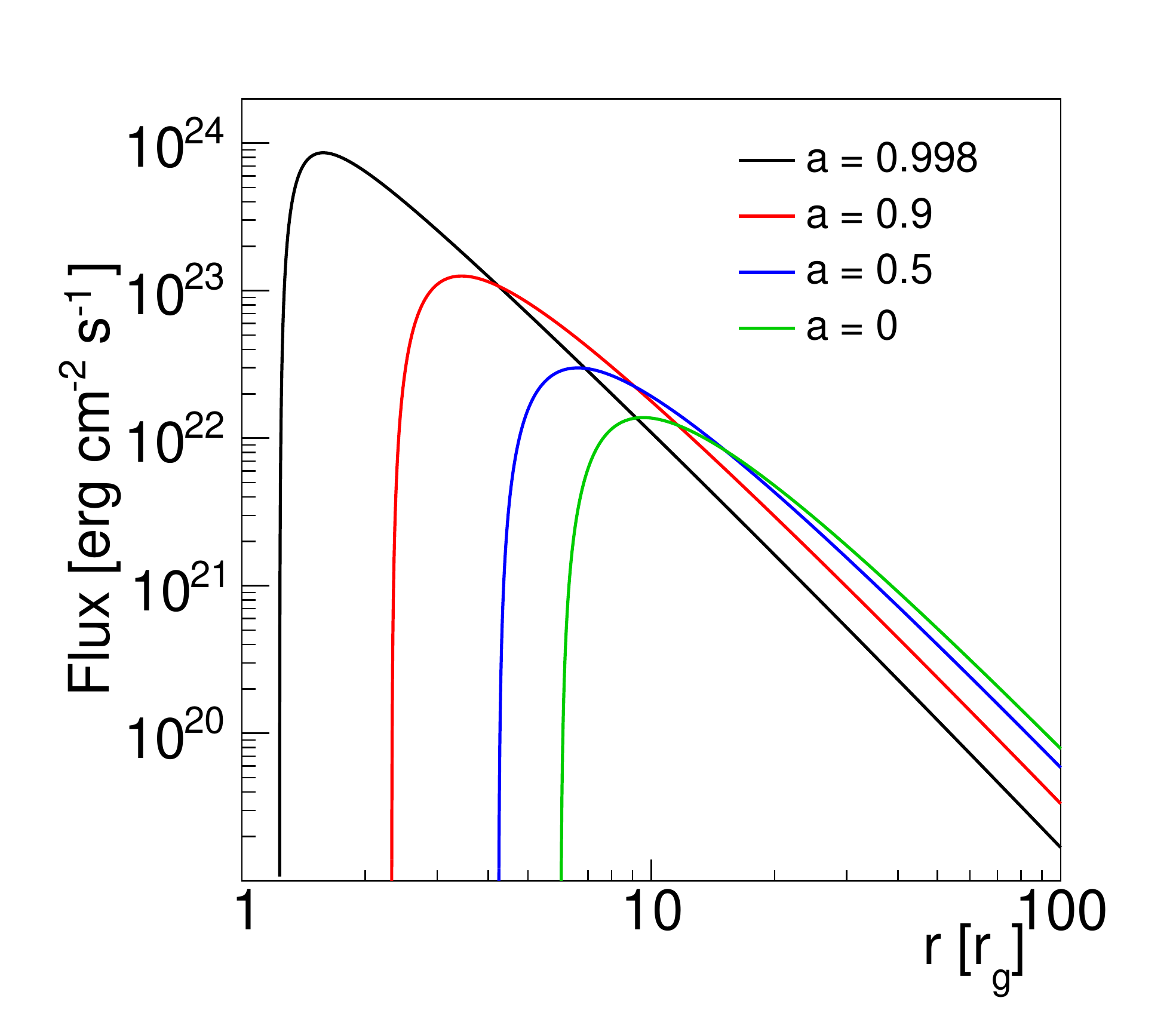}
   			\caption{\label{fig:radialFlux} Radial flux defined in Equation \ref{eq:flux} for BHs of various spins.}
   	    \end{subfigure}
   		\quad
       \begin{subfigure}[b]{0.65\textwidth}
   		 \includegraphics[width=\textwidth]{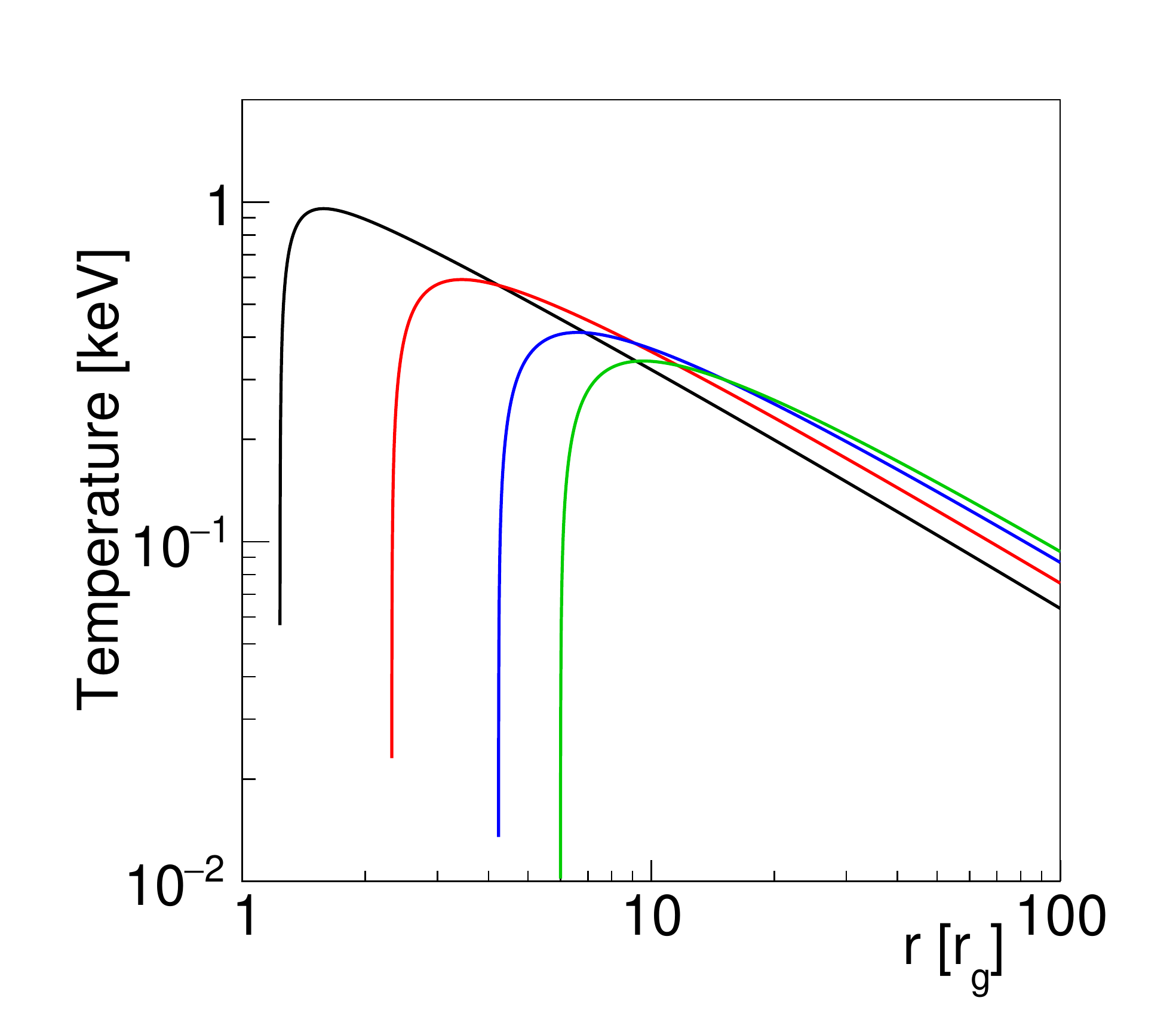}
   		 \caption{\label{fig:radialTemp} Disk temperature as defined by Equation \ref{eq:temp} for BHs of various spins (colored as above).}
   		\end{subfigure}
    \end{center}
       \caption{Disk flux (Figure \ref{fig:radialFlux}) and temperature (Figure \ref{fig:radialTemp}) for various spins.}
\end{figure}

The thermal X-ray emission from the accretion disk is modeled following the methodology outlined in \citet{Page1974} (PT74). All axially symmetric metrics can be written in the form
\begin{equation}
	ds^2 = - e^{2\nu} dt^2 + e^{2 \psi} (d\phi - \omega dt)^2 + e^{2\mu} dr^2 + dz^2.
\end{equation}
where $\nu, \psi, \mu$, and $\omega$ can be determined for any given metric through comparison with this form.  Using this notation, PT74 derive an equation for the radial flux (energy per unit proper time per unit proper area) coming out of the upper surface of the accretion disk using conservation of energy, momentum, and mass. This is given by
\begin{equation} \label{eq:flux}
	F(r) = \frac{\dot{M_0}}{4 \pi} e^{-(\nu+\psi+\mu)}f(r)
\end{equation}
where
\begin{equation}
	f(r)\equiv \frac{-p^t_{,r}}{p_\phi} \int_{r_{ISCO}}^r \frac{p_{\phi,r}}{p^t}dr.
\end{equation}
The temperature of the disk is then determined using 	
\begin{equation} \label{eq:temp}
	T_{eff}= \left( \frac{F}{\sigma_{SB}} \right) ^{1/4}
\end{equation}
where $\sigma_{SB}$ is the Boltzman constant.   Figure \ref{fig:radialFlux} shows the radial flux derived by PT74 for various spins and Figure \ref{fig:radialTemp} shows the corresponding temperatures.

\subsection{Weighting}
\begin{figure}
		\begin{center}
        		\vspace{0pt}
   				 \includegraphics[width=.8\textwidth]{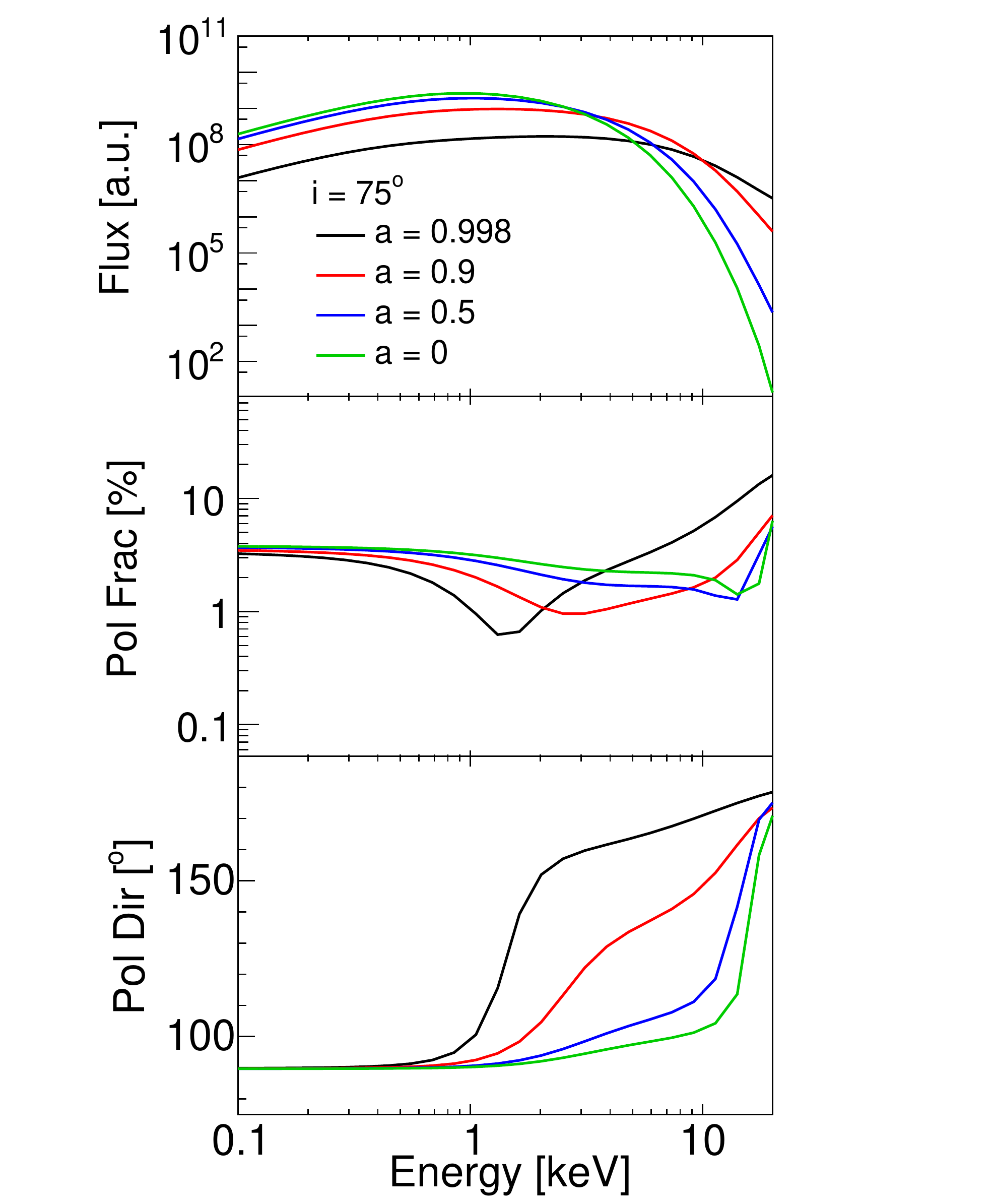}
   			\vspace{10pt}
        		\caption{\label{fig:ContinuumPolarization} Thermal flux, given by $E dN/dE$ (top), polarization fraction (middle), and polarization direction (bottom) of the thermal disk emission for various spins at an inclination of $75^{o}$.} 
        	\end{center}
	\end{figure}

In order to model the thermal emission, photons are emitted from the disk ($\theta = \pi/2$). Because the disk is axisymmetric it is only necessary to simulate one value of $\phi$, i.e. $\phi = 0$.
The statistical weight for each scattering is given by
\begin{equation}
	w_{sc} = \frac{2  \pi \mu I}{\pi \mu_0 F}.
\end{equation}
In the numerator the 2$\pi$ converts probability per solid angle to probability and $\mu$ converts the intensity $I$ into outgoing flux per both solid angle and accretion disk area.  In the denominator, $\mu_0$ converts $\pi$ $F$ into the incoming energy flux per accretion disk area.  The values for $F$ and $I$ are obtained using Table XXV of \citet{Chandrasekhar1960}.  Each photon contributes a statistical weight of 
\begin{equation}
	w_{st} = 2 \pi \Delta r \frac{dN}{dt dr d\phi} w_{em} w_{sc}
\end{equation}
where $w_{em}$ = flux/(comoving $\Delta$r and $\Delta$t) is the emission weight and $\Delta$r is the width of the bin. In order to calculate $dN/dt dr d\phi$ it is necessary to start using the relationship that, in the PF, the number of photons emitted per $d\hat{A}$ and $d\hat{t}$ is given by the relationship 
\begin{equation} \label{dn/dV}
	\frac{dN}{d\hat{A}d\hat{t}} = \frac{dN}{d^3\hat{V}} = \frac{F}{\langle \hat{E} \rangle}.
\end{equation}
Because $dN$ and the proper three volume $d^3 \hat{V} =  \sqrt{-g_{tr\phi}} dt dr d\phi$ are both invariant, Equation \ref{dn/dV} can be written such that
\begin{equation}
    \frac{dN}{dt dr d\phi} = \sqrt{-g_{tr\phi}} \frac{F}{\langle \hat{E} \rangle}
\end{equation}
which describes the number of photons emitted per GC time, radius, azimuthal angle at the given radius.  The mean PF energy per photon can be written as
\begin{equation}
	\langle \hat{E} \rangle \approx 2.7 f_h k_B T_{eff}
\end{equation}
assuming a diluted blackbody spectrum with a hardening factor of $f_h = $ 1.8.  Figure \ref{fig:ContinuumPolarization} shows the results of the simulations for the flux and polarization of the thermal emission from BHs of various spins.  As the inclination of the BH decreases the polarization fraction decreases and the polarization angle decreases with the greatest differences between inclinations seen at higher energies.
			
\section{Power-Law Emission from a Lamp-Post Corona}

\subsection{Emission}	
In order to model the reflected emission from the corona including the broadened Fe-K$\alpha$ line and the Compton hump, I implemented a lamp-post corona model.  The lamp-post is the most commonly used model for the corona (see e.g.\citet{Dovciak2004,Dovciak2012, Matt1991, Wilkins2012}) where unpolarized, power-law photons ranging in energy from 1-100 keV are emitted isotropically from a point source above the BH. The lamp-post is slightly offset from the rotation axis to avoid the singularity that occurs there and photons are ejected away from the BH using the following methodology.  Given a random angle $\theta_{rand}$ such that $\mu_{rand} = \cos \theta_{rand}$ the trajectories of the photons leaving the lamp-post are given in the quasi-Boyer-Lindquist coordinate system by
\begin{eqnarray}
u_t & = & E_i/ \sqrt{-g_{tt}} \\ \nonumber
u_r & = & E_i \mu_{rand} / \sqrt{g_{rr}} \\ \nonumber
u_\theta & = & E_i \sqrt{1-\mu_{rand}^2}/\sqrt{g_{\theta\theta}} \\ \nonumber
u_\phi & = & 0 \nonumber
\end{eqnarray}
where $E_i$ is the initial energy of the photon.  When the photon hits the disk there is the option for it to be reflected, create an iron line and form the Compton Hump. Upon scattering, the polarization is affected in the same way as described above.

		\begin{figure}
			\begin{center}
        		\vspace{0pt}
   			 \includegraphics[width=1\textwidth]{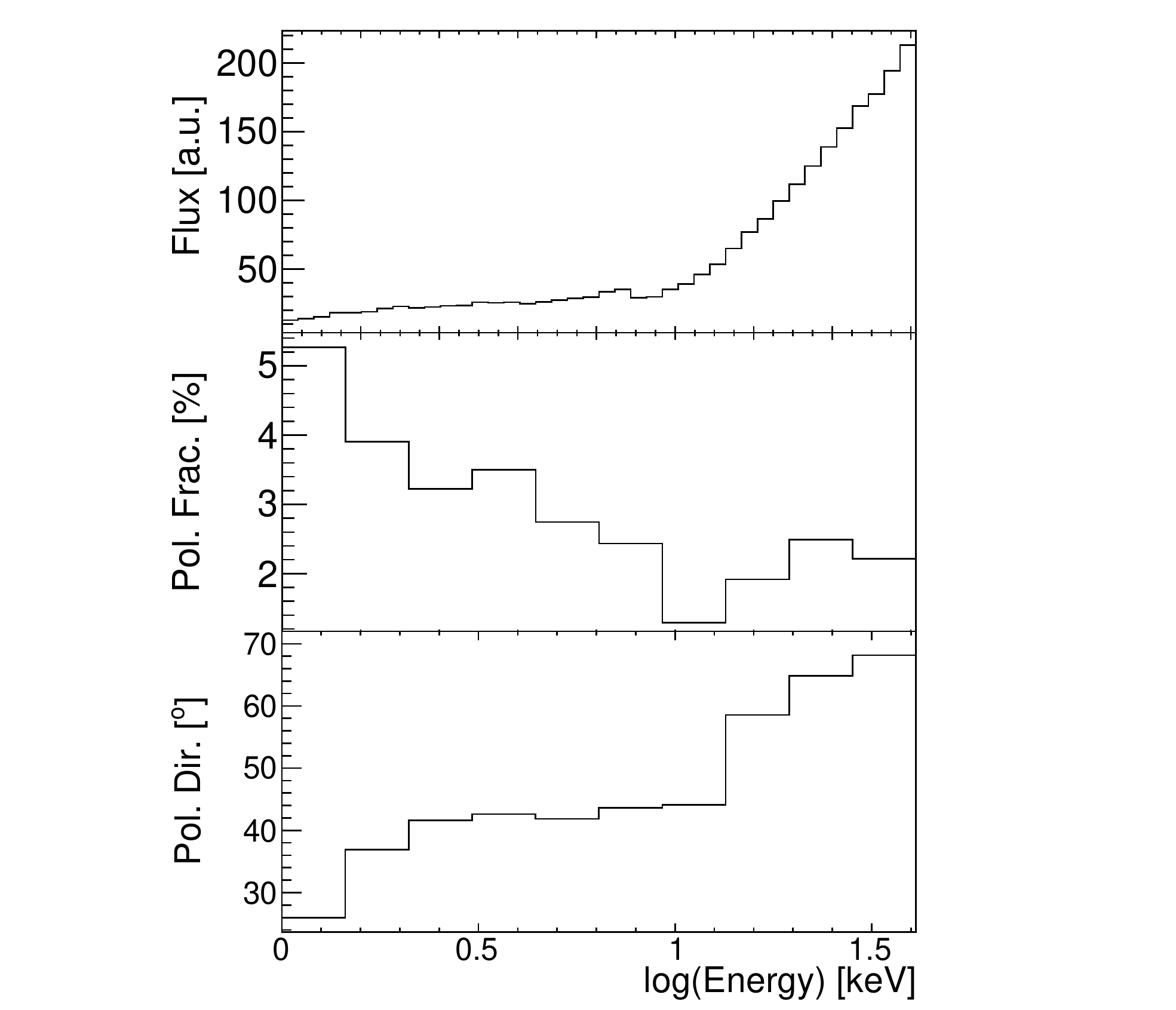}
   			\vspace{10pt}
        		\caption{\label{fig:PLPolarization} Flux, given by $E^2 dN/dE$ (top), polarization fraction (middle), and polarization direction (bottom) of the reflected power-law emission from a lamp-post at the height $h$ = 7$r_g$, $i$ = $60^{o}$ and $a$ = 0.9.} 
        		\end{center}
		\end{figure}
				
\subsection{Weighting}
The results presented in \citep{George1991} are used to determine the likelihood that an incident photon will contribute to the Fe-K$\alpha$ line or to the Compton hump.  The effective fluorescence yield is a function of the probability of whether or not a photon will be absorbed by an iron atom and the probability of whether or not the emitted photon will be able to escape from the disk.  The specific albedo is similarly calculated for the continuum photons describing the chances a photon will be reflected off of the slab or be absorbed. These values are dependent on the incident angle and energy of the photon impinging on the disk.  In the ray-tracing code, a photon is said to generate a Fe-K$\alpha$ line if it hits the disk with an energy above the threshold energy for the creation of this line at 7.1 keV and following the probability of creating an iron line versus Compton scattering by looking at the ratio of the effective fluorescence yield and the specific albedo.  When calculating the spectra of AGN the photon is further weighted by the sum of these two values in order to accurately determine the relative amplitudes of the iron line and the Compton hump.  Figure \ref{fig:PLPolarization} shows the results from the simulations for the reflected power-law emission from a lamp-post corona showing the iron line and Compton hump in the flux and a dip in polarization fraction around the iron line energy.  In order to study effect of the ionization of the disk on the X-ray reverberation signatures, the simulations can be weighted phenomenologically by the location of the scattering off the disk in order to model radially dependent ionization.  This method will be described in detail in Chapter 4.

%%% Local Variables: 
%%% mode: latex
%%% TeX-master: "thesis-main"
%%% End: 

\chapter{Testing the No-Hair Theorem of General Relativity}
\label{AlternativeMetrics}
The majority of the text from this chapter is taken from the paper \cite{Hoormann2016} which describes how the results from the ray-tracing code run for both GR and non-GR metrics can be used to constrain the behavior of gravity around BHs.  I wrote 90\% of the text of the paper and contributed all of the numerical and analytical results.  The figures are identical to those shown in the paper with the exception of some changes after updating the code.	

\section{Introduction \label{sec:intro}}
The work presented in this chapter makes use of several recently proposed metrics that contain additional terms which violate the No-Hair theorem including those of \cite{Johannsen2011a, Glampedakis2006, Aliev2005,Pani2011}. These metrics are used to quantify the degree to which spectroscopic, polarimetric, and timing X-ray observations can constrain deviations from the Kerr metric.
Several authors have used the alternative BH metrics to find observational signatures of non-GR effects.
The following types of observations have been studied:
(i) fitting of the thermal X-ray emission from BHBs \citep{Bambi2011,Bambi2012a,Pun2008};
(ii) fitting of the Fe-K$\alpha$ line emission from BHBs and SMBHs \citep{Bambi2013a,Bambi2013b, Johannsen2013b,Psaltis2012}; 
(iii) spectropolarimetric  observations of stellar mass BHs \citep{Krawczynski2012, Liu2015};
(iv) observations of QPOs \citep{Johannsen2011b, Bambi2012b, Johannsen2014};
(v) X-ray reverberation observations \citep{Jiang2015};
(vi) observations of the radiatively inefficient accretion flow around Sgr $A^*$ \citep{Broderick2014}. 
The studies showed that it is very difficult to observationally distinguish between the Kerr spacetime and 
the non-Kerr BH spacetimes as long as the BH spin and the parameter describing the deviation from the Kerr spacetime
are free parameters that both need to be derived from the observations. 

Several approaches have been discussed  to break the degeneracy between the BH spin and 
deviation parameter(s). In \cite{Bambi2012a,Bambi2012c}, for example, it is proposed that the BH spin can be measured
independently from the accretion disk properties based on 
measuring the jet power, although this method faces several difficulties in practice \citep{Narayan2012}.
Furthermore, the observed results depend on the physics of the accretion disk, the radiation transport around the
BH, the physics of launching and accelerating the jet, and the physics of converting the mechanical and
electromagnetic jet energy into observable electromagnetic jet emission. 

This chapter follows up on the work of \cite{Krawczynski2012}.  The thermal emission from a geometrically thin,
optically thick accretion disk is modeled self-consistently for the Kerr metric and the alternative metrics,
and observational signatures are derived with the help of a ray-tracing code that tracks photons from their origin
to the observer, enabling the modeling of repeated scatterings of the photons off the accretion disk. 
This chapter adds to the previous work by (i) covering the Kerr metric, the metric of Johannsen and Psaltis \citep{Johannsen2011a} and
three additional metrics, (ii) by modeling not only the thermal disk emission but also the emission from a lamp-post corona 
and the reprocessing of the coronal emission by the accretion disk, and (iii) by considering many observational channels.
We analyze the multi-temperature continuum emission from the accretion disk, 
the energy spectra of the reflected emission (including the Fe K-$\alpha$ line and the Compton hump), 
the orbital periods of matter orbiting the BH close to the $r_{\rm ISCO}$, the time lags between the
Fe K-$\alpha$ emission and the direct corona emission, and the size and shape of the BH shadows.

The rest of the chapter is organized as follows.  Section 3.2 begins with a summary of the alternative spacetimes used in this chapter and goes on to 
discuss the model for both the thermal and coronal emission.  
In Section 3.3 we compare the observational signatures of the Kerr and the non-Kerr metrics finding that the observational
differences are rather small given the uncertainties about the properties of astrophysical accretion disks.
We summarize the results in Section 3.4 and emphasize that even though the Kerr and non-Kerr metrics 
can produce similar observational signatures for some regions of the respective parameter spaces, 
we can use X-ray observations of BHs from the literature to rule out large regions of the parameter 
space of the non-Kerr metrics. 

\section{Alternative Metrics}
	As a way to test the No-Hair theorem of general relativity, several non-Kerr metrics have been introduced which contain additional parameters apart from the BH's mass and spin. In this chapter we employ the use of four non-GR metrics including two phenomenological metrics \citep{Johannsen2011a,Glampedakis2006} and two which are solutions to alternative theories of gravity \citep{Aliev2005,Pani2011}. All metrics are variations of the Kerr metric in (quasi) Boyer-Lindquist coordinates $x^\mu = (ct, r, \theta, \phi)$. 
	
	The phenomenological metric of Johannsen and Psaltis, 2011 \citep{Johannsen2011a} (JP) reads:
	\begin{eqnarray} \label{jpmetric}
	ds^2    &=& -[1+h(r,\theta)]\left(1-\frac{2Mr}{\Sigma}\right)dt^2-\frac{4aMr\sin^2\theta}{\Sigma} \times [1+h(r,\theta)]dtd\phi \\ \nonumber
		&+&    \frac{\Sigma[1+h(r,\theta)]}{\Delta + a^2\sin^2\theta h(r,\theta)}dr^2
	       + \Sigma d\theta ^2 + \bigg[ \sin ^2 \theta \left( r^2+a^2+\frac{2a^2Mr\sin^2\theta}{\Sigma}\right) \\ \nonumber
	        &+& h(r,\theta) \frac{a^2(\Sigma+2Mr)\sin^4\theta}{\Sigma} \bigg] d\phi^2 \nonumber
	\end{eqnarray}
	with
	\begin{equation} \label{sigma}
	\Sigma \equiv r^2 + a^2 \cos^2 \theta\ \\
	\end{equation} 
	\begin{equation} \label{delta}
	\Delta \equiv r^2 -2Mr +a^2.
	\end{equation}
	The metric was derived by modifying the temporal and radial components of the Schwarzschild line element by
	a term $h(r,\theta)$. The metric {\it does not exhibit any pathologies outside the event horizon} 
	and can be used for slowly and rapidly spinning BHs. 
	Asymptotic flatness constrains the leading terms of the expansion of $h$ in powers of $r$ 
	and the lowest order correction reads:
	\begin{equation}
	h(r,\theta)= \epsilon_3 \frac{M^3 r}{\Sigma ^2}.
	\end{equation}
	In the limit as $\epsilon_3 \rightarrow 0$ this metric reduces to the Kerr solution in Boyer-Lindquist coordinates. 
	
	\citet{Glampedakis2006} (GB) introduced a metric for 
	slowly spinning BHs ($a \lesssim 0.4$).  This quasi-Kerr metric in Boyer-Lindquist coordinates is
	\begin{equation} \label{gbmetric}
	g_{ab}= g_{ab}^K + \epsilon h_{ab}
	\end{equation}
	where $g_{ab}^K$ is the Kerr metric and $h_{ab}$ is given by
	\begin{eqnarray}
	h^{tt} &=& \left(1-\frac{2M}{r}\right)^{-1}[(1-3\cos^2\theta)\mathcal{F}_1(r)] \\ \nonumber
	h^{rr} &=& \left(1-\frac{2M}{r}\right)[(1-3\cos^2\theta)\mathcal{F}_1(r)] \\ \nonumber
	h^{\theta\theta} &=& -\frac{1}{r^2}[(1-3\cos^2\theta)\mathcal{F}_2(r)] \\ \nonumber
	h^{\phi\phi} &=& -\frac{1}{r^2\sin^2 \theta}[(1-3\cos^2\theta)\mathcal{F}_2(r)] \\ \nonumber
	h^{t\phi} &=& 0 \nonumber
	\end{eqnarray} 
	where $\mathcal{F}_1(r)$ and $\mathcal{F}_2(r)$ are defined in Appendix A in \cite{Glampedakis2006}.  It is clear to see that Equation \ref{gbmetric} reduces to the Kerr metric when $\epsilon \rightarrow 0$.  The details of the JP and GB space times are described in \cite{Johannsen2013a}.

	Another solution is presented by Aliev and G{\"u}mr{\"u}k{\c c}{\"u}o{\v g}lu, 2005 \citep{Aliev2005}  describing an axisymmetric, stationary metric for a rapidly rotating BH which is on a 3-brane in the Randall-Sundrum braneworld. 
	The metric turns out to be identical to GR's Kerr-Newman metric of an electrically charged spinning BH, the only
	difference that $\beta$ is not the electrical charge but a ``tidal charge''.  Interpreting this as a charged BH implies that $\beta = Q^2/ 4 \pi \epsilon_0$ where $Q$ is the electrical charge.
	The metric (referred to as KN-metric in the following) is given by:	
	\begin{equation}
	\begin{aligned} \label{agmetric}
	ds^2 & = - \left( 1-\frac{2Mr-\beta}{\Sigma}\right) dt^2 -\frac{2a(2Mr-\beta)}{\Sigma}\times \sin^2 \theta dt d\phi +  \frac{\Sigma}{\Delta}dr^2 \\
	     &  +\Sigma d\theta^2 + \bigg( r^2+a^2 +\frac{2Mr-\beta}{\Sigma} a^2 \sin^2\theta \bigg) \sin^2\theta d\phi^2 
	\end{aligned}
	\end{equation}
	with $\Sigma$ being the same as Equation \ref{sigma} and with $ \Delta = r^2+a^2-2Mr+\beta $.  
	
	\citet{Pani2011} (PMCC) give a family of solutions for slowly rotating BHs derived augmenting the Einstein-Hilbert action by quadratic and algebraic curvature invariants coupling to a single scalar field. The action is given by the expression:
	\begin{equation}
	\begin{aligned}
	S & =  \frac{1}{16\pi}\int \sqrt{-g}d^4x [R-2 \nabla_a\phi \nabla^a\phi-V(\phi) +f_1(\phi)R^2 +f_2(\phi)R_{ab}R^{ab} \\
	    & +f_3(\phi)R_{abcd}R^{abcd} + f_4(\phi)R_{abcd} \,  ^* R^{abcd}] +S_{mat}[\gamma(\phi)g_{\nu\mu}, \Psi_{mat}]
	\end{aligned}
	\end{equation}
	where
	\begin{equation}
	f_i(\phi)=\eta_i+\alpha_i \phi +\mathcal{O}(\phi^2) 
	\end{equation}
	for $i=1-4$, $S_{mat}$ is the matter action containing a generic non-minimal coupling, and $V(\phi)$ is the scalar self potential.  When $\alpha_3 =0$ the metric reduces to the one for Chern-Simons gravity and when $\alpha_4=0$ it becomes the Gauss-Bonnet solution.

\section{Results}
\begin{figure}
		       \centering 
		       \begin{subfigure}[b]{0.6\textwidth}
		                \includegraphics[width=\textwidth]{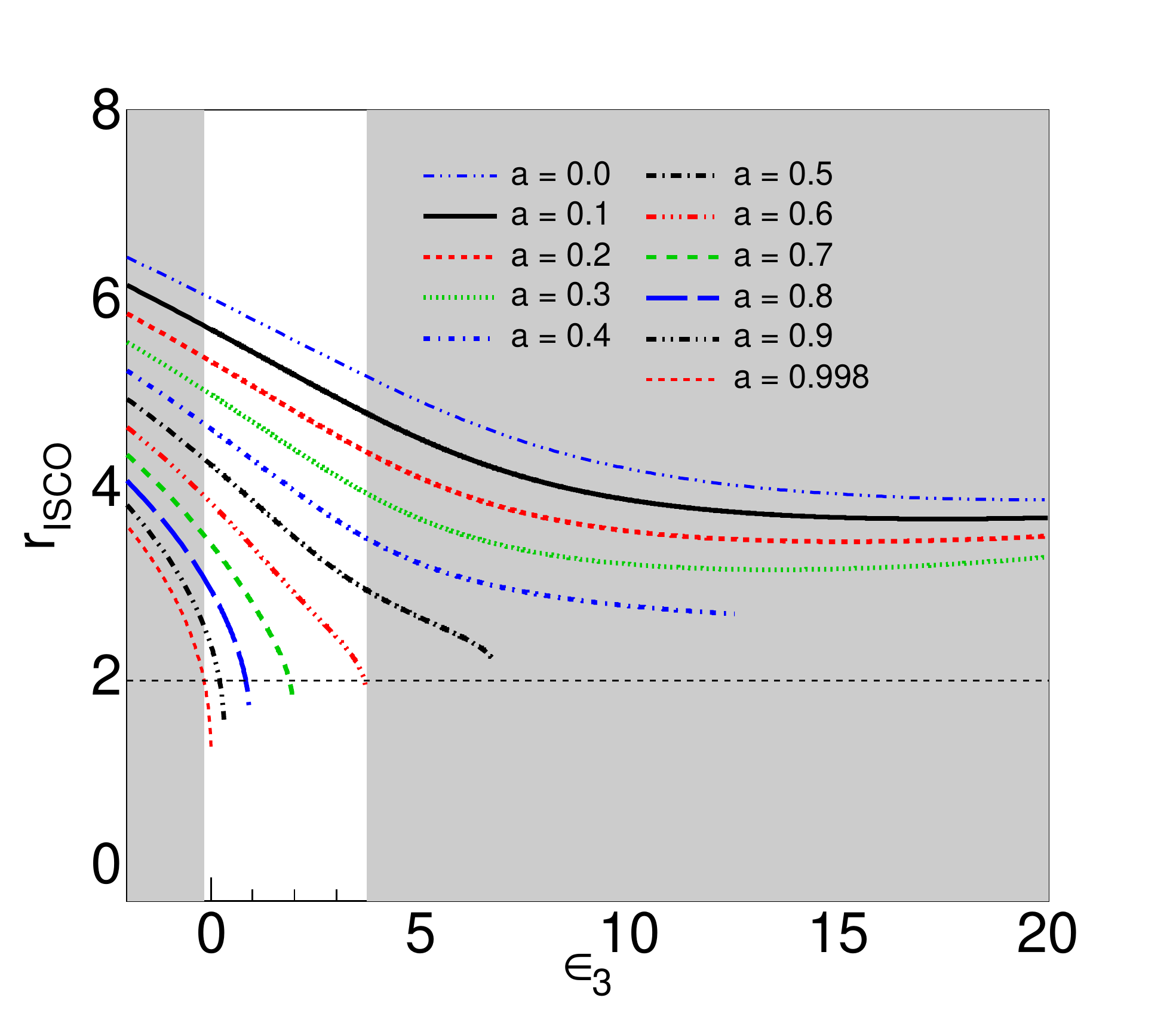}
		                \caption{\label{fig:familyJP}}
		        \end{subfigure}
		        \begin{subfigure}[b]{0.6\textwidth}
		                \includegraphics[width=\textwidth]{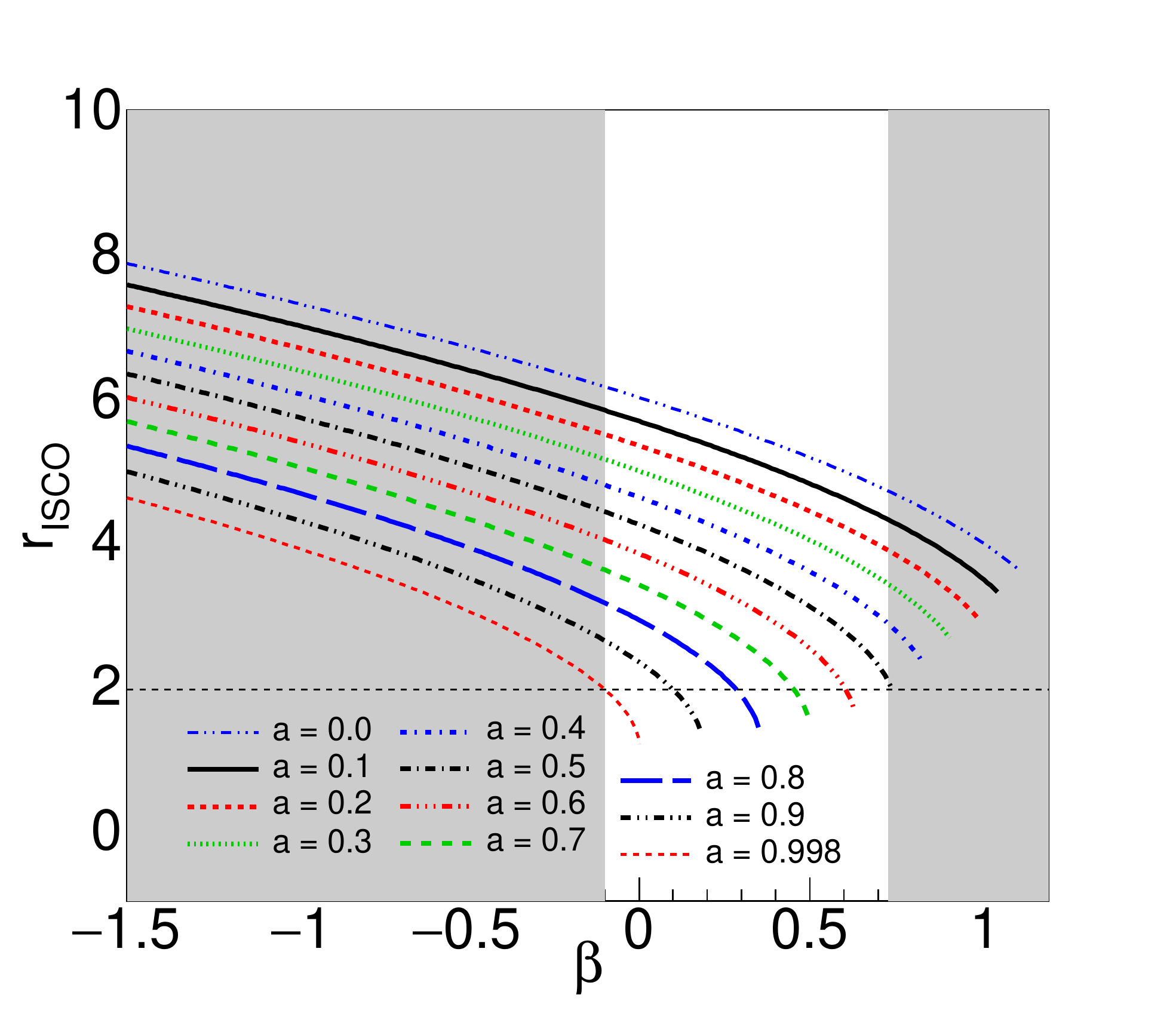}
		                \caption{\label{fig:familyKN}} 
		        \end{subfigure}
		       \caption{\label{fig:family} $r_{ISCO}$ as a function of deviation parameter for the JP metric (a) and the KN metric (b) with different spins illustrating the degeneracies within the metrics. The regions shaded grey indicate the portion of the parameter space excluded by observations of Cyg X-1 where the observed $r_{ISCO}$ \citep{Gou2011} is represented by the horizontal dashed black line.}
			\end{figure}

Previous analyses have shown that the Kerr and non-Kerr metrics give similar spectral and spectropolarimetric 
signatures if the parameters are chosen to give the same $r_{\rm ISCO}$ \citep[e.g.][]{Krawczynski2012,Bambi2013a,Jiang2015,Johannsen2014,Johannsen2013b,Kong2014}.  
Figure \ref{fig:family} 
shows $r_{ISCO}$ as a function of the BH spin $a$ and the parameter	
characterizing the deviation from
the Kerr metric for the JP and KN metrics.  We see that for all JP and KN metrics, we can always find one and only one 
Kerr metric with the same $r_{\rm ISCO}$. The mapping is not unambiguous the other way around: the JP and KN metrics 
can give one $r_{\rm ISCO}$ for several different combinations of the BH spin $a$ and the deviation parameter. 
	
In the following we focus on comparing ``degenerate'' models which give the same $r_{ISCO}$. 
We consider a slowly spinning Kerr BH ($a = 0.2, r_{ISCO} = 5.33 r_g$) and a rapidly spinning Kerr BH ($a = 0.9, r_{ISCO}=2.32 r_g$) and JP and KN models giving the same $r_{ISCO}$ (see Table \ref{fig:table}).

\begin{table}
\begin{center}
\caption{\label{fig:table} List of metric parameters used in the simulations.}
\begin{tabular}{|c|cccc|}
\hline 
\textrm{\rule{0pt}{3ex}Metric}&
\textrm{Spin}&
\textrm{Deviation}&
\textrm{$r_{ISCO}$}&
\textrm{$\dot{M}$ (g/s)}\\
\hline
\rule{0pt}{3ex}Kerr & 0.9 & none & 2.32 & 8.98$\times 10^{17}$ \\ 
JP & 0.5 & $\epsilon_3 =$ 6.33 & 2.32 & 7.51$\times 10^{17}$\\
KN &0.5 & $\beta=$0.69 & 2.32 & 9.86$\times 10^{17}$\\ \hline
\rule{0pt}{3ex}Kerr & 0.2 & none & 5.33 & 2.16$\times 10^{18}$ \\
GB & 0.25 & $\epsilon=0.12$ & 5.33 & 2.15$\times 10^{18}$\\
PMCC & 0.29 & $\alpha_3=0$, $\alpha_4=2.07$ & 5.33 & 2.11$\times 10^{18}$\\
\hline 
\end{tabular}
\end{center}
\end{table}

In the following we show the Kerr, GB and PMCC results for the low-spin case, 
and the Kerr, KN and JP results for the high-spin case. We adjusted the accretion rates to give the same 
accretion luminosity (extracted gravitational energy per unit observer time) for all considered metrics.  This is done by normalizing the accretion rate by the efficiency which is not corrected for the fraction of photons escaping to infinity.
	\begin{figure}
        \centering
        \begin{subfigure}[b]{0.46\textwidth}
                \includegraphics[trim=35mm 0mm 30mm 0mm ,width=\textwidth]{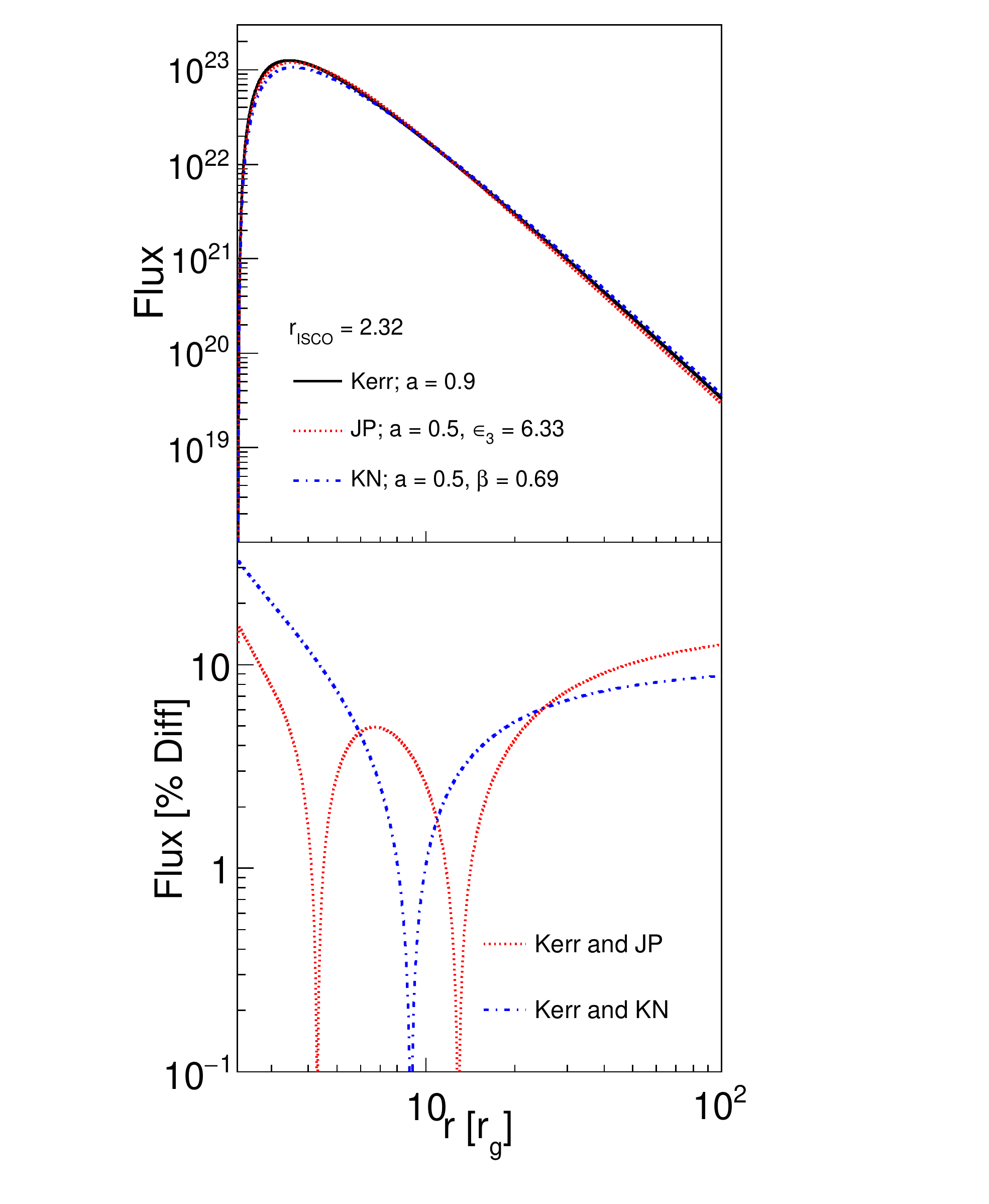}
                \caption{\label{fig:flux}}
        \end{subfigure}
         \quad
        \begin{subfigure}[b]{0.46\textwidth}
                \includegraphics[trim=30mm 0mm 35mm 0mm ,width=\textwidth]{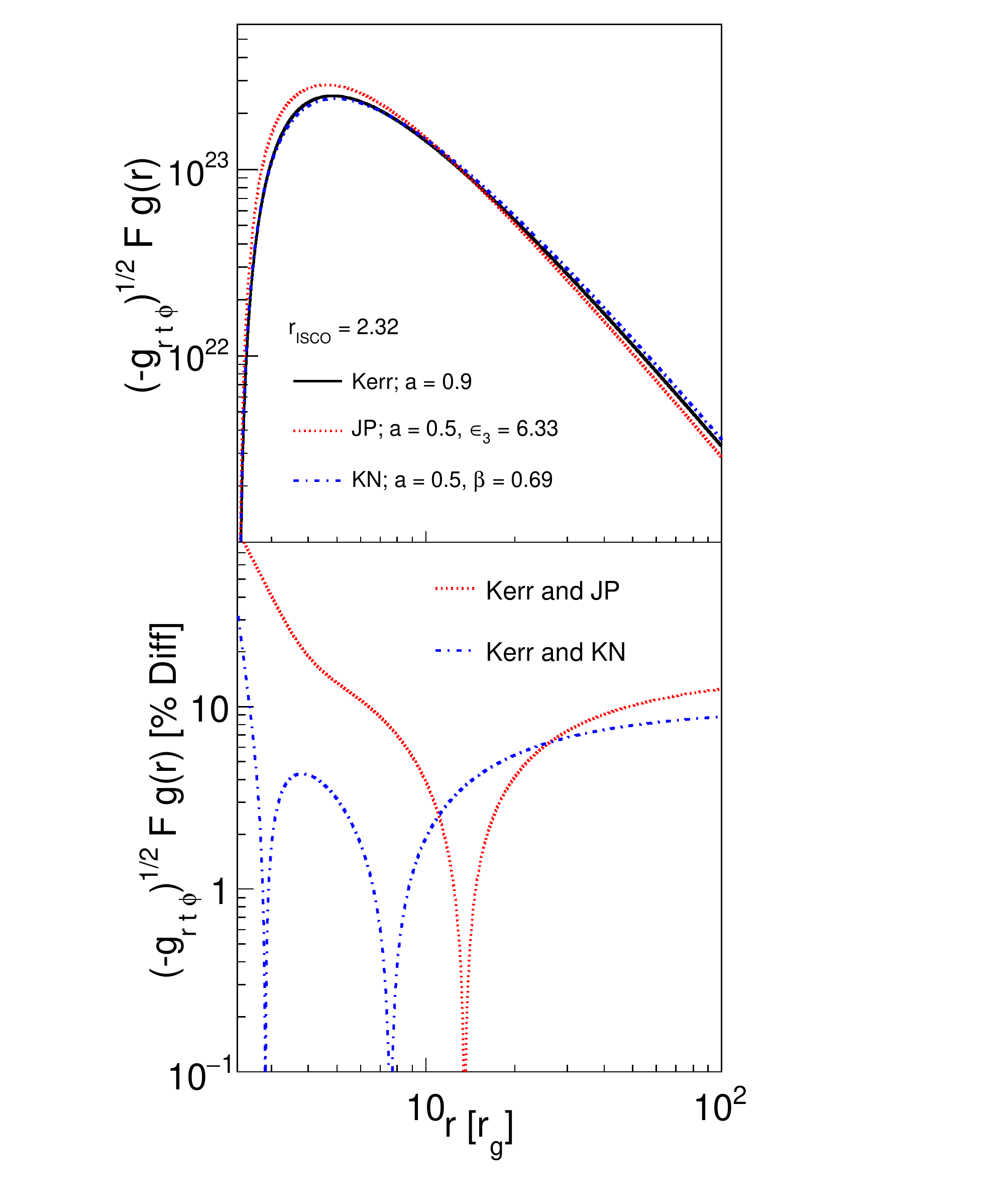}
                \caption{\label{fig:power}} 
        \end{subfigure}
        \caption{\label{fig:fluxPlots}Radial flux (Equation \ref{eq:flux}) (Figure \ref{fig:flux}) and power (Figure \ref{fig:power}) for the JP, KN, and Kerr metrics which all give $r_{ISCO}=$2.32 $r_g$. The bottom panels show the comparison of the alternative metrics to the Kerr metric.}
	\end{figure}
	
		\begin{figure}
        \centering
        \begin{subfigure}[b]{0.46\textwidth}
                \includegraphics[trim=35mm 0mm 30mm 0mm ,width=\textwidth]{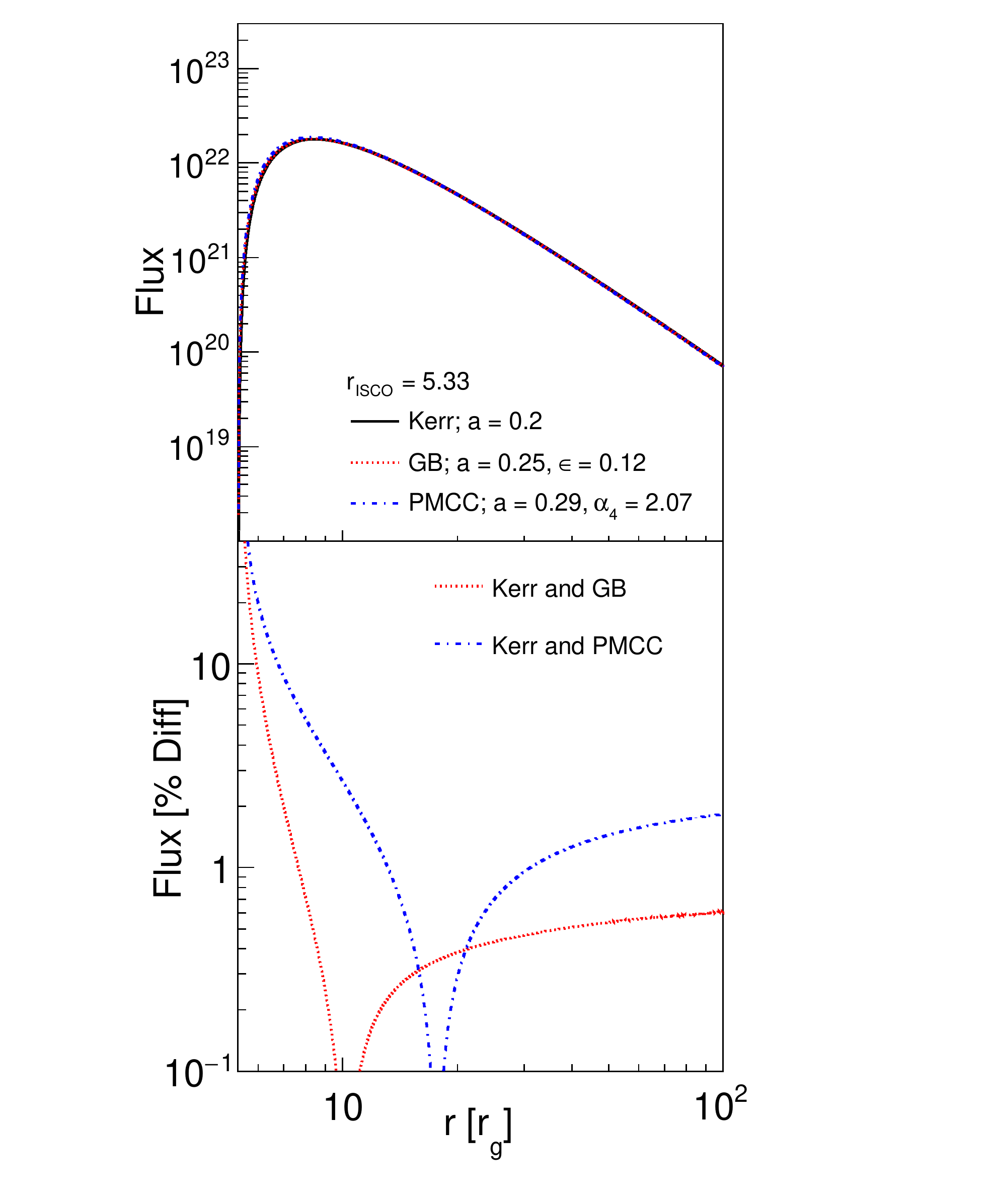}
                \caption{\label{fig:fluxLS}}
        \end{subfigure}
         \quad
        \begin{subfigure}[b]{0.46\textwidth}
                \includegraphics[trim=30mm 0mm 35mm 0mm ,width=\textwidth]{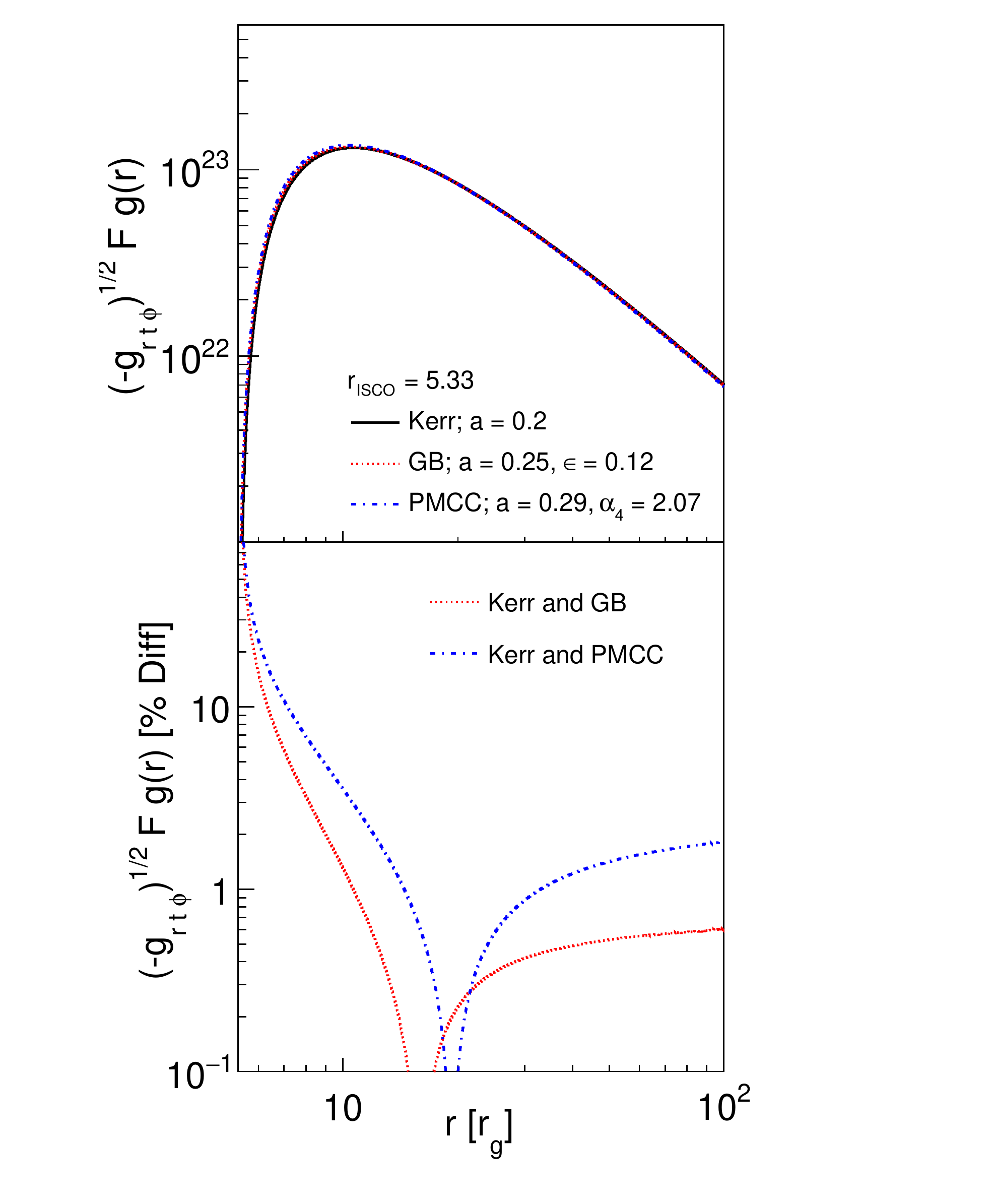}
                \caption{\label{fig:powerLS}} 
        \end{subfigure}
        \caption{\label{fig:fluxPlotsLS} Radial flux (Equation \ref{eq:flux}) (Figure \ref{fig:flux}) and power (Figure \ref{fig:power}) for the GB, PMCC, and Kerr metrics which all give $r_{ISCO}=$5.33 $r_g$.  The bottom panels show the comparison of the alternative metrics to the Kerr metric.}
	\end{figure} 
	
The left panels of Figs. \ref{fig:fluxPlots} and \ref{fig:fluxPlotsLS} compare the fluxes $F(r)$ emitted in the plasma frame for the different metrics. For the rapidly spinning BHs (Figure \ref{fig:flux}), the fractional differences in $F(r)$ are typically a few percent.
The difference is larger for the innermost part of the accretion flow with the Kerr $F(r)$ exceeding the values of the non-Kerr metrics by up to 30\%.  

The right panels of the figures show the power $P$ emitted per unit Boyer-Lindquist time and per Boyer-Lindquist 
radial interval $dr$:   
\begin{equation}
\frac{dP}{dr}(r)\,=\,\sqrt{-g_{t r \phi}} F(r) g_{\rm em}^{\rm obs}
\end{equation}
The factor $\sqrt{-g_{t r \phi}}$ is the $t-r-\phi$ dependent part of the metric and is used to transform 
the {\it number} of emitted photons per plasma frame $d\hat{t}$ and $d\hat{r}$ into that emitted 
per Boyer-Lindquist $dt$ and $dr$ \citep{Kulkarni2011}. 
The last factor corrects for the frequency change of the photons between their emission in the plasma rest frame and their detection by an observer at infinity. We estimate the effective redshift between emission and observation by assuming photons are emitted in the
upper hemisphere with the dimensionless wave vector 
$\hat{k}^{\mu}\,=\,(1,0,-1,0)$ in the plasma frame. After transforming $\hat{k}$ 
into the wave vector $k$ in the Boyer-Lindquist frame we calculate the photon energy at infinity $E_{\gamma}$ from the 
constant of motion associated with the time translation Killing vector $(1,0,0,0)$:
\begin{equation}
E_{\gamma}\,=\,-k_t   
\end{equation}
and set $g_{\rm em}^{\rm obs}\,=\,E_{\gamma}$.
The different metrics exhibit very similar $dP/dr$-distributions with typical fractional differences of  $<$ 10\%. Again, the largest deviations are found near the $r_{ISCO}$.
Overall, the different metrics lead to very similar $F(r)$ and $dP/dr$-distributions because (i) we compare models
with identical $r_{\rm ISCO}$-values (leading radial profiles with a similar $r$-dependence), 
and (ii) we use fine-tuned accretion rates $\dot{M}$ to compensate for the different accretion 
efficiencies where $\eta = 1-E_{ISCO}$ (i.e. the different fractions of the rest mass energy that can be extracted when matter 
moves from infinity to $r_{\rm ISCO}$). In the following we focus on the rapidly spinning BH simulations, 
as the observables depend more strongly on the assumed background spacetime 
than for slowly spinning BHs.

	\begin{figure}
        \centering
        \begin{subfigure}[b]{0.45\textwidth}
                \includegraphics[trim=35mm 0mm 25mm 0mm, width=\textwidth]{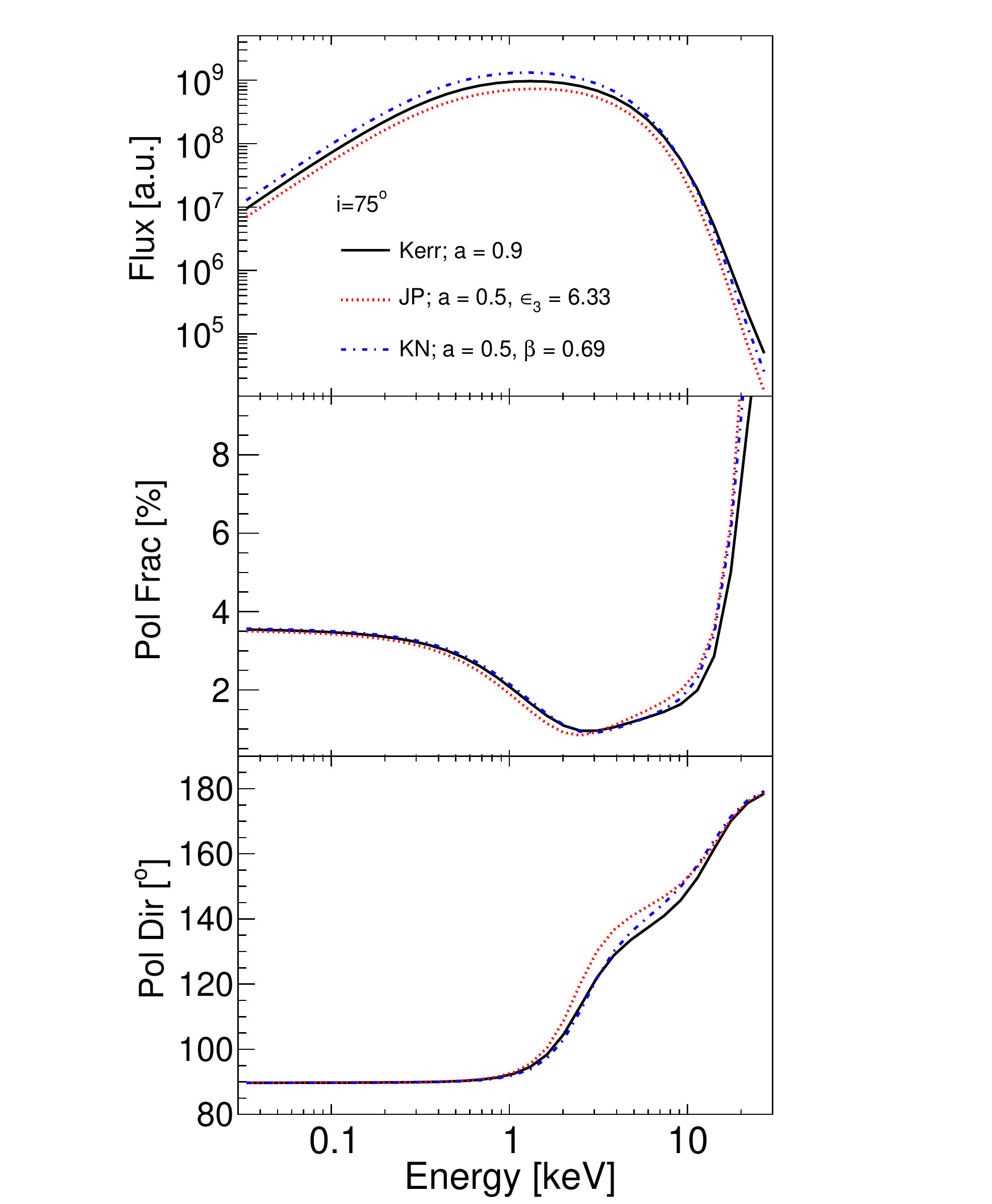}
                \caption{\label{fig:thermalHS}}
        \end{subfigure}
         \quad
        \begin{subfigure}[b]{0.45\textwidth}
                \includegraphics[trim=25mm 0mm 35mm 0mm, width=\textwidth]{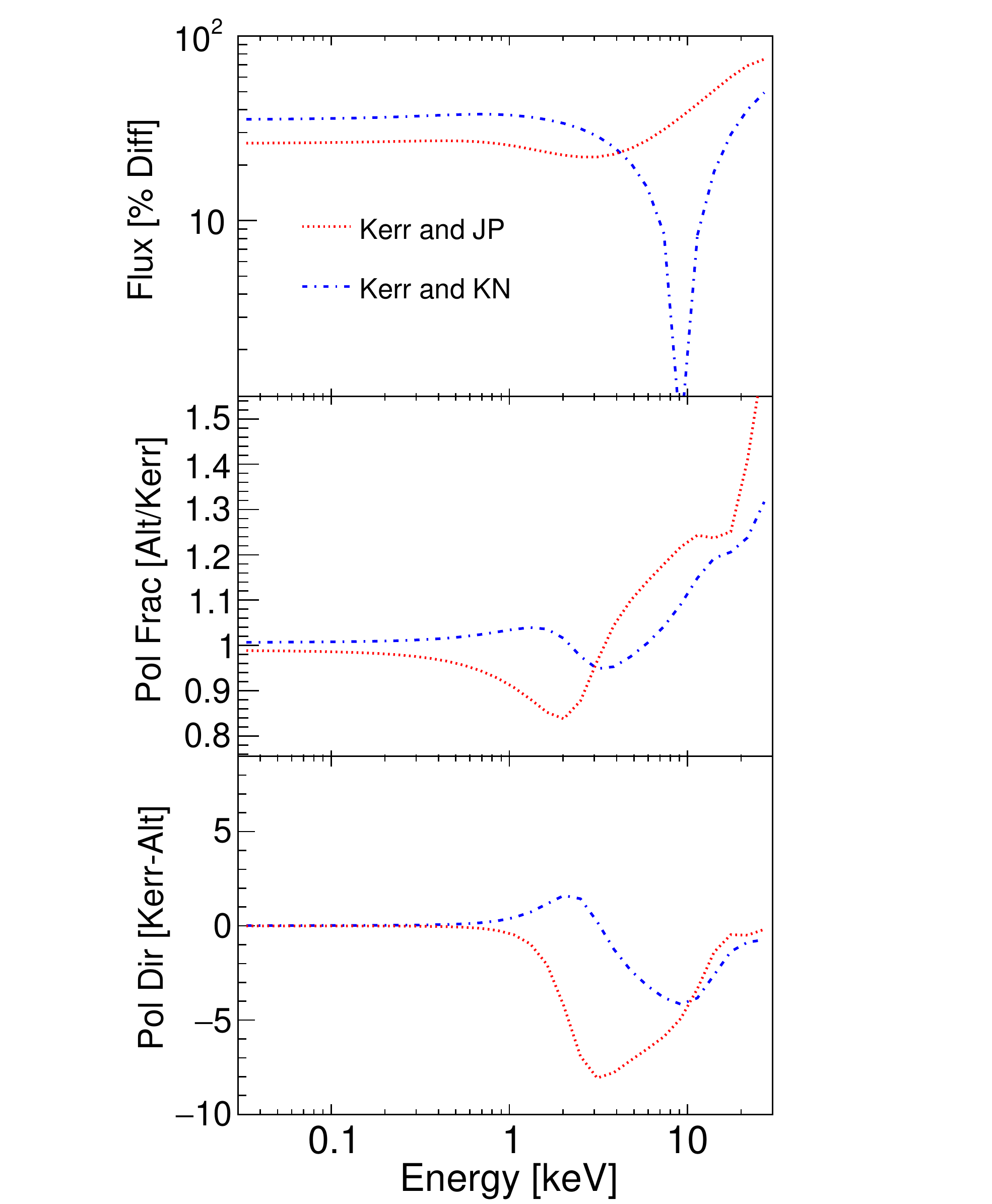}
                \caption{\label{fig:thermalHSPD}} 
        \end{subfigure}
       \caption{\label{fig:thermal} Flux (top panel), polarization fraction (middle panel), and polarization 	direction (bottom panel) of the thermal disk emission for the JP, KN, and Kerr metrics (Figure \ref{fig:thermalHS}) and corresponding comparisons of the alternative metrics with respect to the Kerr metric (Figure \ref{fig:thermalHSPD}).}
	\end{figure}
The analysis presented here focuses on specific choices for matching the Kerr metric with non-Kerr counterparts.
This matching is not unique as the non-Kerr metrics give the same $r_{\rm ISCO}$ for a continuous family of different
metrics. We simulated a few non-Kerr metrics giving the same $r_{\rm ISCO}$, and found that the differences between
these metrics and the Kerr metric are all comparable to the differences shown above.

Figure \ref{fig:thermal} shows the flux, polarization fraction, and polarization direction of the thermal disk emission of a mass accreting stellar mass BH as shown for an observer at an inclination of 75$^{\circ}$.We only show the results for the rapidly spinning Kerr, JP and KN BHs.  While there are some differences in these spectra the overall shapes are similar.   At the highest energies, deep in the Wien tail of the multi-temperature energy spectrum, the fluxes show more differences owing to the different orbital velocities and thus Doppler boost of the emission and different fractions of photons reaching the observer versus photons falling into the BHs. The different metrics also lead to very similar polarization fractions and polarization angles.  Overall, the main conclusion is that once we choose models with identical $r_{\rm ISCO}$ and correct for the different accretion efficiencies, the observational signatures depend only very  weakly on the considered metric.  Assuming that the background spacetime is described by the Kerr metric, 
the thermal energy spectrum and the polarization properties can be used to fit $r_{\rm ISCO}$ and the BH inclination $i$ \citep{Li2009, Schnittman2009}. The results presented so far
indicate that the fitted $r_{\rm ISCO}$ and $i$ values will not depend strongly on the assumed background spacetime. 

We now turn to the properties of the reflected corona emission from an AGN, assuming the lamp-post corona emits 
unpolarized emission with a photon power law index of $\Gamma = $ 1.7 from a height $h = $ 3 $r_g$ above the BH
(Figure \ref{fig:Diagram}).  Again, the flux and polarization energy spectra are almost the same for all considered
metrics and with some of the differences due to numerical limitations. The KN metric shows slightly larger deviations from the Kerr metric in flux and polarization fraction than the JP metric.

		\begin{figure}
        \centering
        \begin{subfigure}[b]{0.45\textwidth}
                \includegraphics[trim=35mm 0mm 25mm 0mm,width=\textwidth]{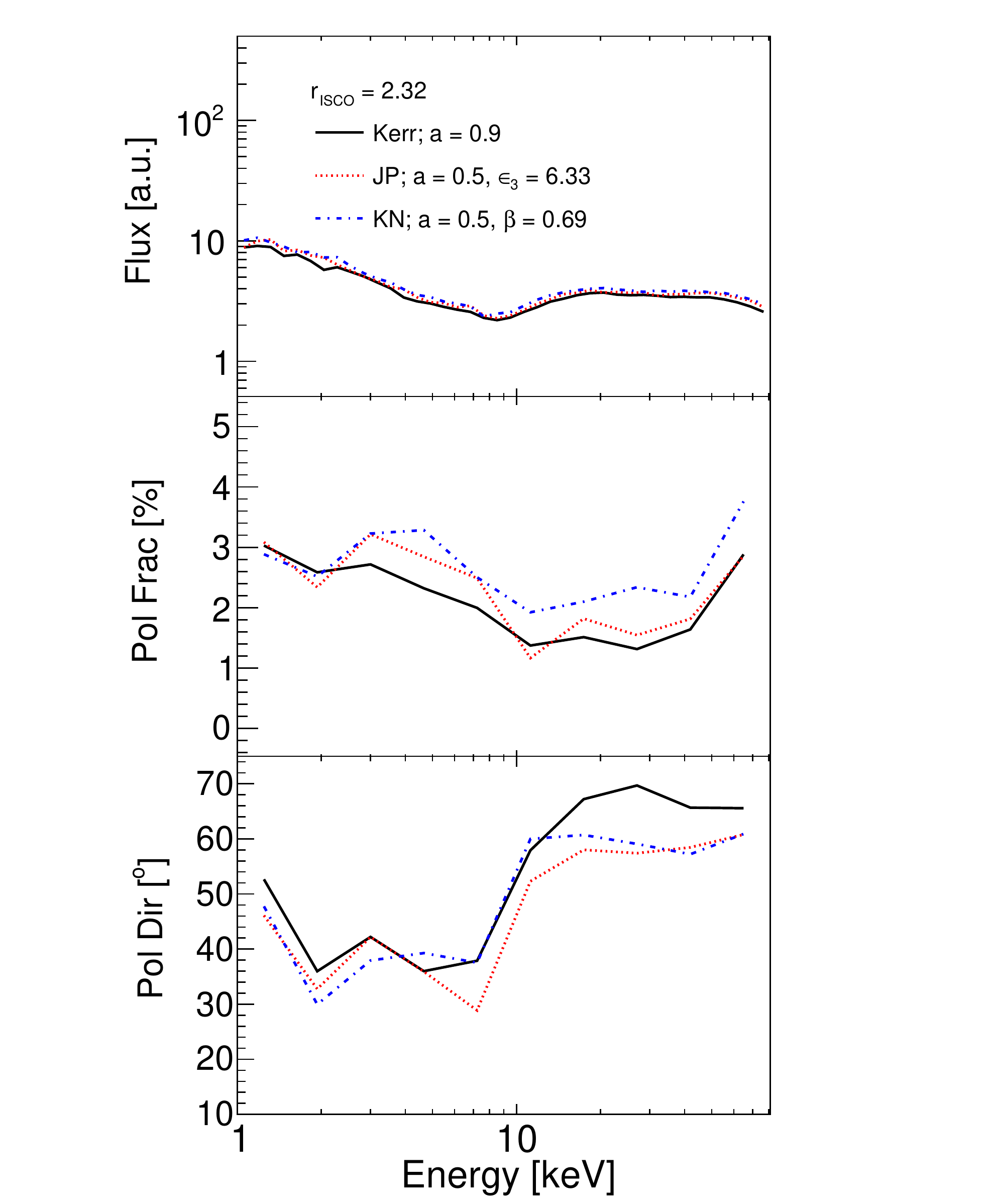}
                \caption{\label{fig:polI}}
        \end{subfigure}
         \quad
        \begin{subfigure}[b]{0.45\textwidth}
                \includegraphics[trim=25mm 0mm 35mm 0mm,width=\textwidth]{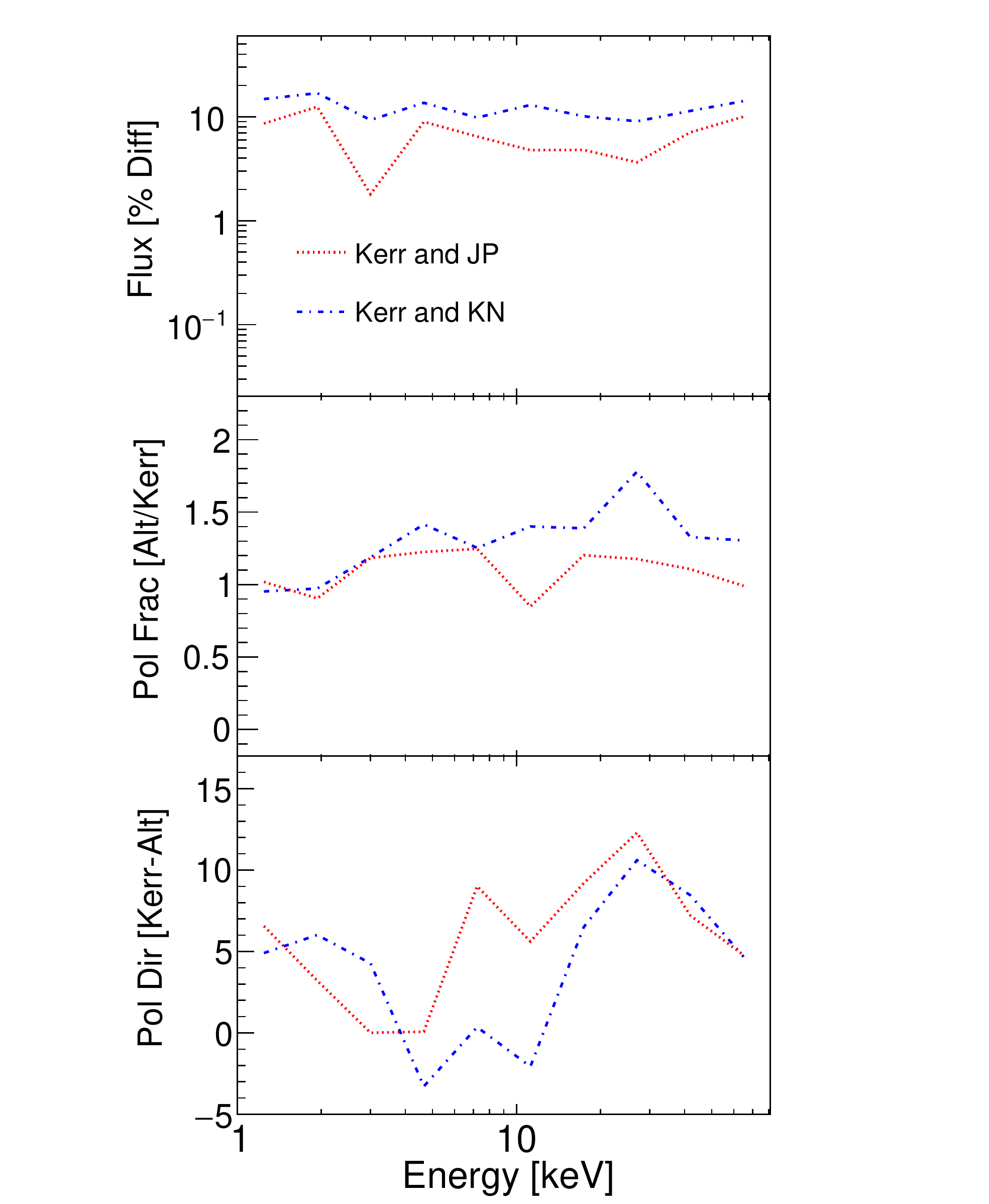}
                \caption{\label{fig:polIPD}} 
        \end{subfigure}
       \caption{Flux (top panel), polarization fraction (middle panel), and polarization direction (bottom panel) for the reflected spectrum showing the Fe-K$\alpha$ line of an AGN with $h=3 r_g$ for the JP, KN, and Kerr metrics (Figure \ref{fig:polI}) and corresponding comparisons of the alternative metrics with respect to the Kerr metric (Figure \ref{fig:polIPD}).}
	\end{figure}

Some accreting BHs exhibit QPOs, i.e. peaks in the Fourier transformed power spectra.
The orbiting hot spot model \citep{Schnittman2004,Schnittman2005,Stella1998, Stella1999, Abramowicz2001, Abramowicz2003} explains the HFQPOs 
of accreting stellar mass BHs with a hot spot orbiting the BH close to the $r_{\rm ISCO}$. 
If we succeeded to confirm the model (e.g. through the observations of the phase resolved energy spectra and/or polarization
properties \citep{Beheshtipour2016}), one could use HFQPO observations to measure the orbital periods close to the $r_{\rm ISCO}$. 
In the case of AGNs, tentative evidence for periodicity associated with the $r_{\rm ISCO}$ has been found for several objects. 
Examples include orbital periodicity on the time scale of a few days as seen in the blazar OJ 287 \citep{Pihajoki2013} and QPO's at frequencies of O(100 Hz) such as that seen for the microquasar GRO J1655-40 \citep{Strohmayer2001}. 
\begin{figure}
		\centering
        \begin{subfigure}[b]{0.46\textwidth}
                \includegraphics[trim=35mm 0mm 30mm 0mm ,width=\textwidth]{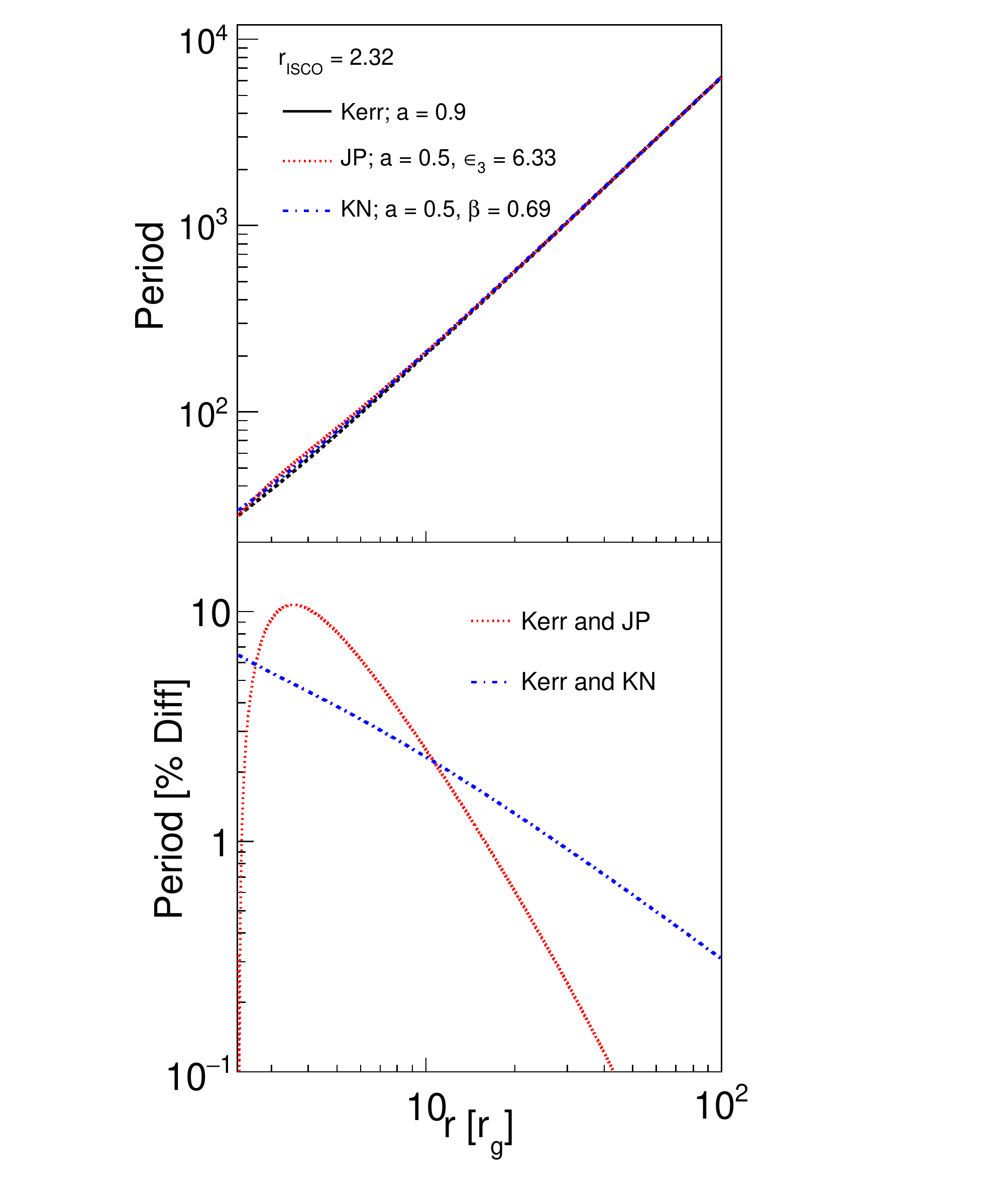}
                \caption{\label{fig:periodbig}}
        \end{subfigure}
         \quad
        \begin{subfigure}[b]{0.46\textwidth}
                \includegraphics[trim=30mm 0mm 35mm 0mm ,width=\textwidth]{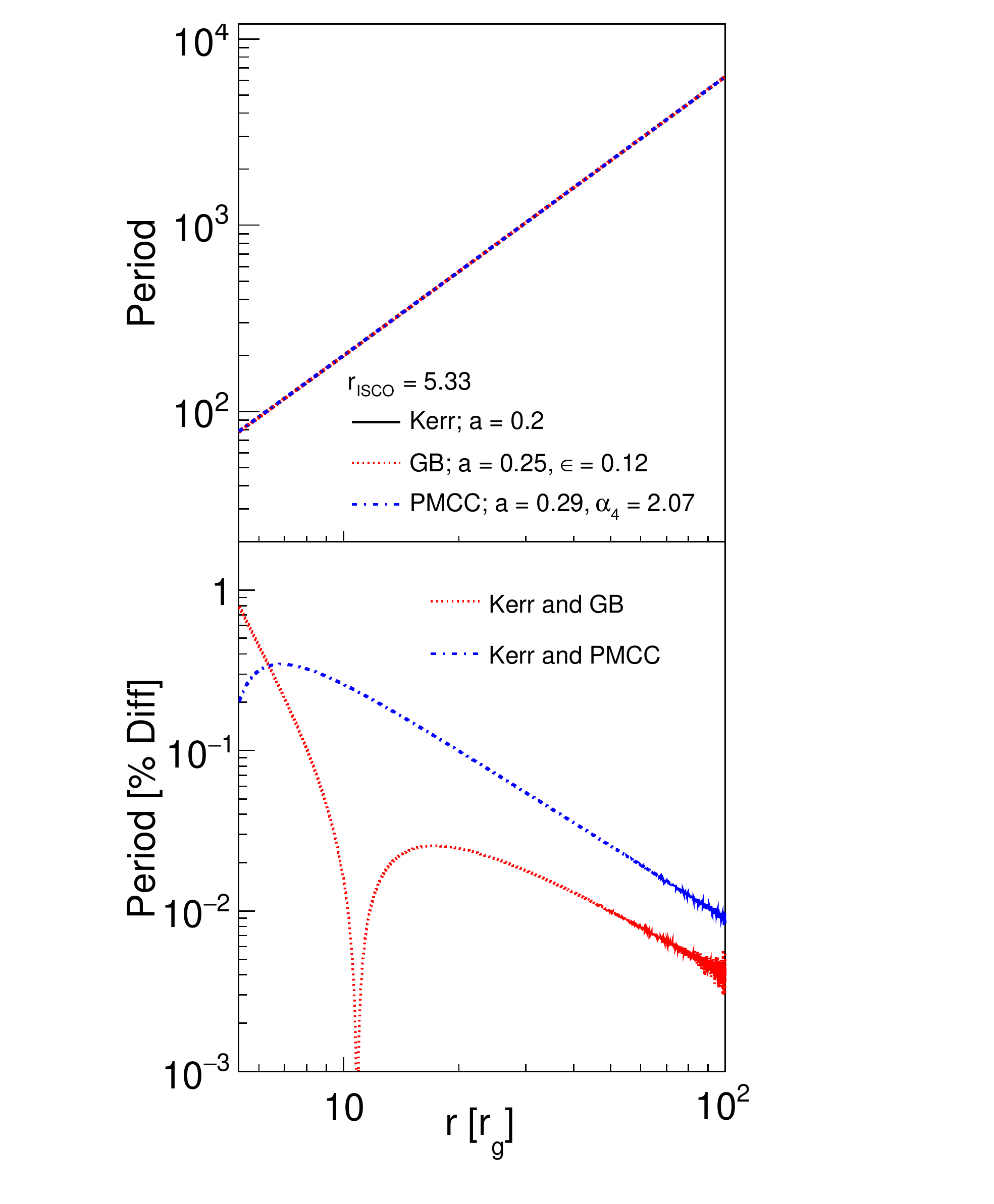}
                \caption{\label{fig:periodsmall}} 
        \end{subfigure}
        \caption{\label{fig:period} Orbital periods (top) for the metrics and the percent difference (bottom) for the JP and KN metrics compared to the Kerr metric (Figure \ref{fig:periodbig}) and the GB and PMCC metrics (Figure \ref{fig:periodsmall}).}
	\end{figure}
	\begin{figure} 
	\centering
 		\includegraphics[trim=25mm 0mm 30mm 0mm ,width=.50\textwidth]{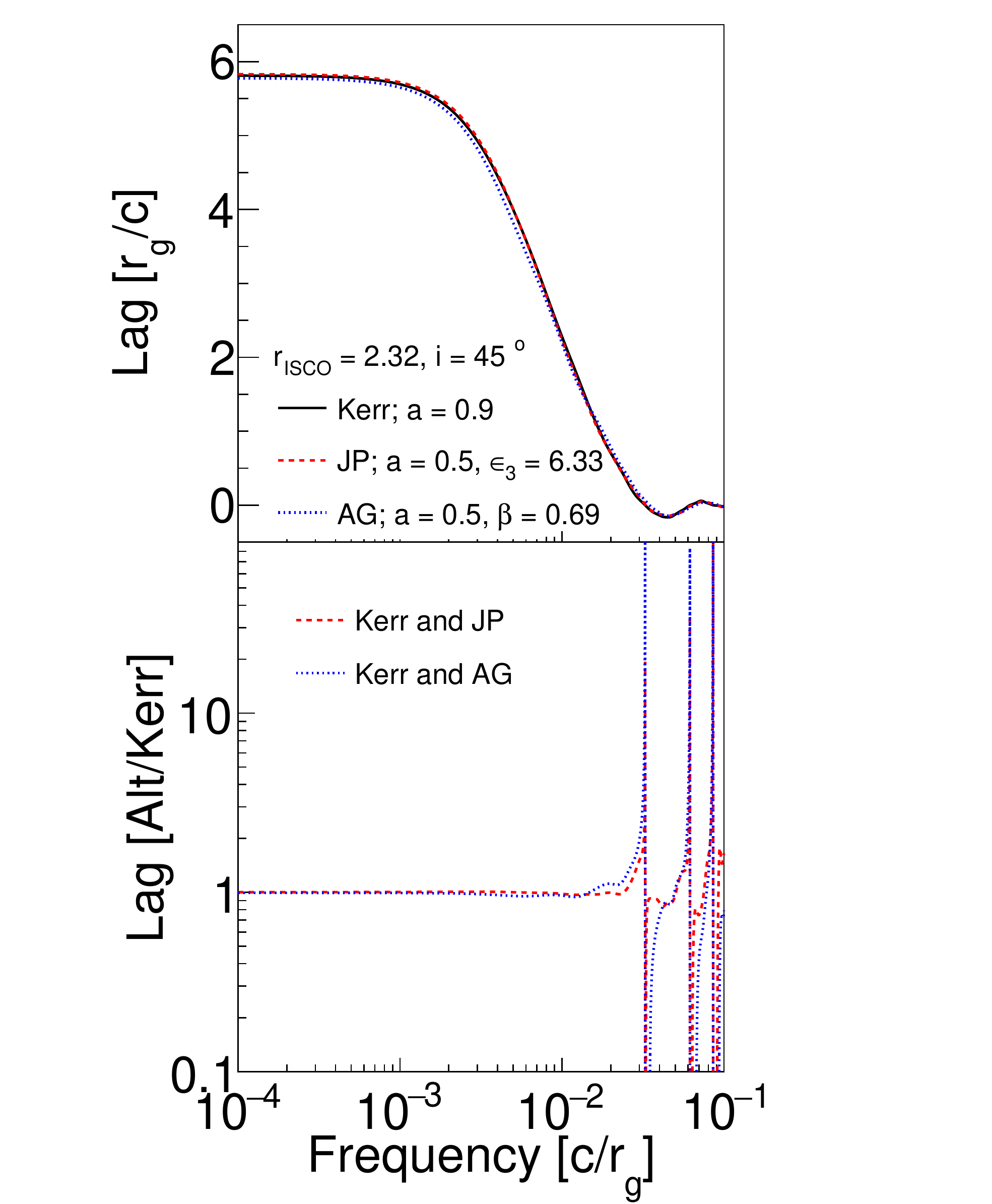} 
 		\caption{\label{fig:lag} Lag-frequency spectrum (top) of an AGN with $h=3 r_g$ for the JP, KN, and Kerr metrics along with the percent difference (bottom) of the KN and JP metrics when compared with the Kerr metric. A positive lag corresponds to the reflected emission lagging behind the direct emission where the direct emission is given in the 1-2 keV band and the reflected emission is in the 2-10 keV band.}
	\end{figure}
Figure \ref{fig:period} shows that different metrics do predict different orbital periods which vary by up to $\sim$ 10\%. 

We investigated if other {\it timing properties} can be used to observationally distinguish between the different metrics by 
analyzing the observable time lags between the direct corona emission and the reflected emission assuming the lamp-post geometry.
We use the standard X-ray reverberation analysis methods described by \cite{Uttley2014}. As expected, 
the 2-10 keV flux variations lag the 1-2 keV flux variations (Figure \ref{fig:lag}). 
	The difference seen in the lags is small, particularly when compared to the other uncertainties in the model.  At frequencies above 0.01 $c/r_g$  the
phase wrapping begins to occur (when the lag changes sign and begins to oscillate around 0) 
leading to the larger differences seen in this range.

\begin{figure*}
        \centering
        \begin{subfigure}[b]{0.4\textwidth}
                \includegraphics[trim=5mm 0mm 5mm 0mm, width=\textwidth]{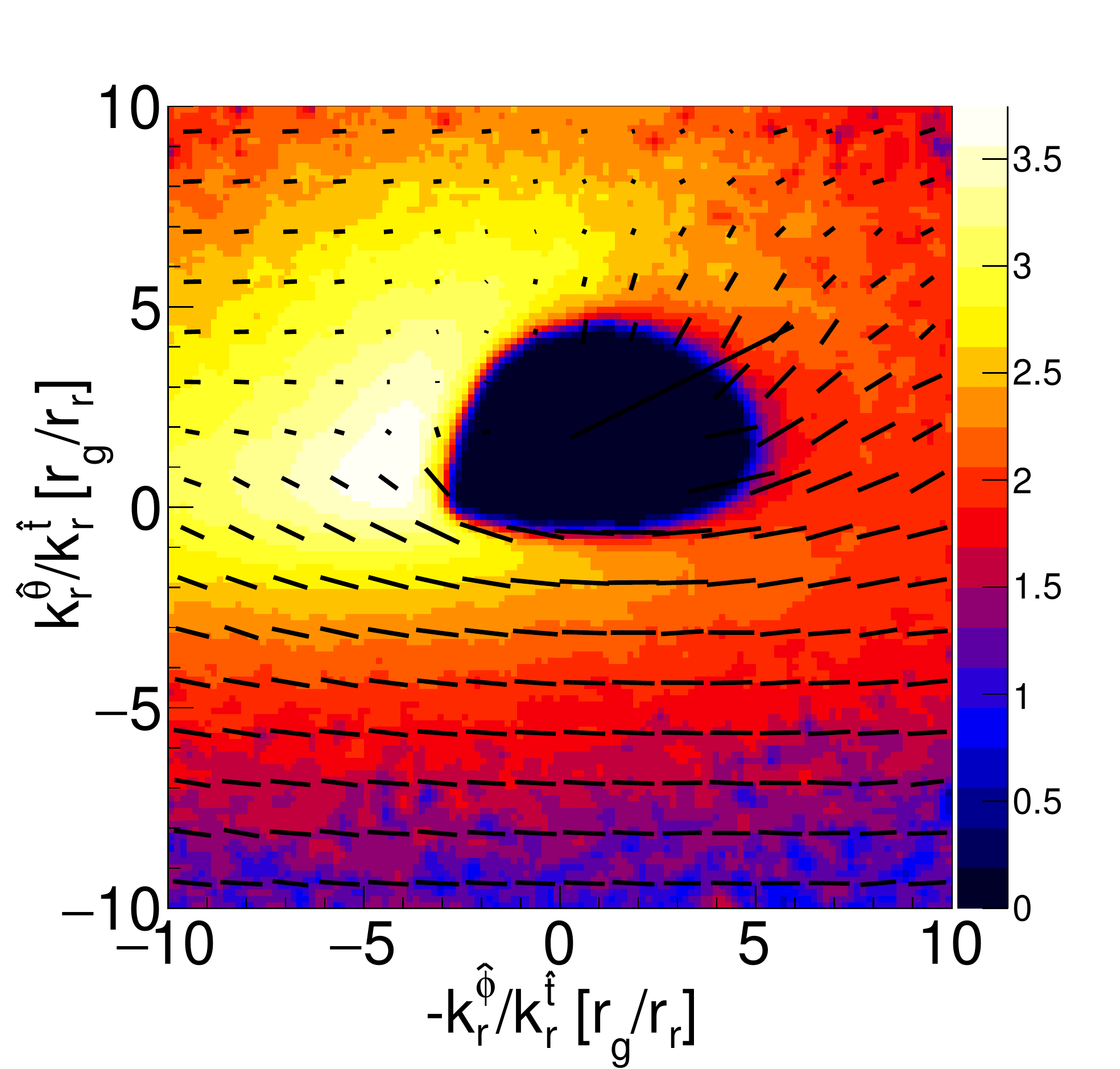}
        \end{subfigure}
         \quad
        \begin{subfigure}[b]{0.4\textwidth}
                \includegraphics[trim=5mm 0mm 5mm 0mm ,width=\textwidth]{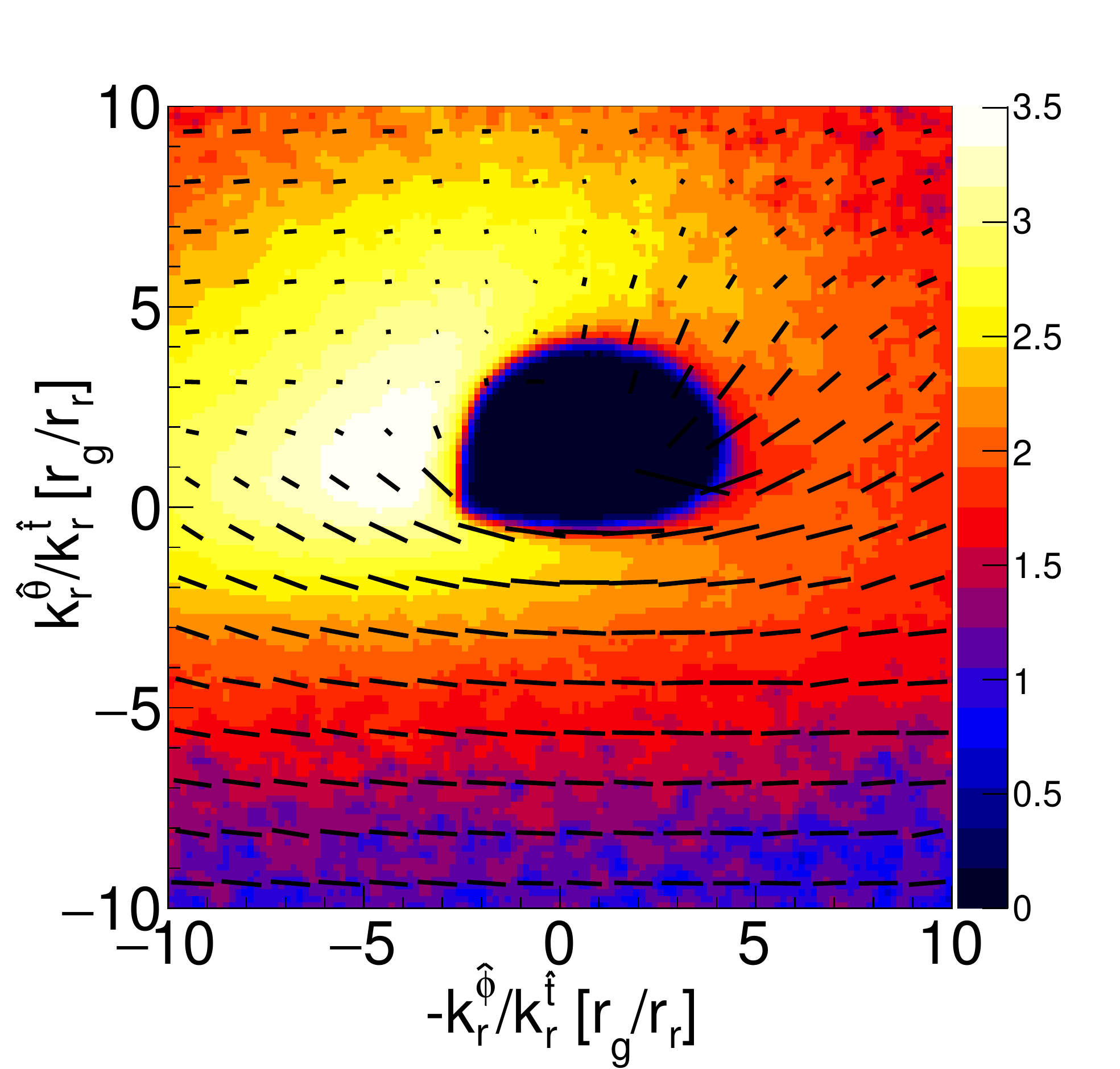}
        \end{subfigure}
        
        \begin{subfigure}[b]{0.4\textwidth}
                \includegraphics[trim=5mm 0mm 5mm 0mm ,width=\textwidth]{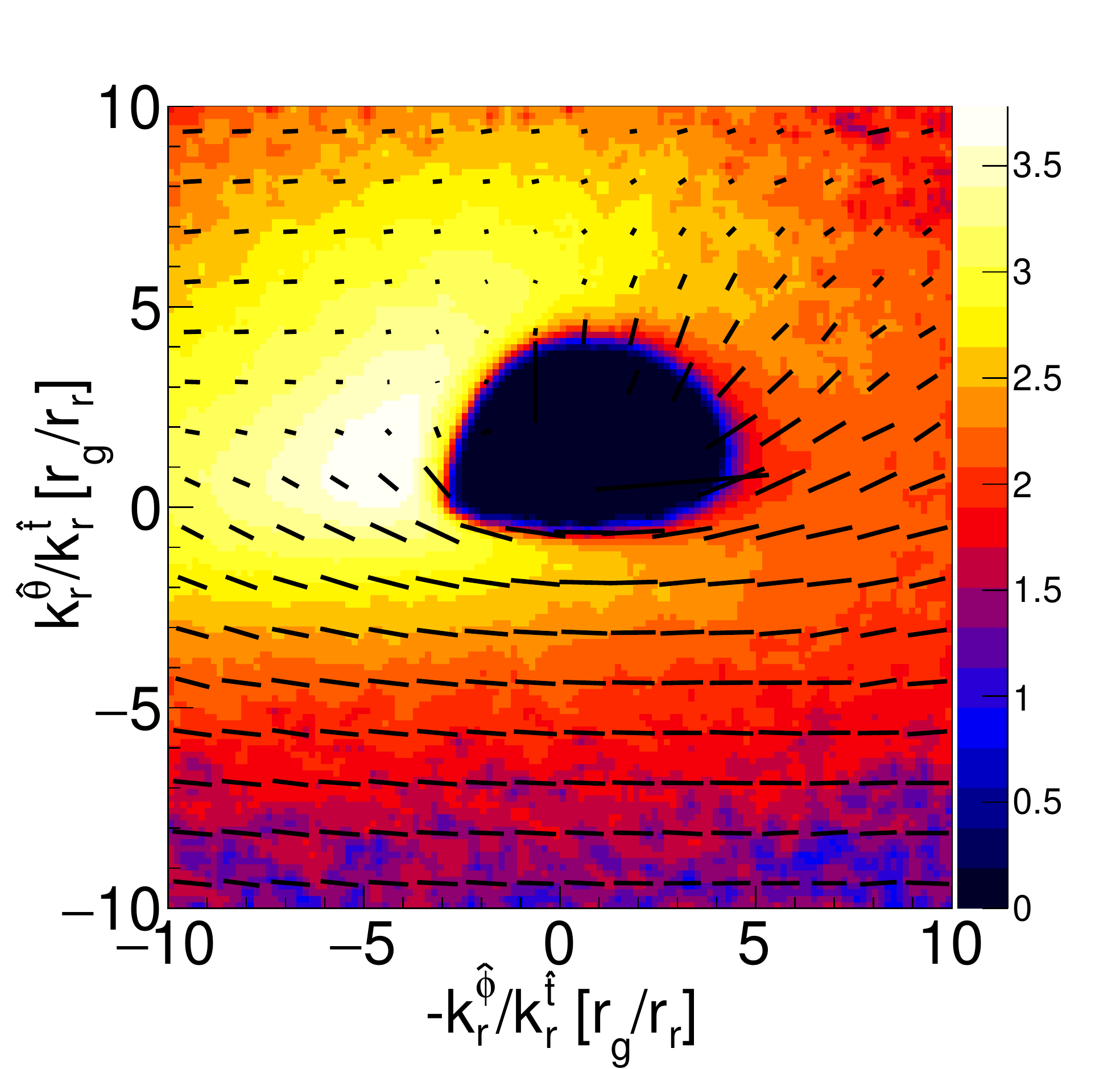}
        \end{subfigure}
        \quad
        \begin{subfigure}[b]{0.4\textwidth}
                \includegraphics[trim=5mm 0mm 5mm 5mm,width=\textwidth]{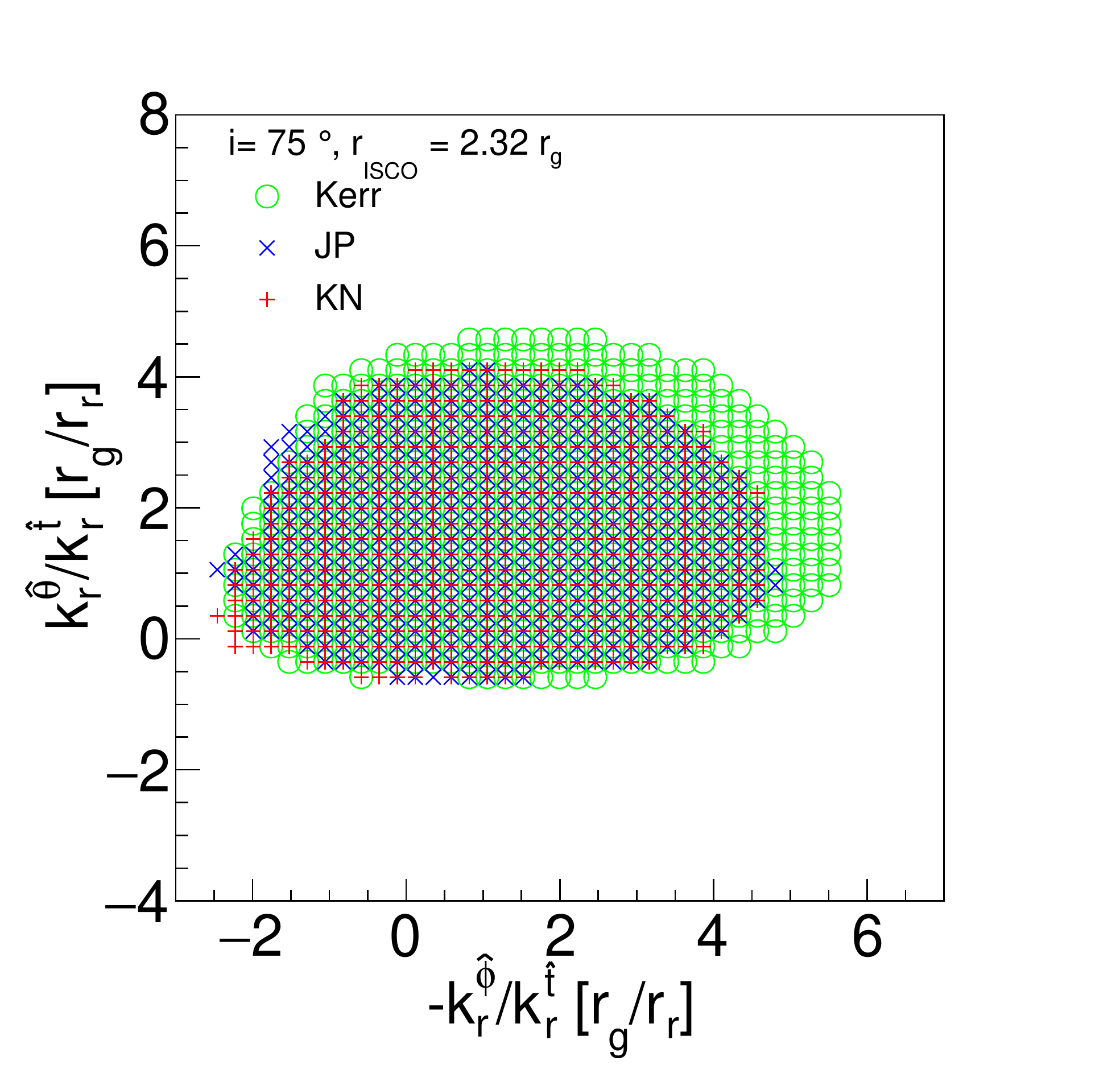}
        \end{subfigure}
        \caption{\label{fig:set1hi} Intensity-Polarization maps of the three rapidly spinning metrics in: Kerr (top left), JP (top right), KN (bottom left) for a BH at a high inclination along with a comparison of their shadows (bottom right).}
	\end{figure*}

\begin{figure}
        \centering
        \begin{subfigure}[b]{0.48\textwidth}
                \includegraphics[trim=15mm 0mm 15mm 0mm ,width=.95\textwidth]{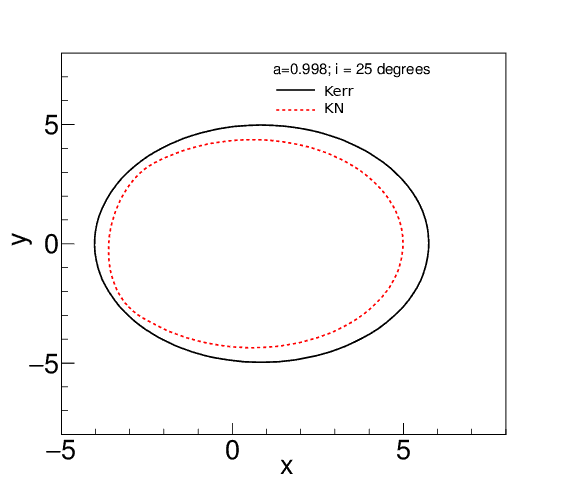}
                \caption{\label{fig:ring25}}
        \end{subfigure}
         \quad
        \begin{subfigure}[b]{0.48\textwidth}
                \includegraphics[trim=15mm 0mm 15mm 0mm ,width=.95\textwidth]{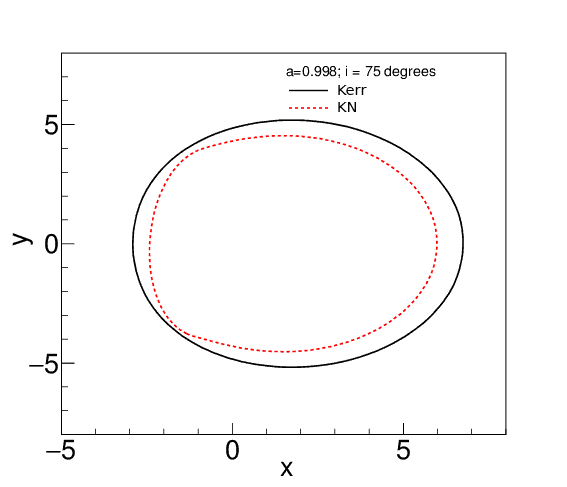}
                \caption{\label{fig:ring75}} 
        \end{subfigure}
       \caption{\label{fig:rings} Photon rings of the Kerr and KN metrics as seen by an observer at both $i = 25 ^{o}$ (\ref{fig:ring25}) and $i = 75^{o}$ (\ref{fig:ring75}).}
	\end{figure}

Although the considered metrics give the same $r_{\rm ISCO}$ in Boyer-Lindquist coordinates, the BH 
shadow may have a different shape and/or size when viewed by an observer at infinity \citep[see][for a related study]{Johannsen2010,Johannsen2013c}. The results for the Kerr, JP, and KN metrics are shown in Figure \ref{fig:set1hi} 
for an inclination of $i = 75^{\circ}$.  The shadow of the KN and JP metrics is $\sim$15\% smaller than that 
of the Kerr metric. Furthermore, the shapes differ slightly. Similarly, the shapes of the photon rings are shown in Figure \ref{fig:rings} which are calculated by following the procedures outlined in Section III of \cite{Bardeen1973}.

\section{Summary and Discussion}
In this chapter we studied the observational differences between accreting BHs in five different background spacetimes, including GR's Kerr spacetime, and four alternative spacetimes. We chose the  parameters of the considered metrics 
as to give identical innermost stable orbits in Boyer-Lindquist coordinates. 
The predicted observational differences are larger for rapidly spinning BHs. 
Overall the observational differences are very small if we adjust the accretion rate to correct for the metric-dependent 
accretion efficiency. 
The measurement of the predicted differences are very small -- especially if one accounts for the astrophysical uncertainties, 
i.e. observational and theoretical uncertainties of the accretion disk properties. 
From an academic standpoint, it is interesting to compare the small differences of the predicted properties, e.g. the 
differences of the thermal energy spectra and the BH shadow images.  The thermal spectrum of the Kerr BH is slightly 
harder than that of the JP and KN BHs (for the same $r_{\rm ISCO}$),
and the Kerr BH shadow is slightly larger than that of the JP and KN BH shadows.
Reducing the spin of the Kerr BH would make the spectral difference smaller, but would increase the mismatch between 
the apparent BH shadow diameters. Thus, in the absence of astrophysical uncertainties, the combined information from various observational channels could be used to distinguish between different metrics.

Although the analysis shows that the differences between the Kerr and non-Kerr metrics are rather subtle (especially
in the presence of uncertainties of the structure of astrophysical accretion disks), we can use existing observations 
to constrain large parts of the parameter space of the non-Kerr metrics (see also \citet{Bambi2014}).  As an example we use the recent observations 
of the accreting stellar mass BH Cyg X-1 \citep{Gou2011}. The observations give a 3\,$\sigma$ 
upper limit of $r_{\rm ISCO}<1.94$. Inspecting Figure \ref{fig:family} we see that the constraints on the $r_{\rm ISCO}$ can only
be fulfilled for JP deviation parameters $\epsilon_3 \in\left[-0.15 ,3.72\right]$ and KN deviation parameters 
$\beta\in \left[-0.01,0.73\right]$, excluding all parameter values outside of these intervals.
Our limits rest on the matching of Kerr to non-Kerr metrics using identical $r_{\rm ISCO}$ values.  Maximally rotating BHs, when $a =$ 0.998 \citep{Thorne1974}, provide the best opportunity to test GR because  observations of these systems can, in principle, be used to exclude all deviations from GR down to a small interval around 0, limited only by the actual spin value, the statistics of the observations, and the astrophysical uncertainties. 
A follow-up analysis could use state-of-the-art modeling of the actual X-ray data with accretion disk and
emission models in the Kerr and non-Kerr background spacetimes.

Further progress will be achieved by continuing to refine our understanding of BH accretion disks based on
General Relativistic Magnetohydrodynamic (GRMHD) and General Relativistic Radiative Magnetohydrodynamic (GRRMHD) simulations \citep[e.g.][]{Schnittman2013a, Schnittman2013b, Sadowski2015} and matching simulated
observations to X-ray spectroscopic, X-ray polarization, and X-ray reverberation observations.  

\section{Other Tests of GR in the Strong Gravity Regime} 

The images of the BH shadow of Sgr A$^*$ with the Event Horizon Telescope can give additional constraints.
Whereas the images of the BH shadow still depend on the astrophysics of the accretion disk, 
imaging of the photon ring would be free of such uncertainties. Of course, much better imaging 
would be required to do so.  Furthermore, observations of Sgr A$^*$ could also potentially be used to test the No-Hair theorem by using stellar orbits to measure its spin and quadrupole moment.   These parameters, as well as mass, could also be measured and used to test GR in the strong gravity regime through pulsar observations if one is found close to Sgr A$^*$ (see \citet{Johannsen2015} for a review of tests using Sgr A$^*$).  Gravitational wave detections could also be used to test GR in this regime.  Deviations from GR could be seen in gravitational wave signatures during the ringdown stage after BHs of similar mass merge and from extreme mass ratio inspirals (see \citet{Yunes2013} and references therein).
	
%%% Local Variables: 
%%% mode: latex
%%% TeX-master: "thesis-main"
%%% End: 

\chapter{Ionization Dependence of X-ray Reverberation Signatures}
\label{Reverberation}

\section{Introduction}
\begin{figure}
	\begin{center}
	 \includegraphics[width=0.5\textwidth]{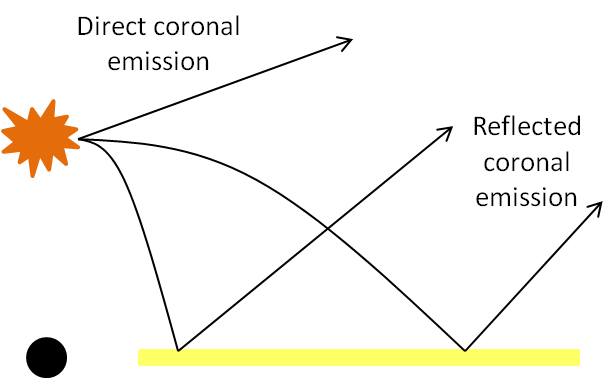}
   			 \caption{\label{fig:diagram}  Diagram illustrating origin of the reverberation signatures with shorter lags resulting from interactions closer to the BH.}
	\end{center}
\end{figure}

\begin{figure}
	\vspace{-50pt}
        \centering
        \begin{subfigure}[b]{\textwidth}
        		\begin{center}
   				 \includegraphics[width=.45\textwidth]{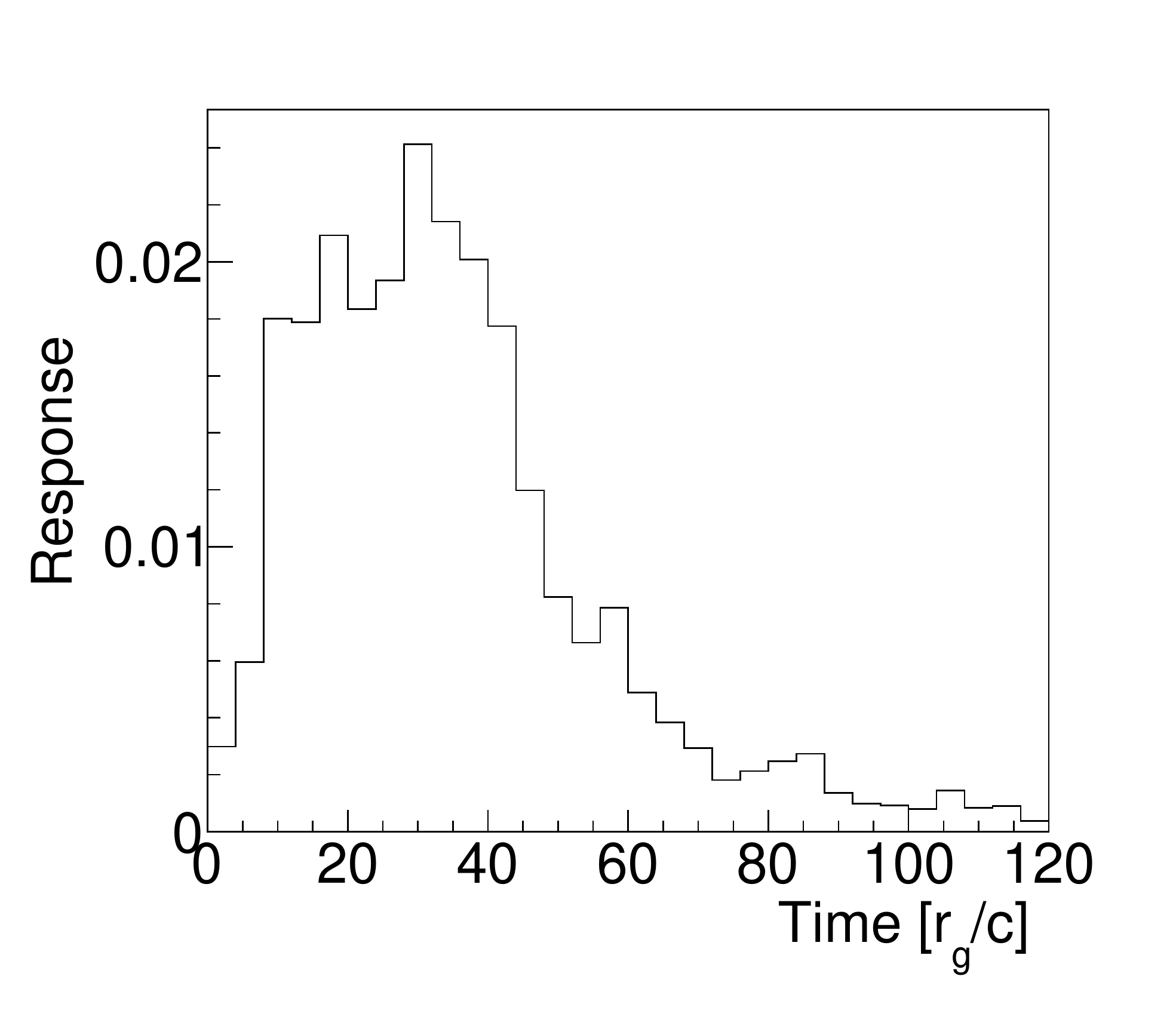}
                \caption{\label{fig:response}} 
             \end{center}
        \end{subfigure}
        \quad
        \begin{subfigure}[b]{\textwidth}
        		\begin{center}
   			 \includegraphics[width=.45\textwidth]{LagFreqThesis.pdf}
              \caption{\label{fig:lagFreq}} 
             \end{center}
        \end{subfigure}
        \quad
        \begin{subfigure}[b]{\textwidth}
       		\begin{center}
   			 \includegraphics[width=.45\textwidth]{LagEnergyThesis.pdf}
              \caption{\label{fig:lagEnergy}}
             \end{center} 
        \end{subfigure}
        \caption{\label{fig:reverb2} Fe-K reverberation signatures calculated for a BH with $a =$ 0.998, $i =$ 45$^\circ$, $h =$ 5 $r_g$ with Figure \ref{fig:response} showing the transfer function, Figure \ref{fig:lagFreq} showing the lag-frequency spectrum, and Figure \ref{fig:lagEnergy} showing the lag-energy spectrum calculated in the (1-2)$\times 10^{-3}$ c/$r_g$ shown in the grey shaded region in Figure \ref{fig:lagFreq}.}
\end{figure}

As discussed in Chapter \ref{Intro:timing},  X-ray reverberation is a powerful tool to study the structure of accretion flow, in particular when looking at the time lag between the direct coronal emission and both the Fe-K line and Compton hump (see \citet{Uttley2014} for a review).  This chapter describes the methods used to study the dependence of the X-ray reverberation signatures on the ionization of the accretion disk. In order to calculate this time lag the cross-spectrum is calculated given two light curves, one characterizing the direct emission and the second characterizing the reflected emission.  The phase difference, which relates directly to the lag, is then found by taking the argument of this cross-spectrum \citep{Nowak1999}. Figure \ref{fig:diagram} shows the difference in the path and therefore the light travel time depending on where the photon interacts with the disk.  This chapter will further discuss the effect of spin, height, inclination, and ionization on the observed lags.
To do this using the simulations described in Chapter \ref{RayTracingCode} the Fe-K reverberation is calculated following the methodology presented in \citet{Cackett2014} which will be outlined below.  Results for each case are shown in Figure \ref{fig:reverb2} for a BH with $a = 0.998$, $i = 45^\circ$, and $h = 5 r_g$.  A light curve is calculated for the reflected emission in the energy range of interest, in this case a band covering the iron line from 2-10 keV.  The transfer function, $\psi(t)$, proportional to the distribution of photon arrival times $t$ for a corona flare at time $t=0$ shown in Figure \ref{fig:response}, describes the response of the disk to the impinging light and is calculated by rescaling the $x$-axis of this curve which signifies the arrival times of the photons using the arrival time of the direct coronal emission (found in the energy band from 1-2 keV) as the zero point.  In order to determine the lag-frequency spectrum characterizing the time delay (or lag) between the direct and reflected emission, the Fourier transform of $\psi(t)$ is calculated using the equation
\begin{equation} \label{eq:FT}
\Psi(f) = R \int_{0}^{\infty} \psi(t) \exp^{-2 \pi f t} dt
\end{equation}
where $R$ is the ratio of the reflected flux to the power-law flux.  This can be found using the ray-tracing code and is set to 1 in the results presented here. The effect of changing this value is discussed in \citep{Cackett2014}.
The phase difference between the direct and reflected emission is calculated by taking the argument of Equation \ref{eq:FT}
\begin{equation}
\phi(f) = \tan^{-1}\left(\frac{Im(\Psi)}{1+Re(\Psi)}\right)
\end{equation}
where the additional factor of 1 in the denominator accounts for the presence of the direct emission in the reflected band.  The lag, $\tau$, at each frequency $f_n$ is calculated by converting the phase difference to linear frequency leading to
\begin{equation}
\tau = \frac{\phi}{2 \pi f_n}
\end{equation}
for each $f_n = n/N dt$ given N frequency bins of width $dt$ and $1 \leq n \leq N$.  Figure \ref{fig:lagFreq} shows the resulting lag-frequency spectrum.  In order to determine the relationship between the lag and the energy of the photons a lag-energy spectrum can be calculated.  This is done by calculating multiple transfer functions in small energy ranges.  In this case transfer functions are calculated in 0.5 keV energy bands ranging from 2-10 keV.  Then, in each band the lag is calculated as before but in this case only considering an isolated frequency range.  This lag-energy spectrum is shown in Figure \ref{fig:lagEnergy} where the lag was calculated for frequencies in the range from (1-2)$\times 10^{-3} \ c/r_{g}$.  This frequency range can be changed depending on the region of the disk you want to examine from seeing the whole disk at low frequencies to focusing on the inner disk at high frequencies.

\section{Modeling Radial Ionization}
\begin{figure}
\begin{center}
\vspace{-40pt}
	\begin{subfigure}[b]{\textwidth}
		\begin{center}
   			 \includegraphics[width=0.8\textwidth]{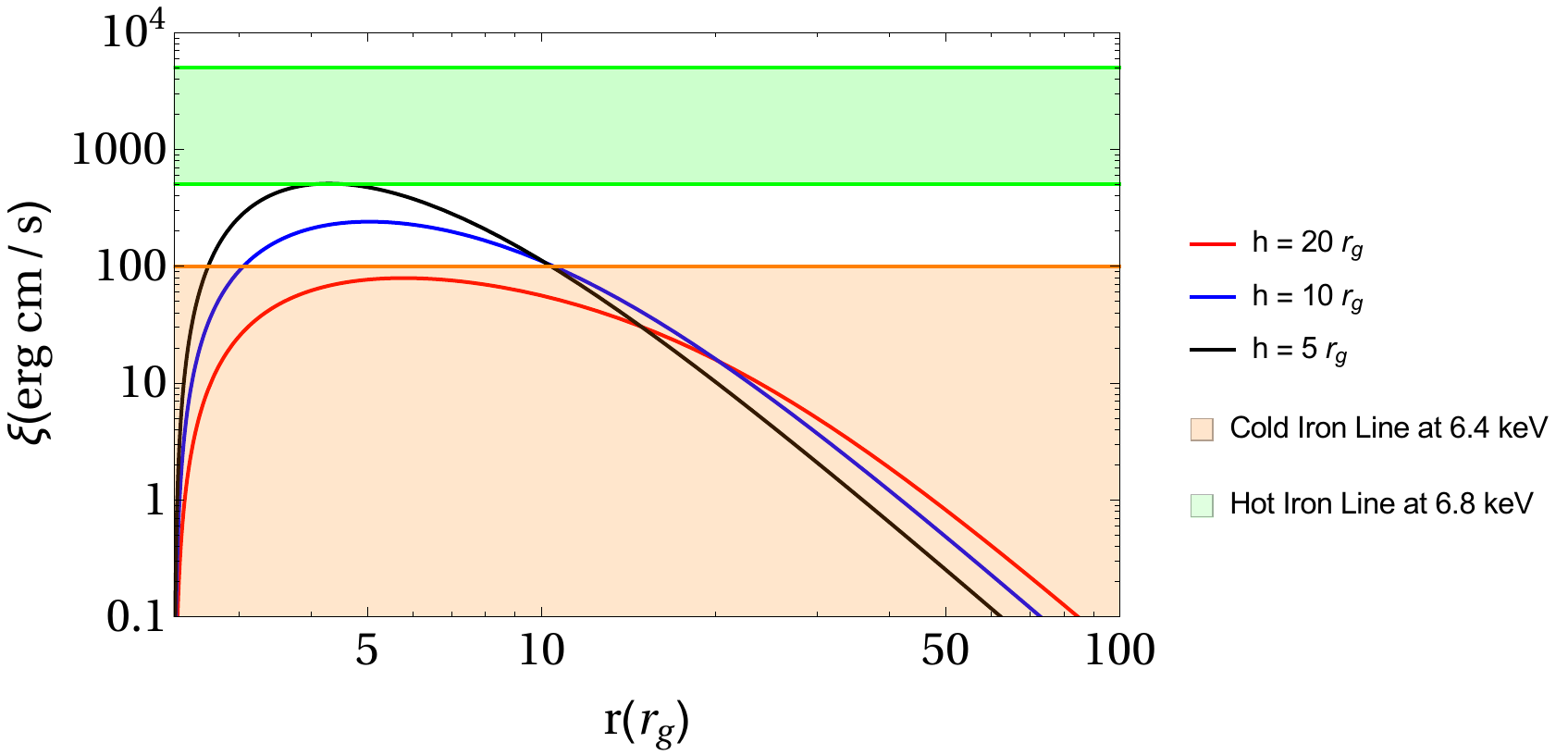}
   			 \caption{\label{fig:mattH} Ionization as a function of radius for a BH with $a = 0.9$, $\dot{m}=0.1$ and varying $h$.}
   		\end{center}
	\end{subfigure}
	\quad
	\begin{subfigure}[b]{\textwidth}
		\begin{center}
   			 \includegraphics[width=0.8\textwidth]{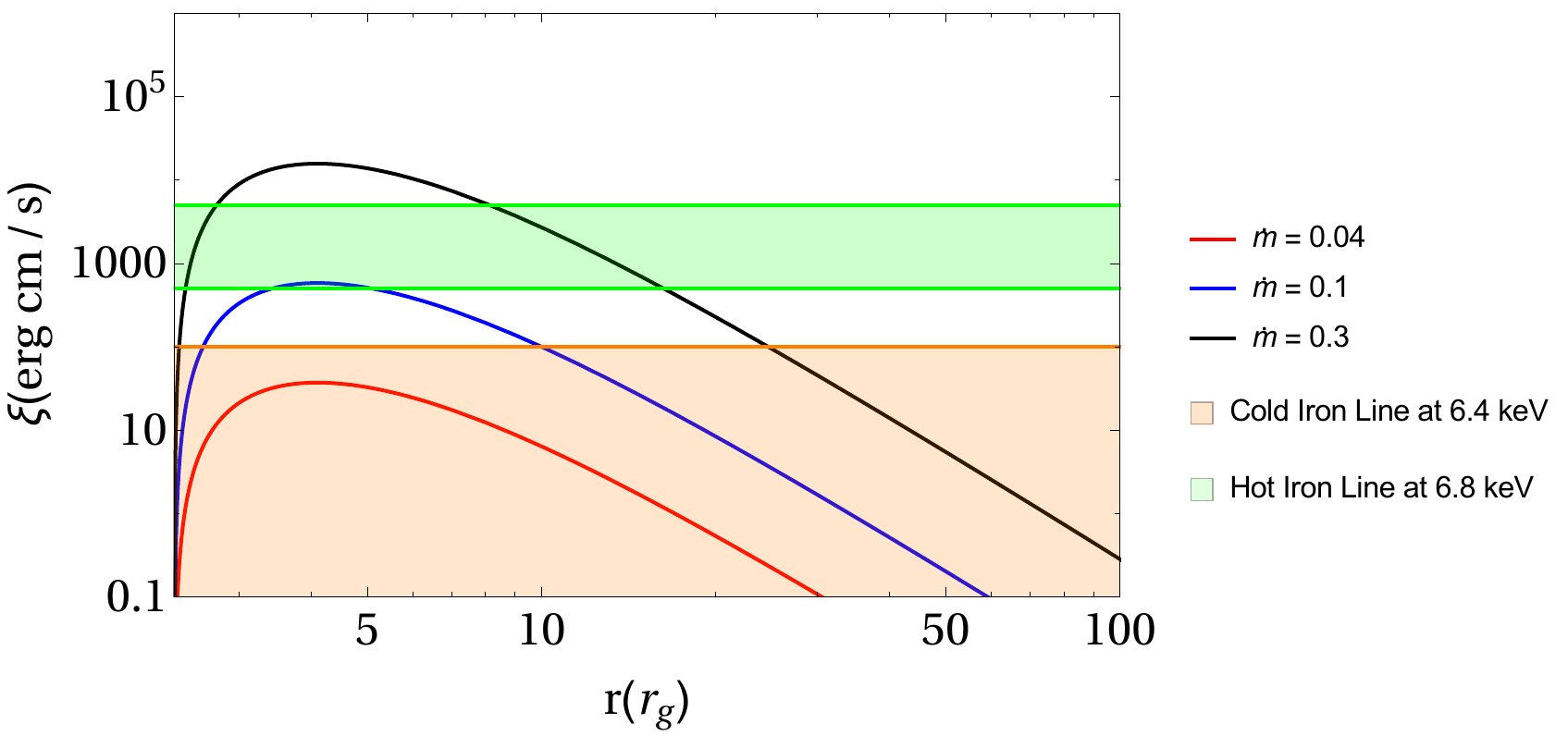}
   			 \caption{\label{fig:mattM} Ionization as a function of radius for a BH with $a = 0.9$, $h = 4r_{g}$ and varying $\dot{m}$.}
   		\end{center}
	\end{subfigure}
	\quad
	\begin{subfigure}[b]{\textwidth}
		\begin{center}
   			 \includegraphics[width=0.8\textwidth]{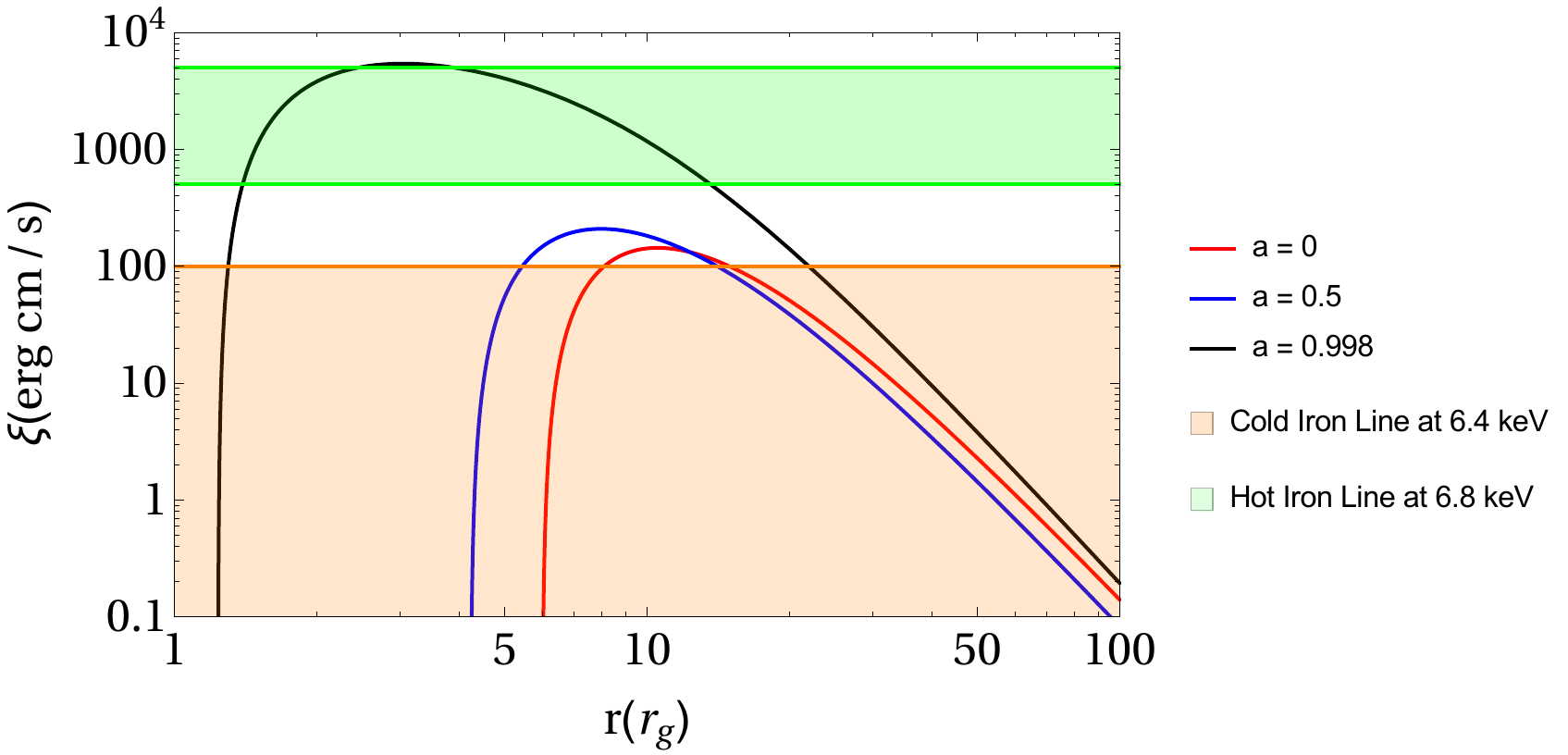}
   			 \caption{\label{fig:mattA} Ionization as a function of radius for a BH with $h = 10 r_g$, $\dot{m} = 0.1$, and varying spin.}
   		\end{center}
	\end{subfigure}
\end{center}
\caption{\label{fig:ionMatt} Ionization as a function of radius as defined by Equation \ref{eq:ionMatt} while varying lamp-post height (Figure \ref{fig:mattH}), accretion rate (Figure \ref{fig:mattM}), and spin (Figure \ref{fig:mattA}).}
\end{figure}

The ionization of the accretion disk, which is used to determine the likelihood that an iron line will be created, is defined as \citep{Matt1993}
\begin{equation}
\xi = \frac{4 \pi F}{n_H}
\end{equation}
where $n_h$ is the hydrogen number density and F is the incident flux. It has been found that the tendency to form an iron line can be broken up into four categories based upon the ionization parameter in the following way
\begin{itemize}
\item $\xi < $100 erg cm s$^{-1}$: cold iron line at 6.4 keV

\item 100 erg cm s$^{-1} < \xi < $500 erg cm s$^{-1}$: very weak iron line

\item 500 erg cm s$^{-1} < \xi < $5000 erg cm s$^{-1}$: hot iron line at 6.8 keV

\item $\xi > $5000 erg cm s$^{-1}$: no iron line.
\end{itemize}
Strong iron lines are seen at low and moderately high ionizations. However, highly ionized gas is unable to produce an iron line and the iron line is very weak for moderately low ionizations as a majority of the photons are subsequently absorbed following emission and trapped in the disk \citep{Matt1993, Matt1996, Fabian2000}.  \citet{Matt1993} derived an expression for the ionization of an accretion disk surrounding a SMBH being illuminated by a lamp-post corona which is given by
\begin{equation} \label{eq:ionMatt}
\xi(r) = 8.97 \times 10^8 \left( \frac{\dot{m}^3 \eta_h \alpha}{\eta^2} \right) f^2(r) r^{-3/2} g(r,h)
\end{equation}
where
\begin{equation}
g(r,h) = \frac{h}{(r^2+h^2)^{3/2}}
\end{equation}
and
\begin{equation}
f(r) = 1-\sqrt{r_{ISCO}/r}.
\end{equation}
Here, $\dot{m}$ is the accretion rate normalized by the Eddington luminosity, $\eta = 1-E_{ISCO}$ is the conversion efficiency, $\eta_h$ describes the fraction between the hard and soft luminosities, and $\alpha$ is the $\alpha$-viscosity value.  The following analysis assumes $\alpha = 0.1$ and  $\eta_h = 1$.  Figure \ref{fig:ionMatt} shows the ionization as defined by Equation \ref{eq:ionMatt} while varying the lamp-post height, accretion rate, and BH spin.  The ionization parameter range yielding hot and cold iron lines are shown by the green and orange shaded regions respectively.

\begin{figure}
	\begin{center}
	 \includegraphics[width=0.75\textwidth]{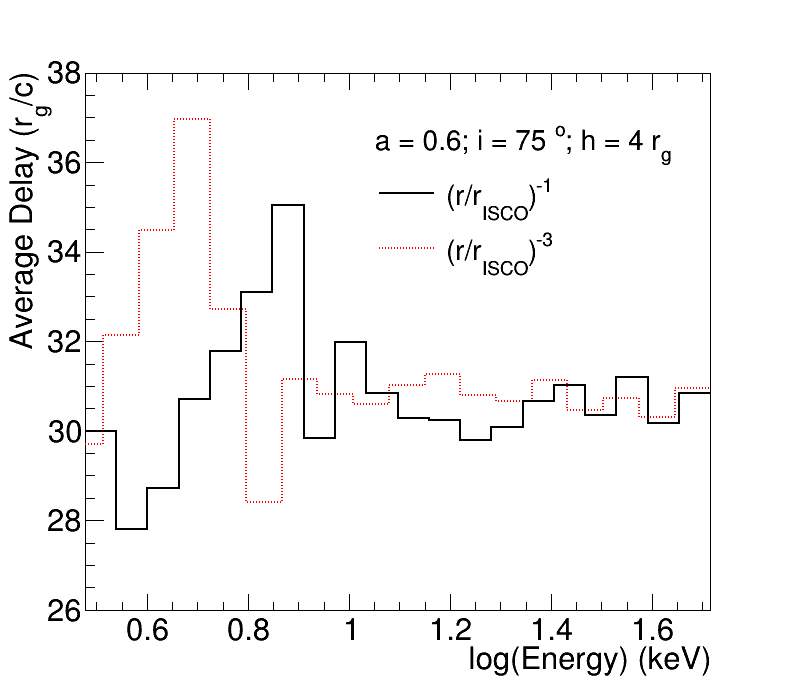}
   			 \caption{\label{fig:arrival}  Average delay of scattered photons for two radial ionization schemes.}
	\end{center}
\end{figure}

In order to see the effect varying the ionization as a function of radius has on the expected reverberation signatures, simulations were run scaling the ionization parameter by the factor $(r/r_{ISCO})^{-\gamma}$
where $\gamma = 1,3$ and using the prescription described above to determine if an iron line should be created.  Using these results the average delay between direct and reflected emission for photons with various energies were determined as shown in Figure \ref{fig:arrival}.  From this, it is seen that ionization does play a role in determining lags, particularly at lower energies.

In order to better understand the effects ionization schemes will have on reverberation signatures the following modifications were made to the code described in Chapter \ref{RayTracingCode}.  The photons are tracked as before; however, the first time they hit the disk there is a 90\% chance that an Fe-K$\alpha$ line will be emitted provided the photon has sufficient energy to create one.  All subsequent scatterings will occur with no chance of creating an Fe-K$\alpha$ line.  Once the simulations are run the events can be weighted using the following methodology to phenomenologically model the radial ionization of the disk.  The emitted photons are scaled as a power-law with 
\begin{equation}
\frac{dN}{dE} = E^{-\Gamma}.
\end{equation}
The probability of scattering versus absorption is described using two arbitrary parameters, $a0$ and $\gamma_1$,  leading to the following weighting, 
\begin{equation}
w = a0 \left( \frac{r_{scatter}}{r_{ISCO}} \right)^{-\gamma_1}.
\end{equation}
 The first time a photon hits the disk it is either multiplied by the weight $w_s$ if it scatters or $w_{ka}$ if it creates an Fe-K$\alpha$ line. These weights are determined by solving the following system of equations:
\begin{equation}
0.1 w_{s} + 0.9 w_{ka} = 1
\end{equation}
to model an equal probability of Fe-K$\alpha$ emission versus scattering and 
\begin{equation}
\frac{0.1w_s}{0.9 w_{ka}} = R(r)
\end{equation}
to add in a radial dependence to this process where 
\begin{equation}
R(r) = r0 \left(\frac{r_{scatter}}{r_{ISCO}} \right) ^{-\gamma_2}.
\end{equation}

\section{Results}
\begin{figure}
\begin{center}
\vspace{-40pt}
	\begin{subfigure}[b]{\textwidth}
		\begin{center}
   			 \includegraphics[width=0.75\textwidth]{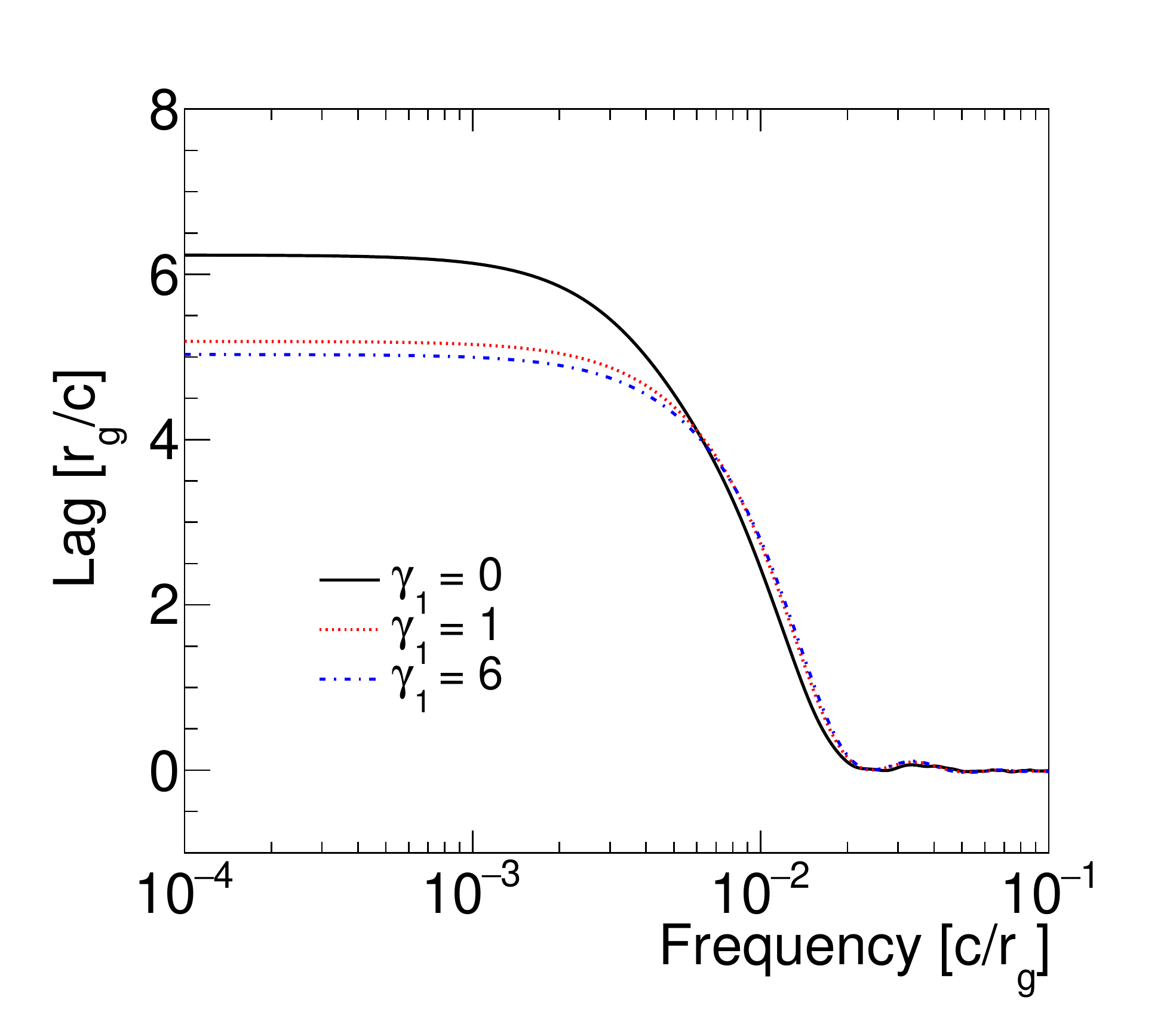}
   			 \caption{\label{fig:i1a}}
   		\end{center}
	\end{subfigure}
	\quad
	\begin{subfigure}[b]{\textwidth}
		\begin{center}
   			 \includegraphics[width=0.75\textwidth]{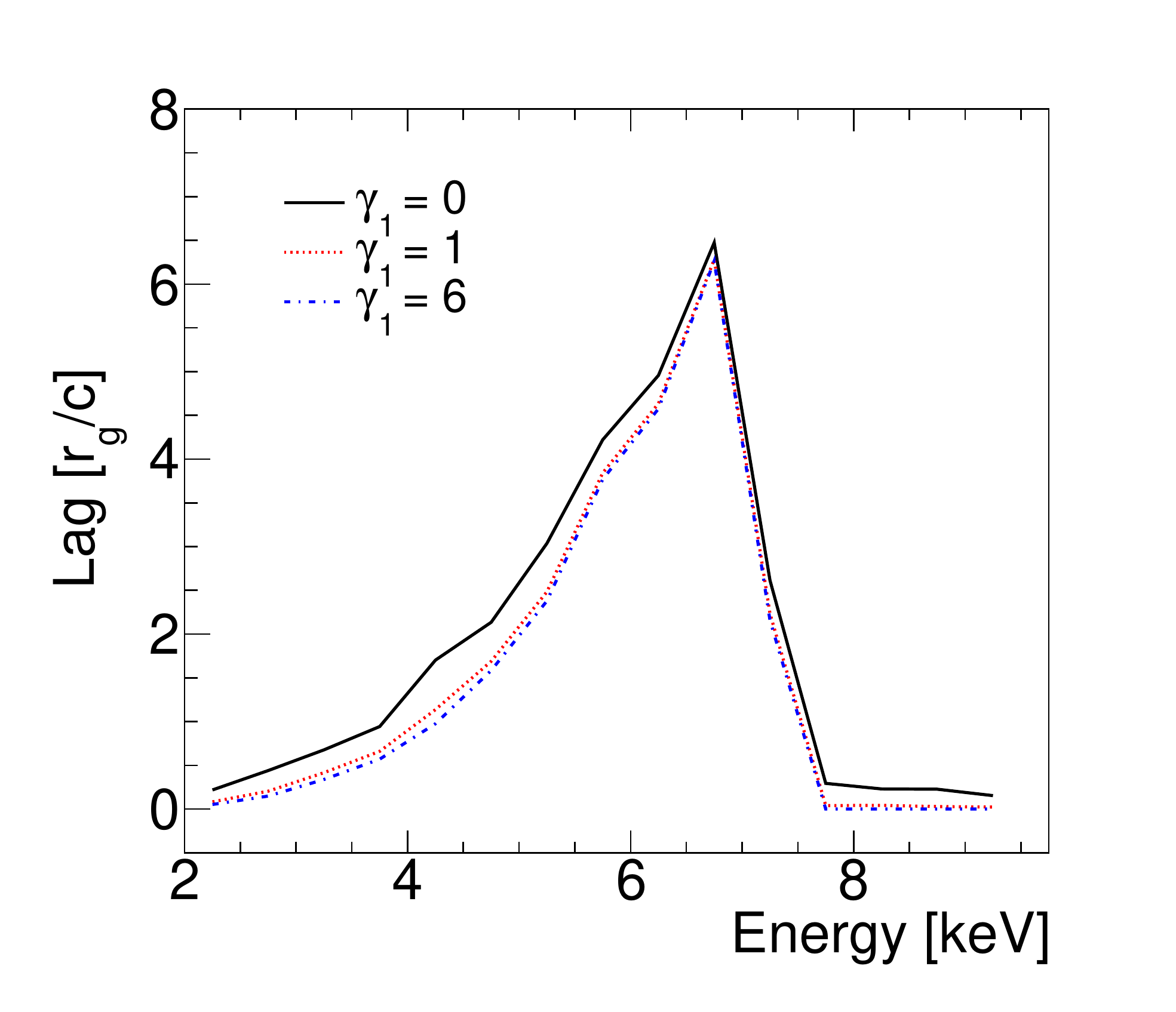}
   			 \caption{\label{fig:i1b}}
   		\end{center}
	\end{subfigure}
	\caption{\label{fig:index1} Lag-frequency spectrum (Figure \ref{fig:i1a}) and lag-energy spectrum (Figure \ref{fig:i1b}) varying $\gamma_1$.}
\end{center}
\end{figure}

\begin{figure}
\begin{center}
\vspace{-40pt}
	\begin{subfigure}[b]{\textwidth}
		\begin{center}
   			 \includegraphics[width=0.75\textwidth]{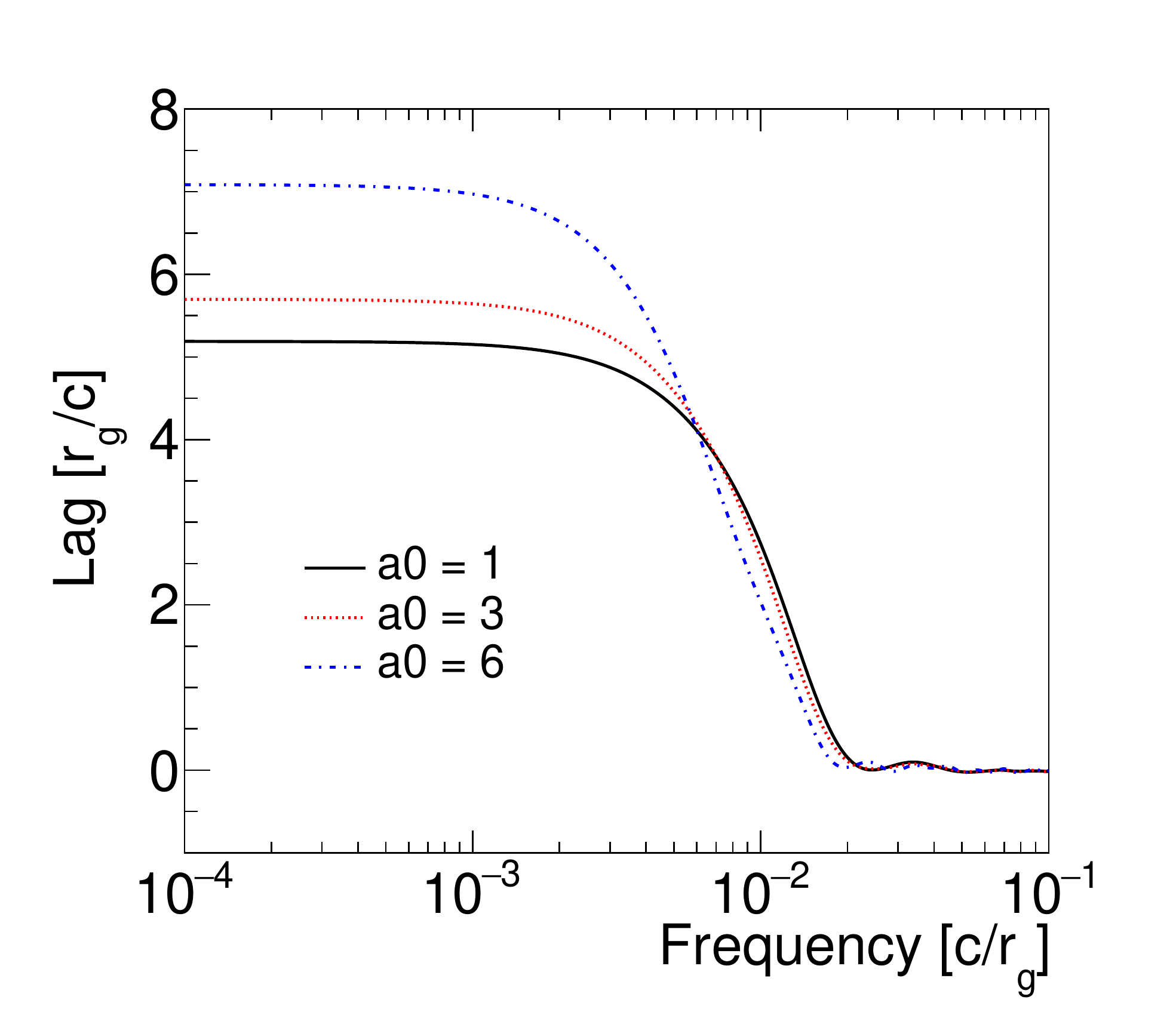}
   			 \caption{\label{fig:a0a}}
   		\end{center}
	\end{subfigure}
	\quad
	\begin{subfigure}[b]{\textwidth}
		\begin{center}
   			 \includegraphics[width=0.75\textwidth]{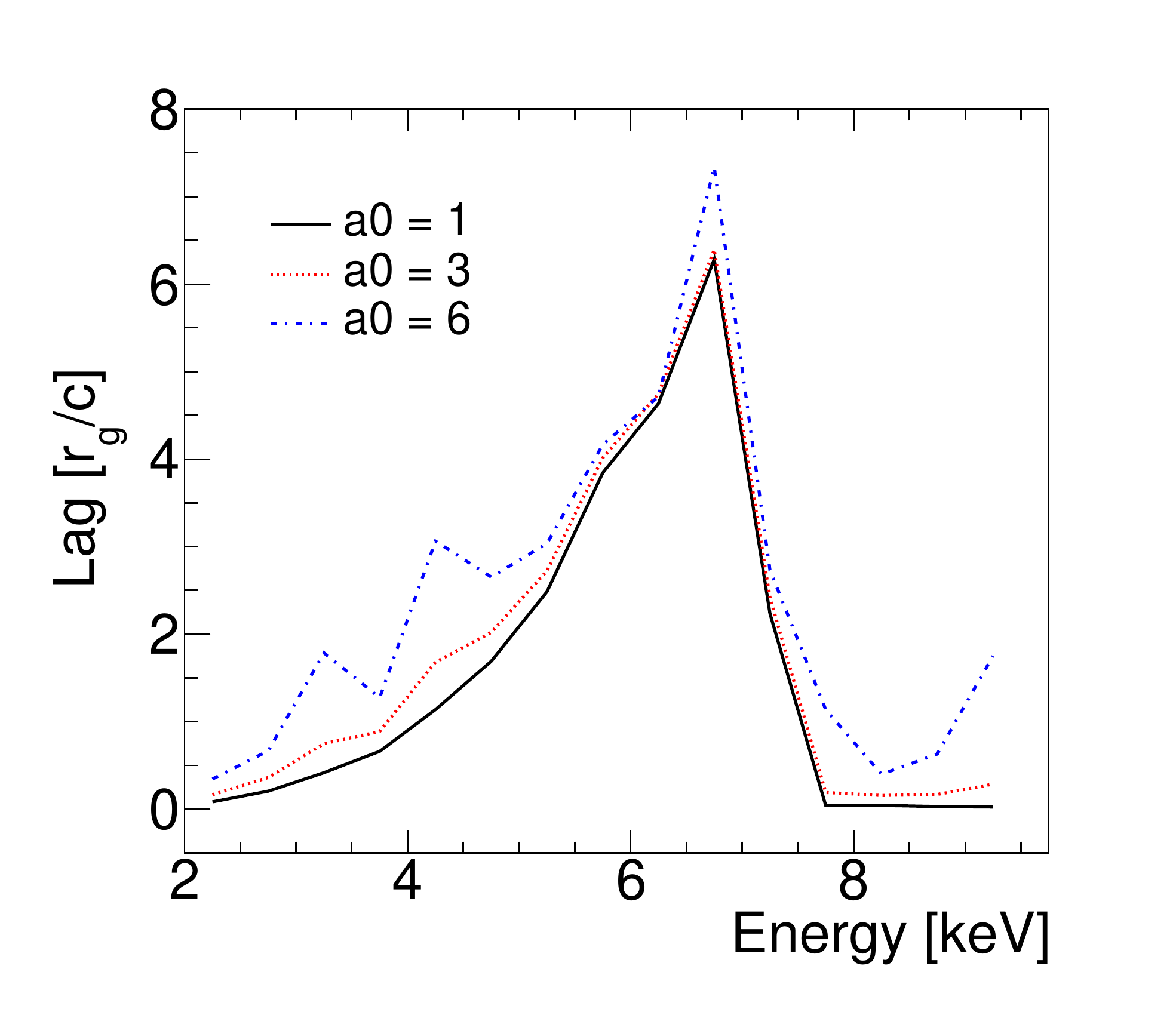}
   			 \caption{\label{fig:a0b}}
   		\end{center}
	\end{subfigure}
\end{center}
	\caption{\label{fig:a0} Lag-frequency spectrum (Figure \ref{fig:a0a}) and lag-energy spectrum (Figure \ref{fig:a0b}) varying $a0$.}
\end{figure}

\begin{figure}
\begin{center}
\vspace{-40pt}
	\begin{subfigure}[b]{\textwidth}
		\begin{center}
   			 \includegraphics[width=0.75\textwidth]{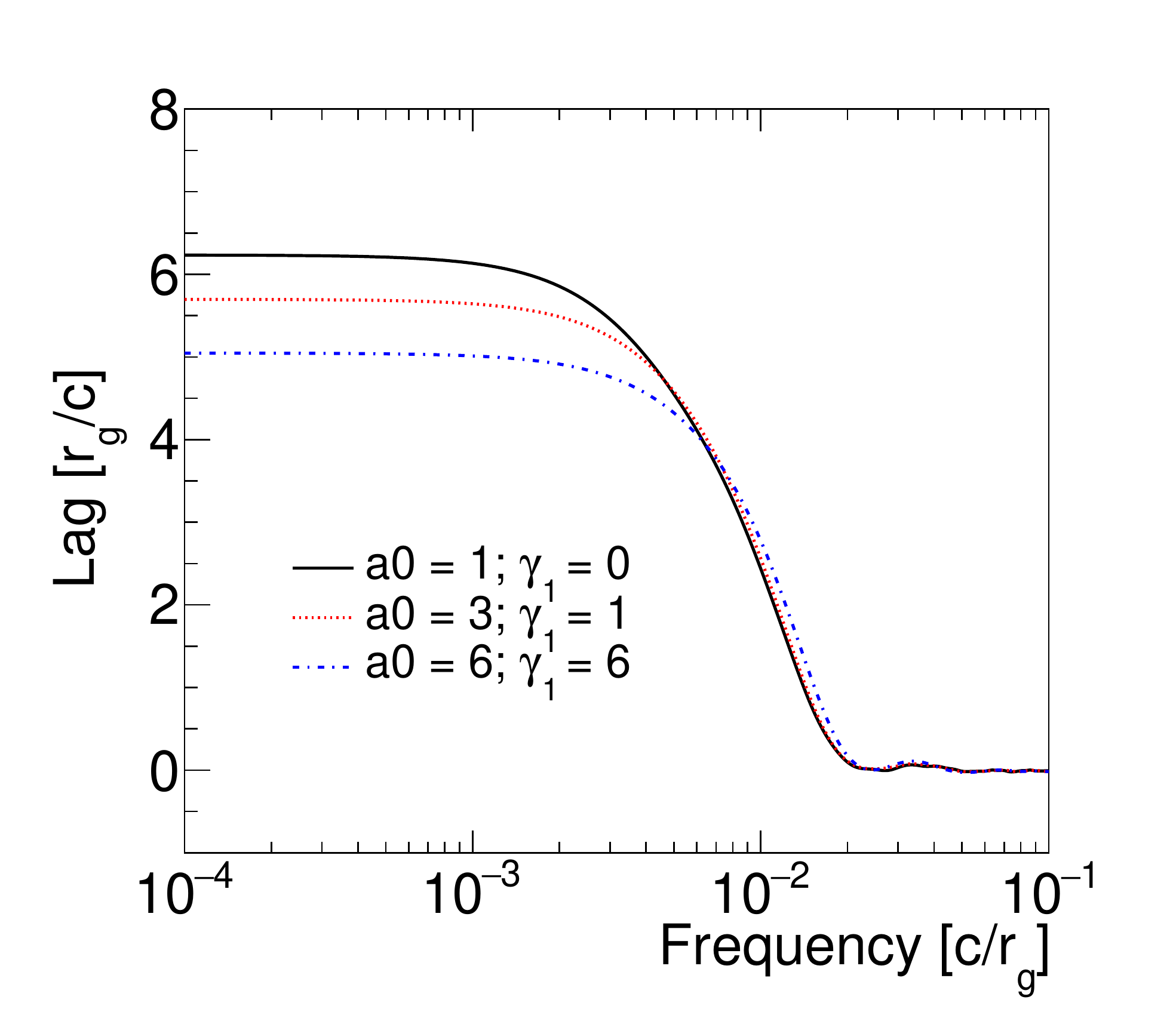}
   			 \caption{\label{fig:v22a}}
   		\end{center}
	\end{subfigure}
	\quad
	\begin{subfigure}[b]{\textwidth}
		\begin{center}
   			 \includegraphics[width=0.75\textwidth]{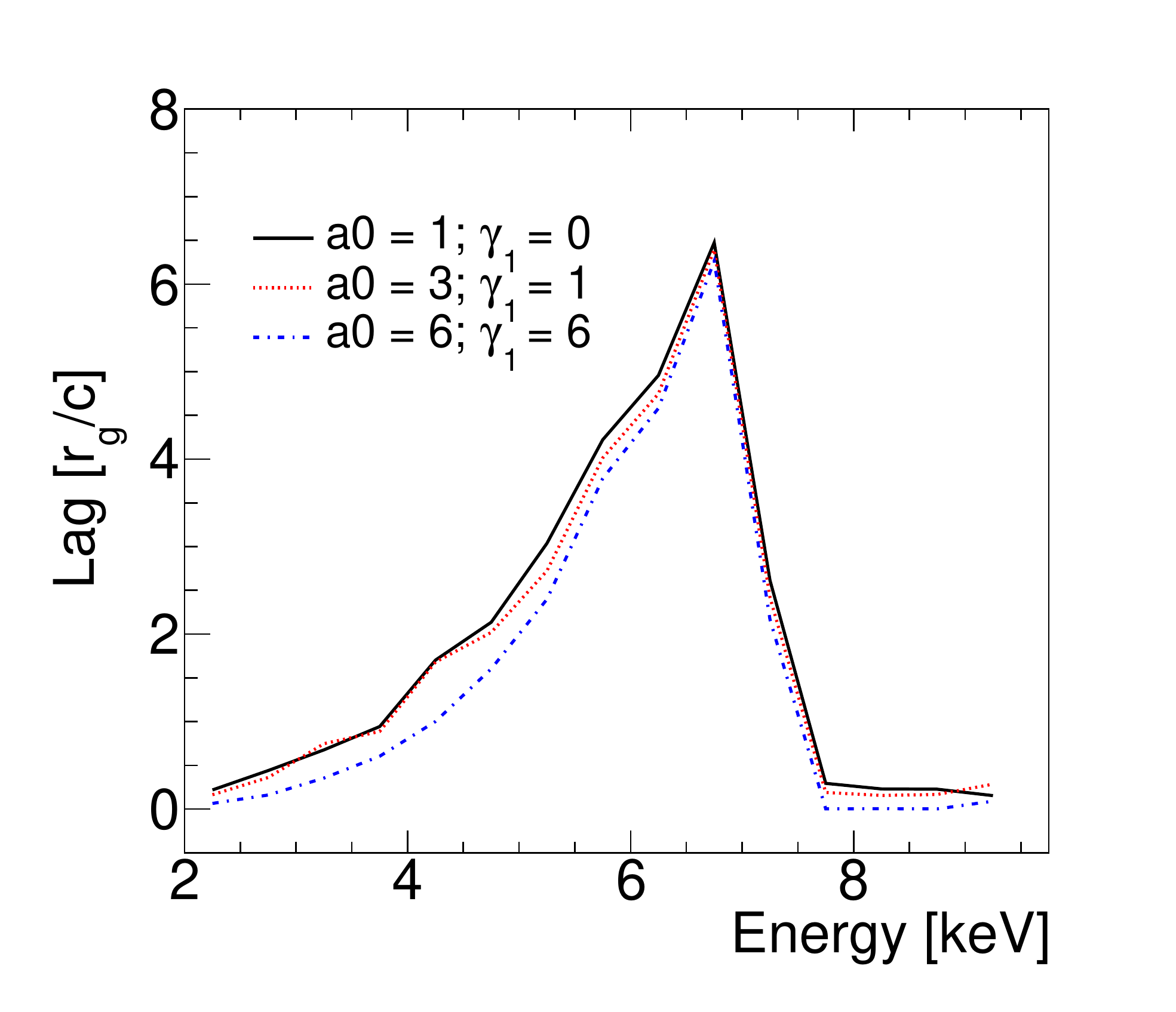}
   			 \caption{\label{fig:v22b}}
   		\end{center}
	\end{subfigure}
\end{center}
	\caption{\label{fig:v22} Lag-frequency spectrum (Figure \ref{fig:v22a}) and lag-energy spectrum (Figure \ref{fig:v22b}) varying $\gamma_1$ and $a0$.}
\end{figure}

\begin{figure}
\begin{center}
\vspace{-40pt}
	\begin{subfigure}[b]{\textwidth}
		\begin{center}
   			 \includegraphics[width=0.75\textwidth]{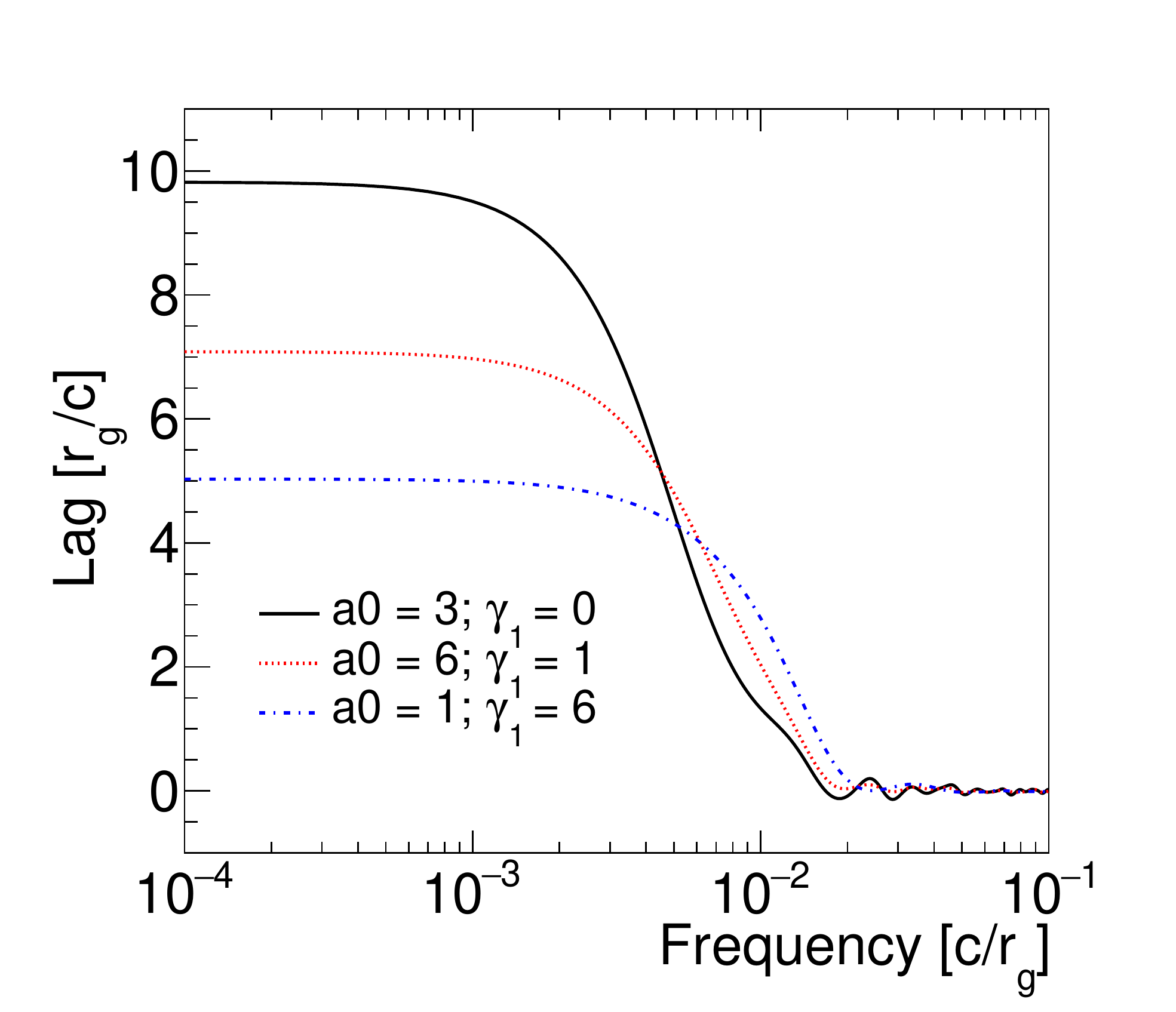}
   			 \caption{\label{fig:v2a}}
   		\end{center}
	\end{subfigure}
	\quad
	\begin{subfigure}[b]{\textwidth}
		\begin{center}
   			 \includegraphics[width=0.75\textwidth]{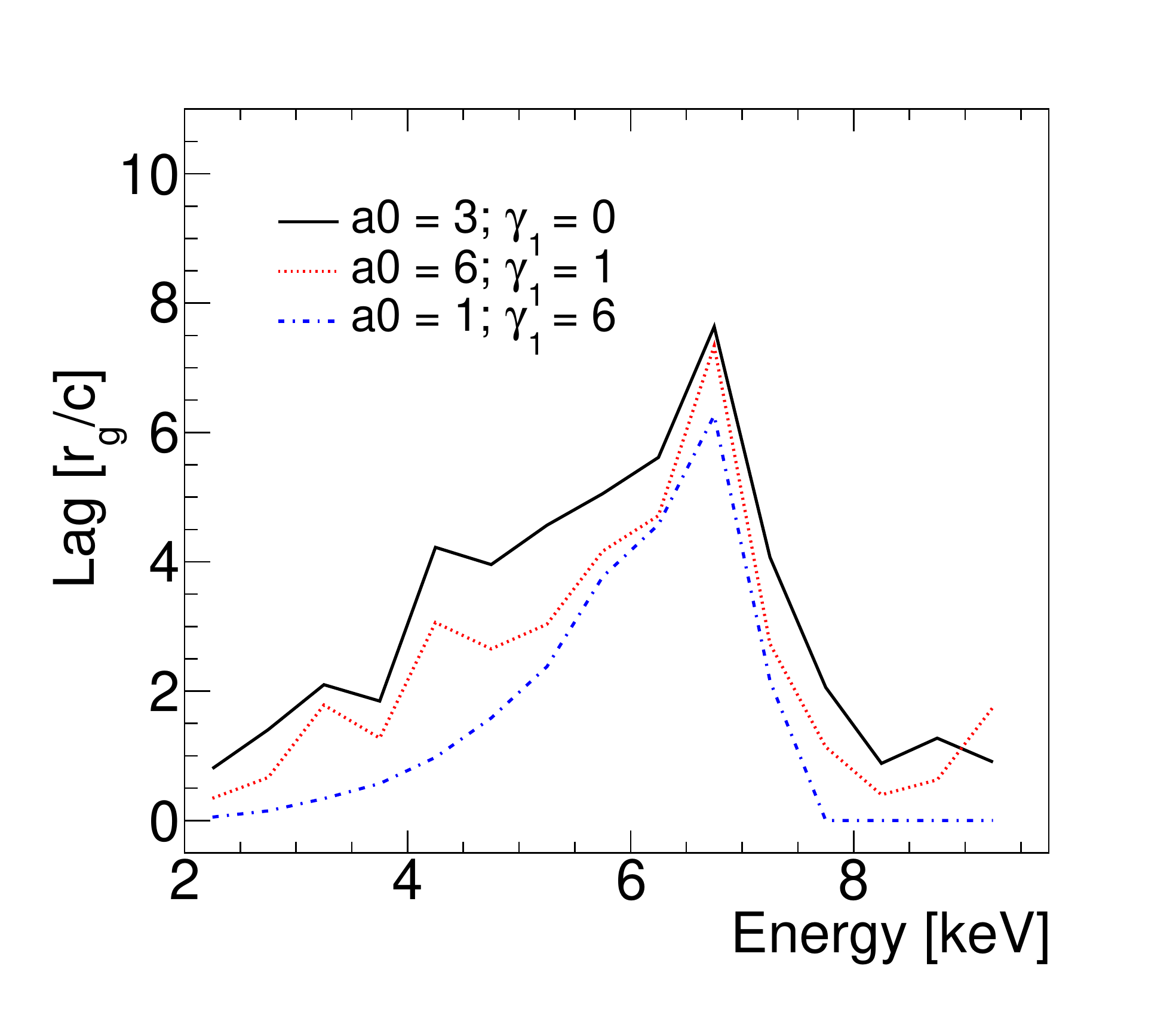}
   			 \caption{\label{fig:v2b}}
   		\end{center}
	\end{subfigure}
\end{center}
	\caption{\label{fig:v2} Lag-frequency spectrum (Figure \ref{fig:v2a}) and lag-energy spectrum (Figure \ref{fig:v2b}) varying $\gamma_1$ and $a0$.}
\end{figure}

\begin{figure}
\begin{center}
\vspace{-40pt}
	\begin{subfigure}[b]{\textwidth}
		\begin{center}
   			 \includegraphics[width=0.75\textwidth]{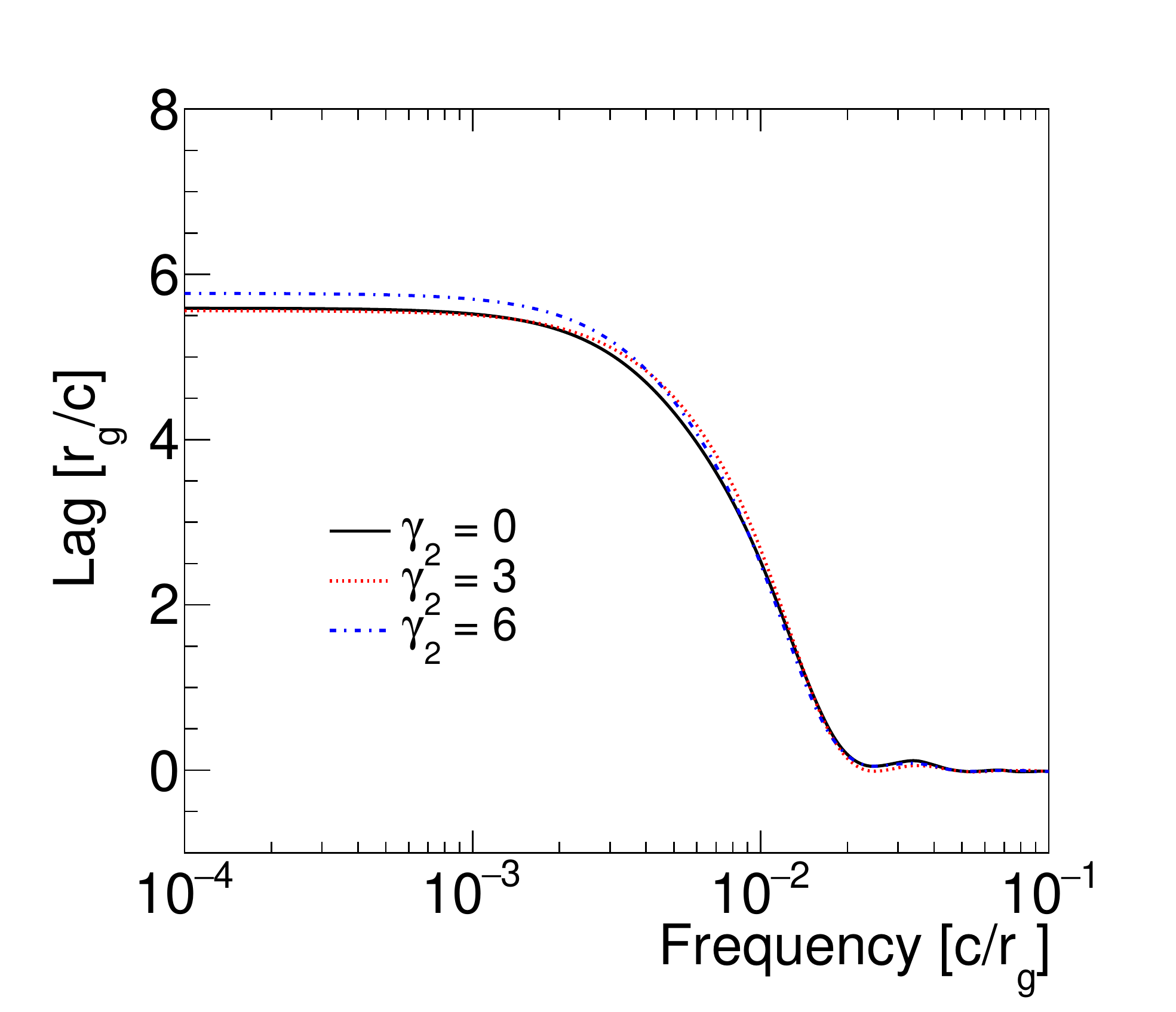}
   			 \caption{\label{fig:i2a}}
   		\end{center}
	\end{subfigure}
	\quad
	\begin{subfigure}[b]{\textwidth}
		\begin{center}
   			 \includegraphics[width=0.75\textwidth]{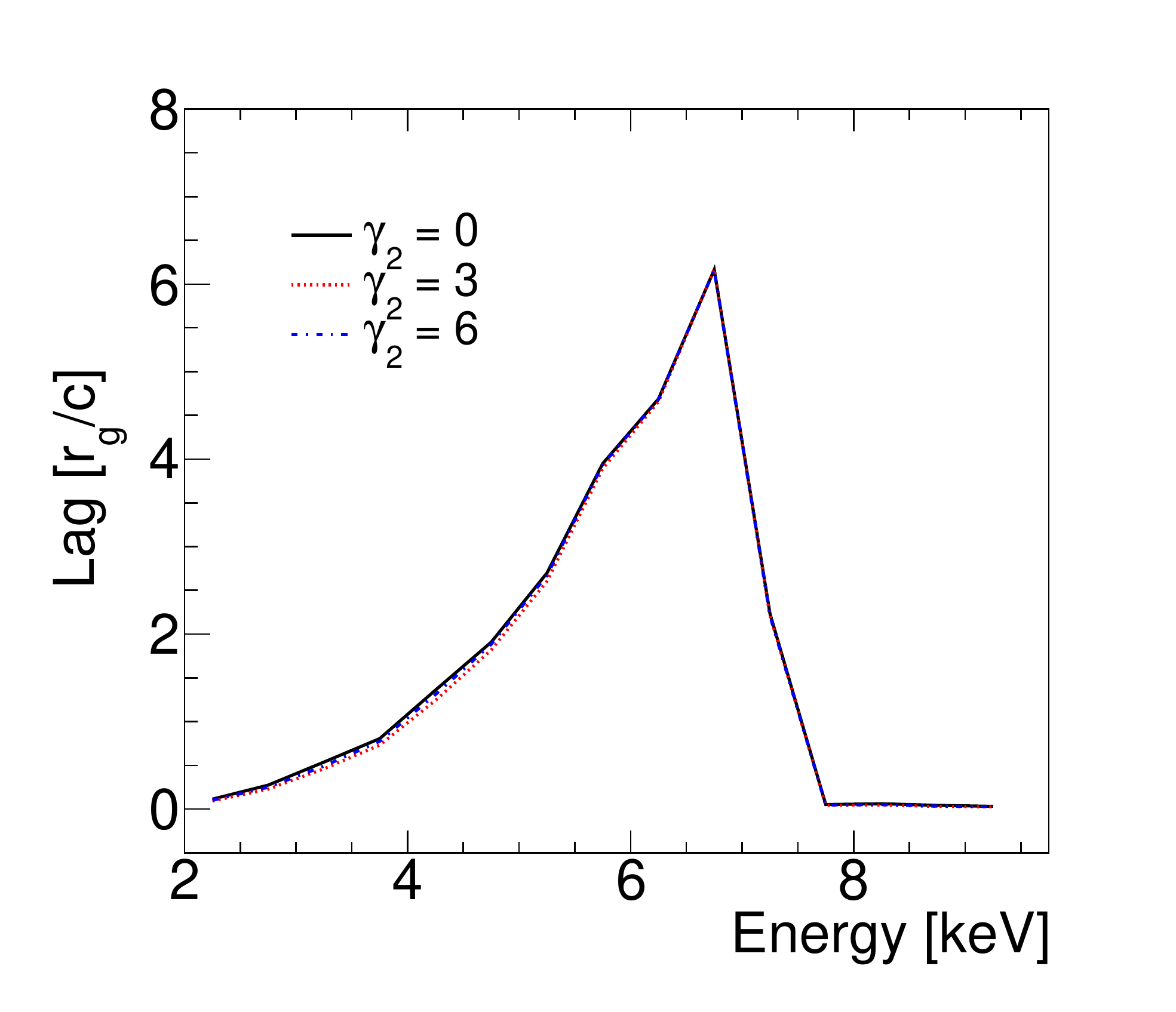}
   			 \caption{\label{fig:i2b}}
   		\end{center}
	\end{subfigure}
\end{center}
	\caption{\label{fig:index2} Lag-frequency spectrum (Figure \ref{fig:i2a}) and lag-energy spectrum (Figure \ref{fig:i2b}) varying $\gamma_2$.}
\end{figure}

\begin{figure}
\begin{center}
\vspace{-40pt}
	\begin{subfigure}[b]{\textwidth}
		\begin{center}
   			 \includegraphics[width=0.75\textwidth]{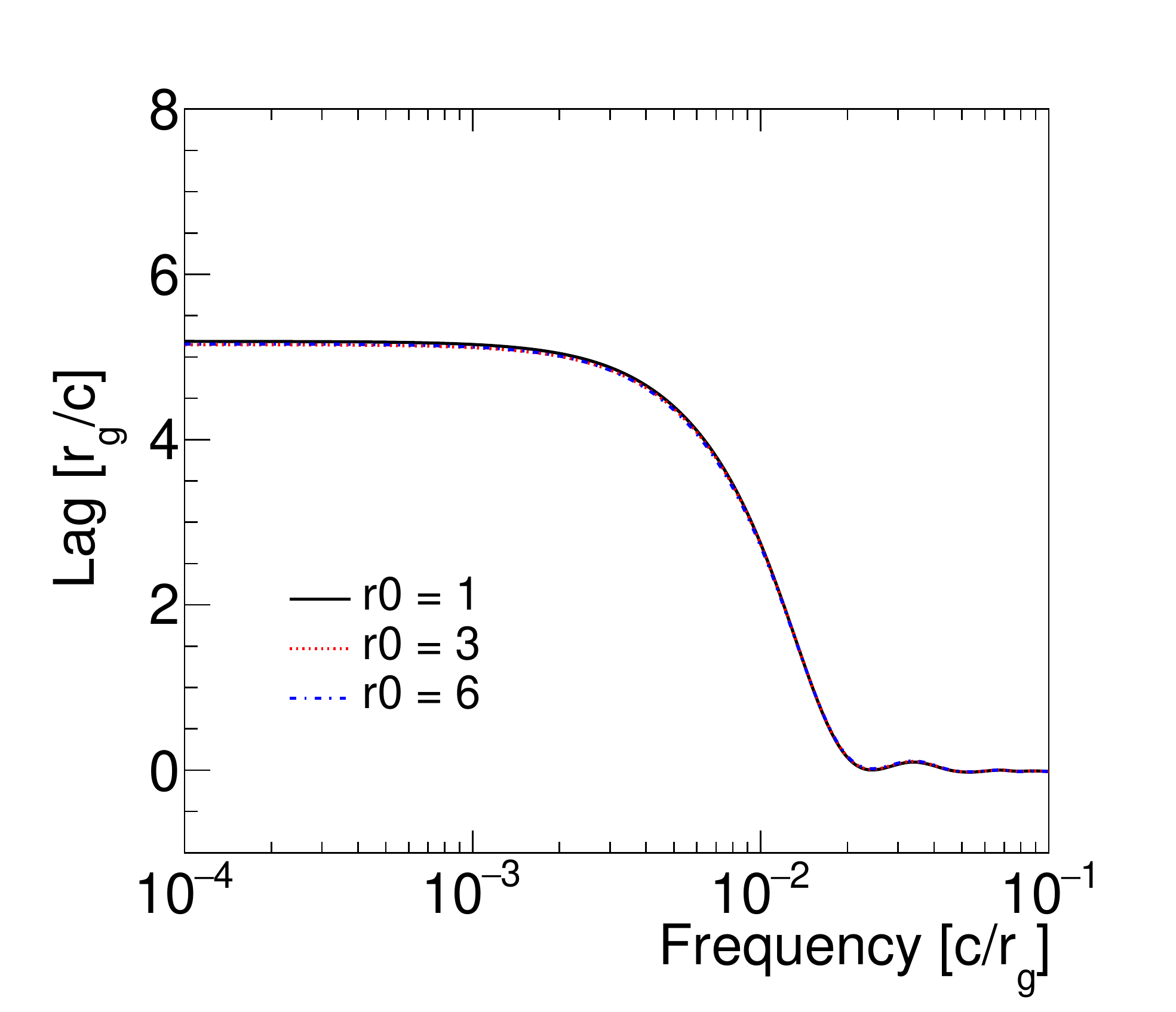}
   			 \caption{\label{fig:r0a}}
   		\end{center}
	\end{subfigure}
	\quad
	\begin{subfigure}[b]{\textwidth}
		\begin{center}
   			 \includegraphics[width=0.75\textwidth]{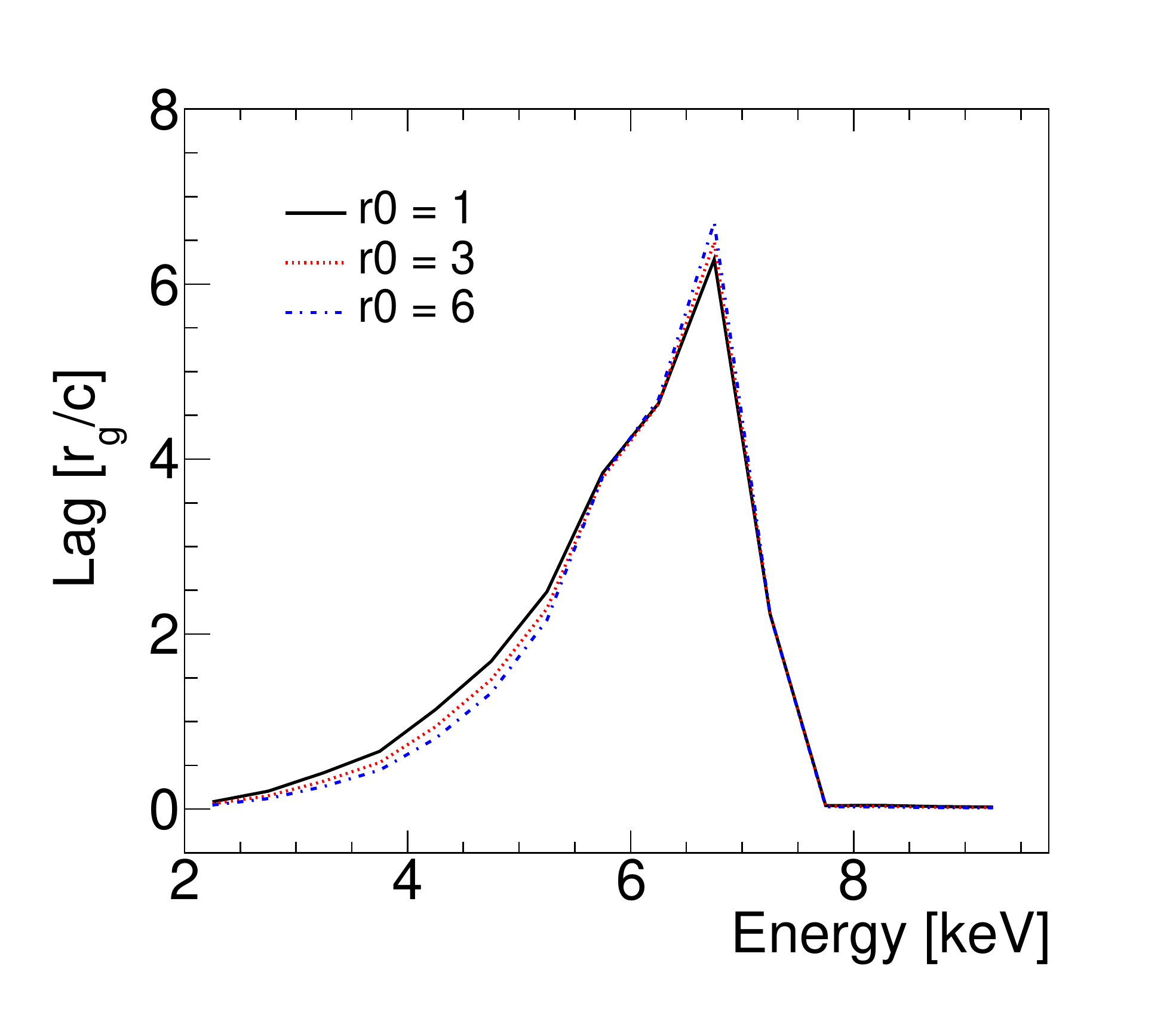}
   			 \caption{\label{fig:r0b}}
   		\end{center}
	\end{subfigure}
\end{center}
	\caption{\label{fig:r0} Lag-frequency spectrum (Figure \ref{fig:r0a}) and lag-energy spectrum (Figure \ref{fig:r0b}) varying $r0$.}
\end{figure}

\begin{figure}
\begin{center}
\vspace{-40pt}
	\begin{subfigure}[b]{\textwidth}
		\begin{center}
   			 \includegraphics[width=0.75\textwidth]{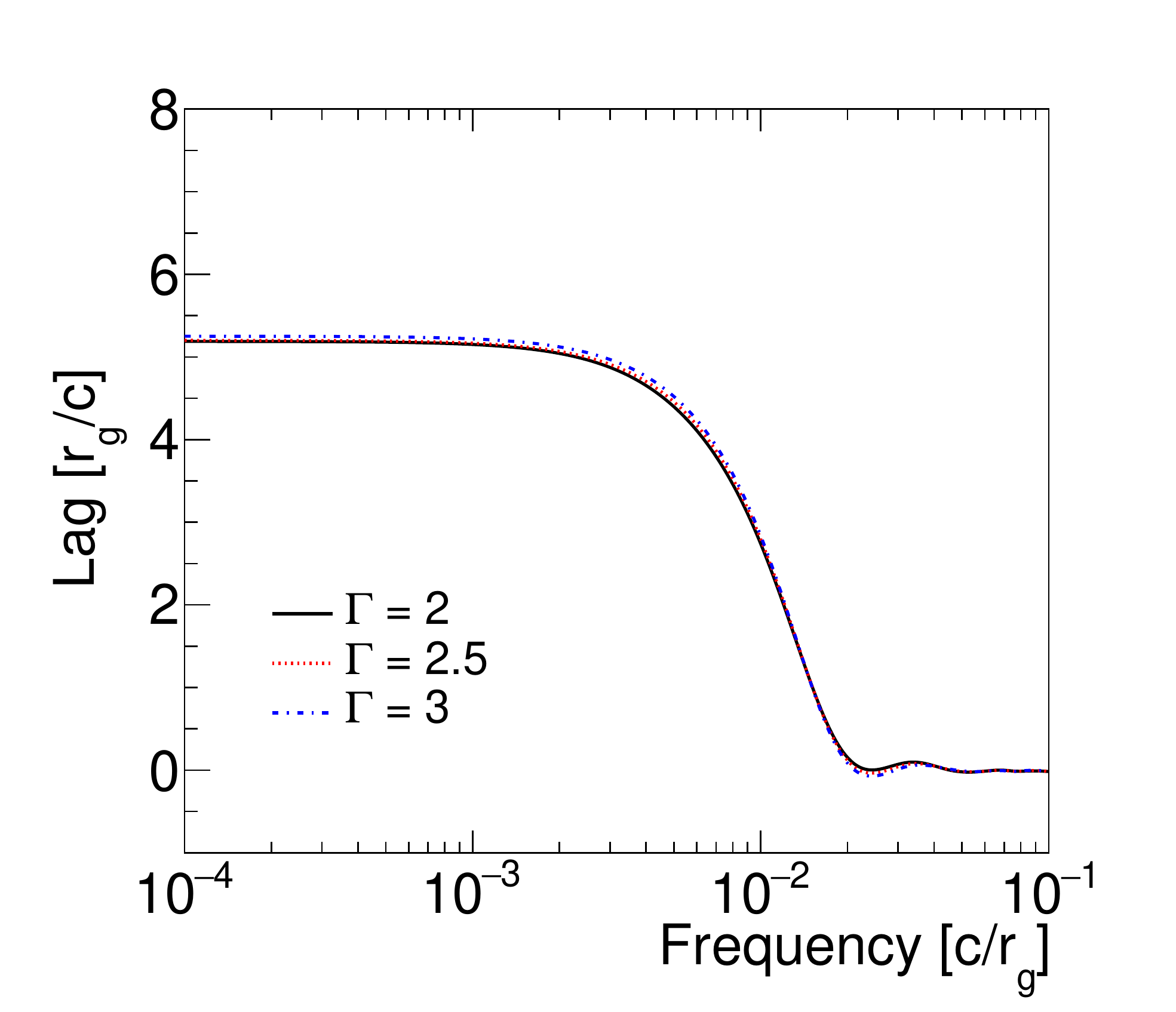}
   			 \caption{\label{fig:ga}}
   		\end{center}
	\end{subfigure}
	\quad
	\begin{subfigure}[b]{\textwidth}
		\begin{center}
   			 \includegraphics[width=0.75\textwidth]{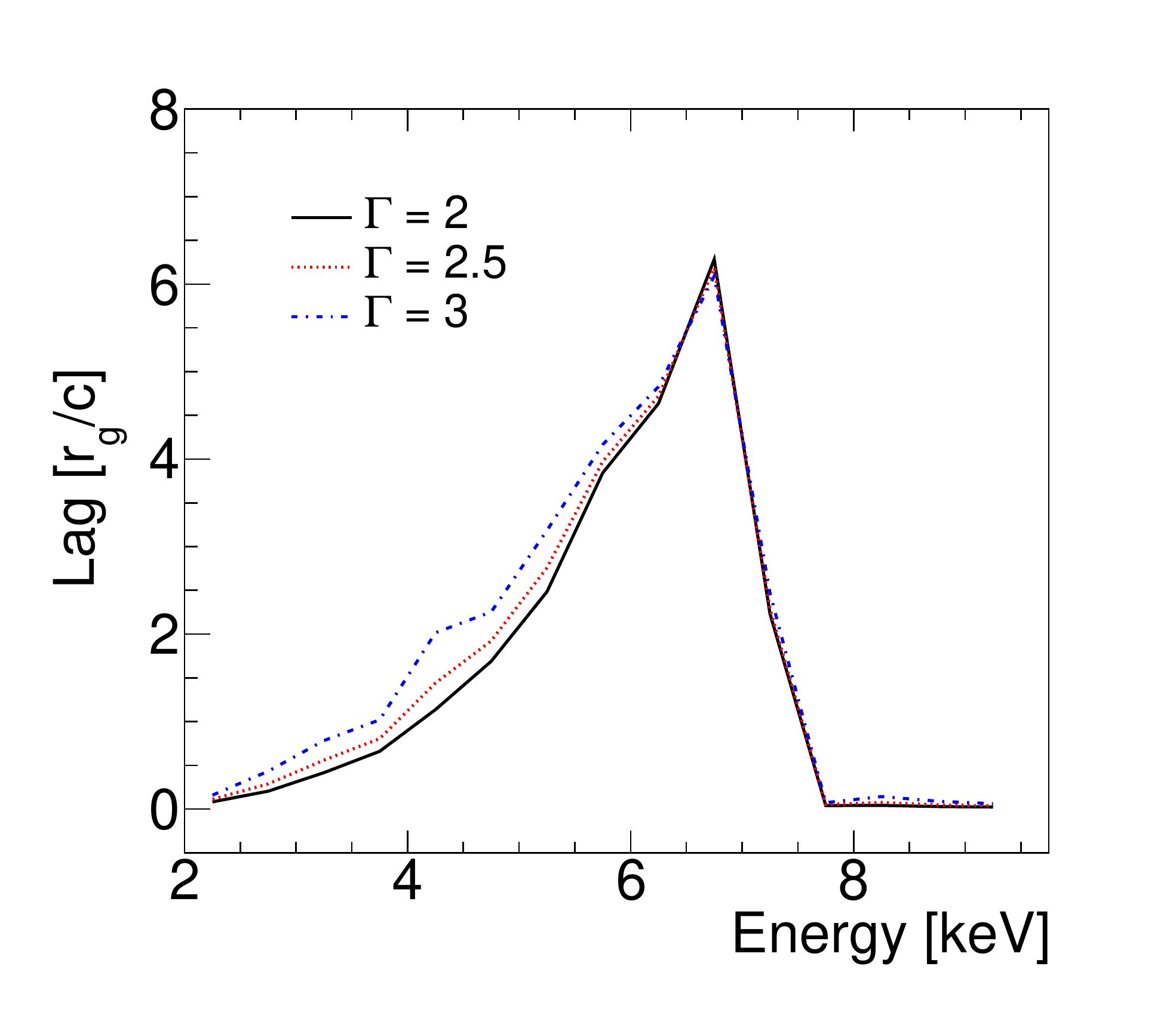}
   			 \caption{\label{fig:gb}}
   		\end
   		{center}
	\end{subfigure}
\end{center}
	\caption{\label{fig:gamma} Lag-frequency spectrum (Figure \ref{fig:ga}) and lag-energy spectrum (Figure \ref{fig:gb}) varying $\Gamma$.}
\end{figure}

\begin{figure}
\begin{center}
\vspace{-40pt}
	\begin{subfigure}[b]{\textwidth}
		\begin{center}
   			 \includegraphics[width=0.75\textwidth]{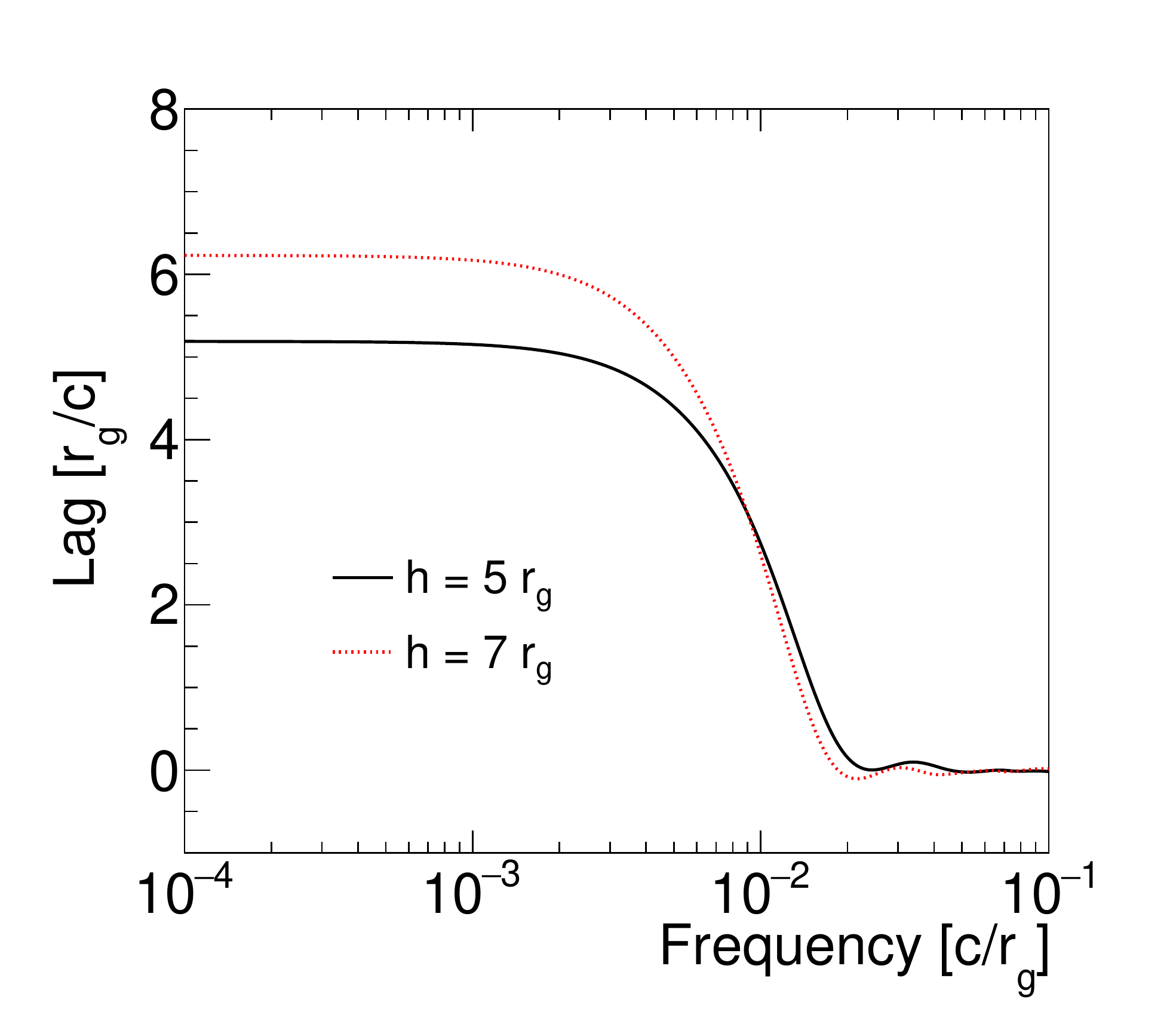}
   			 \caption{\label{fig:ha}}
   		\end{center}
	\end{subfigure}
	\quad
	\begin{subfigure}[b]{\textwidth}
		\begin{center}
   			 \includegraphics[width=0.75\textwidth]{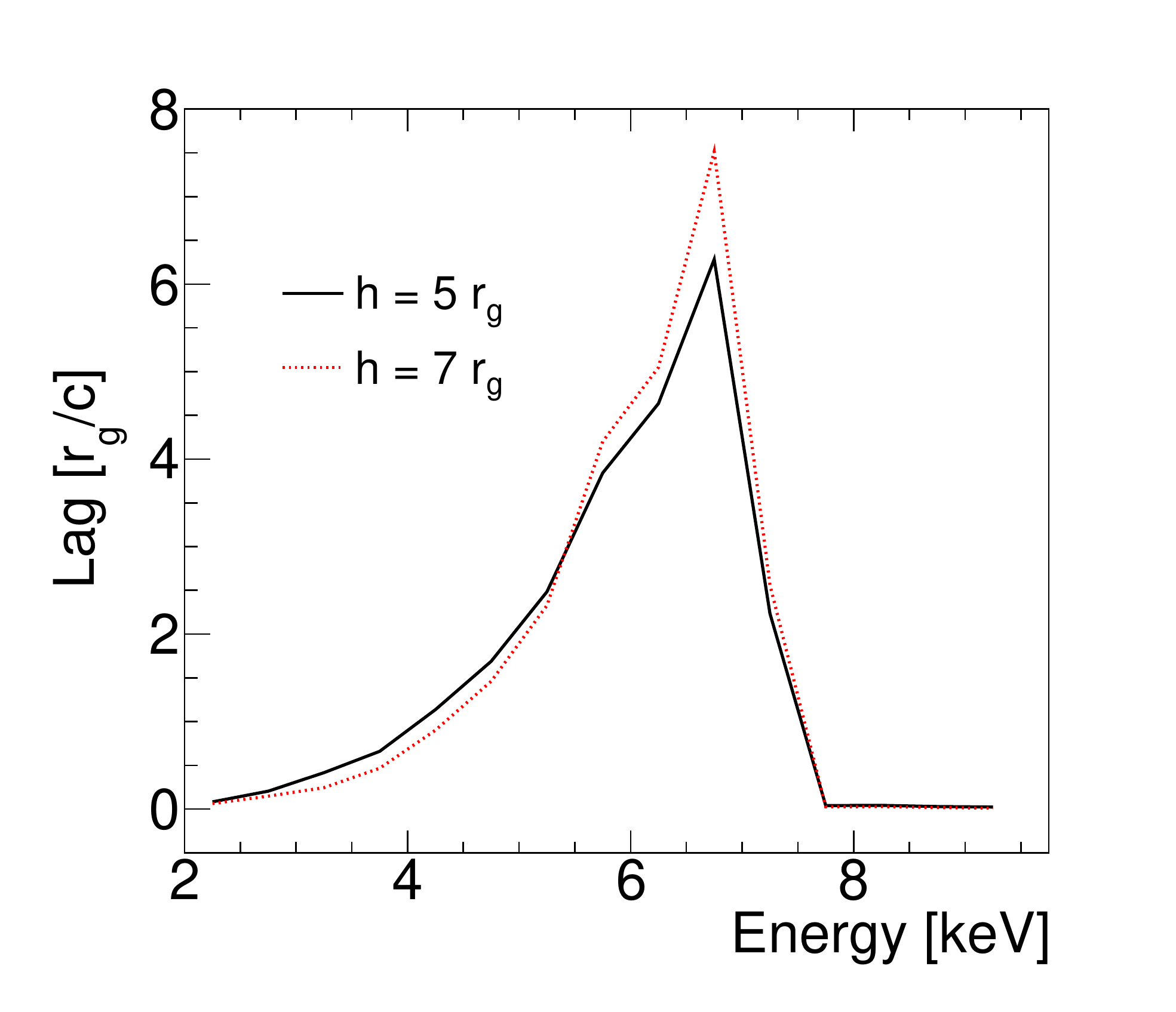}
   			 \caption{\label{fig:hb}}
   		\end{center}
	\end{subfigure}
\end{center}
	\caption{\label{fig:height} Lag-frequency spectrum (Figure \ref{fig:ha}) and lag-energy spectrum (Figure \ref{fig:hb}) varying the lamp-post height.}
\end{figure}

\begin{figure}
\begin{center}
\vspace{-40pt}
	\begin{subfigure}[b]{\textwidth}
		\begin{center}
   			 \includegraphics[width=0.75\textwidth]{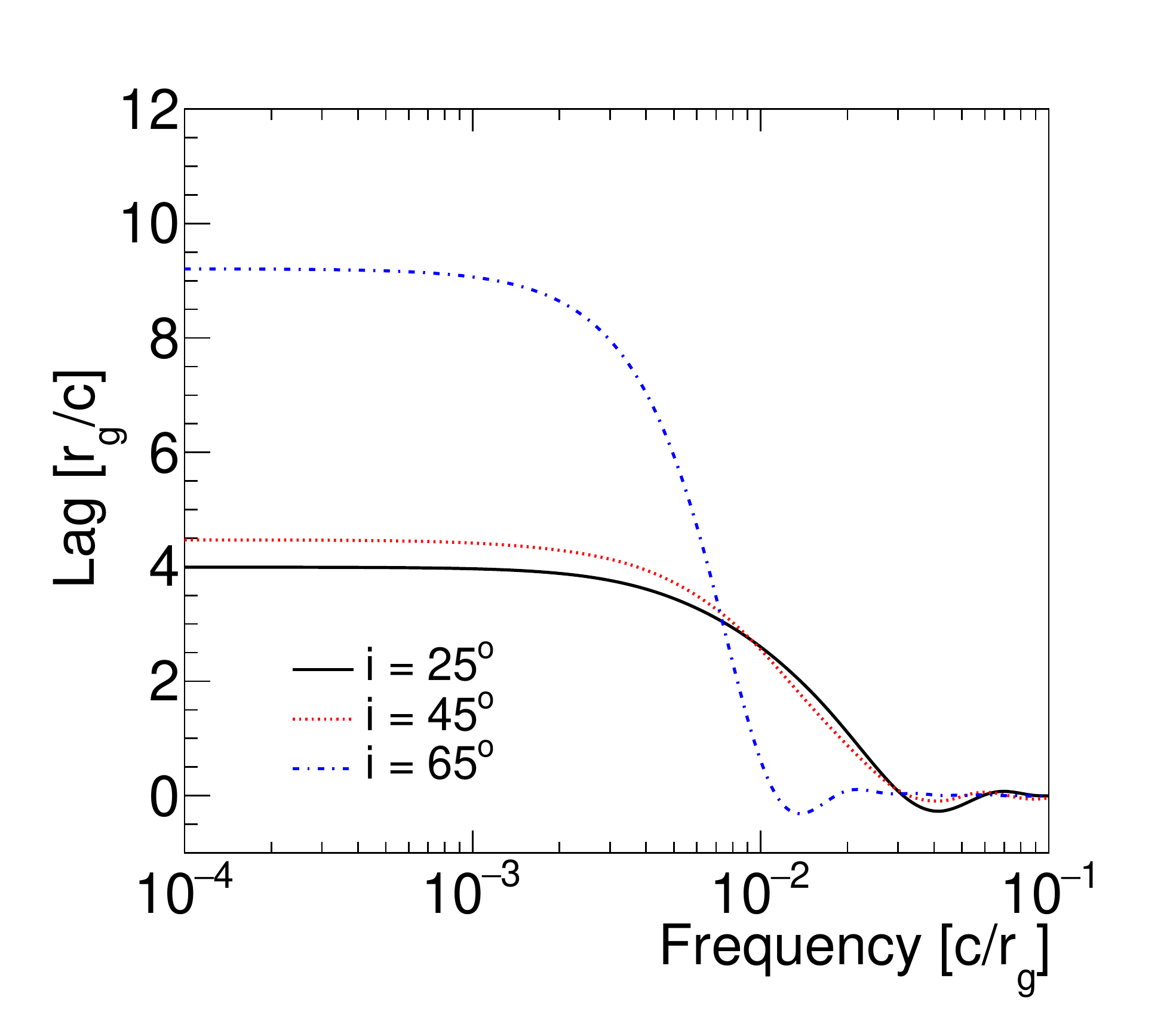}
   			 \caption{\label{fig:ia}}
   		\end{center}
	\end{subfigure}
	\quad
	\begin{subfigure}[b]{\textwidth}
		\begin{center}
   			 \includegraphics[width=0.75\textwidth]{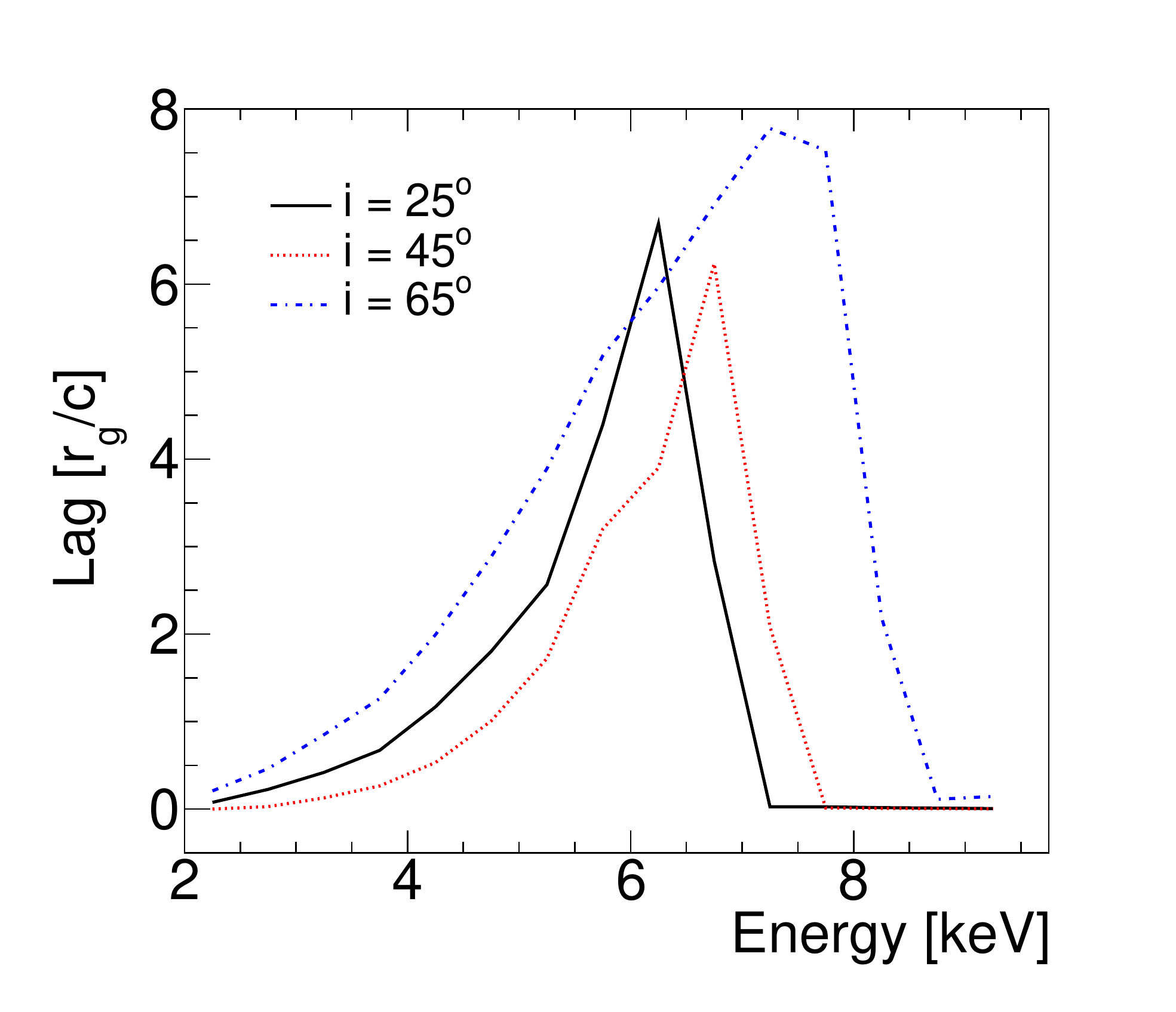}
   			 \caption{\label{fig:ib}}
   		\end{center}
	\end{subfigure}
	\caption{\label{fig:incl} Lag-frequency spectrum (Figure \ref{fig:ia}) and lag-energy spectrum (Figure \ref{fig:ib}) varying the inclination.}
\end{center}
\end{figure}

\begin{figure}
\begin{center}
\vspace{-40pt}
	\begin{subfigure}[b]{\textwidth}
		\begin{center}
   			 \includegraphics[width=0.75\textwidth]{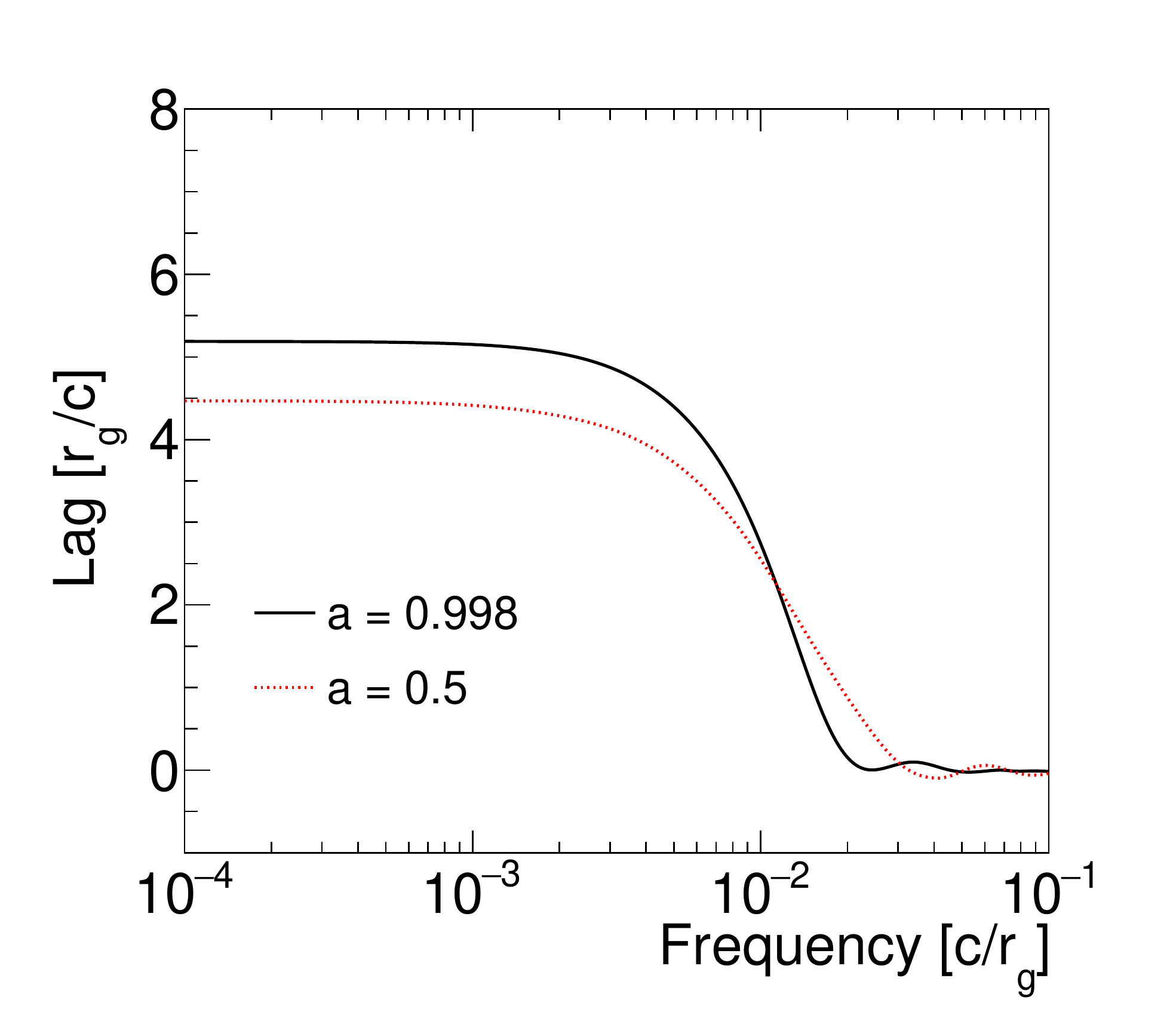}
   			 \caption{\label{fig:aa}}
   		\end{center}
	\end{subfigure}
	\quad
	\begin{subfigure}[b]{\textwidth}
		\begin{center}
   			 \includegraphics[width=0.75\textwidth]{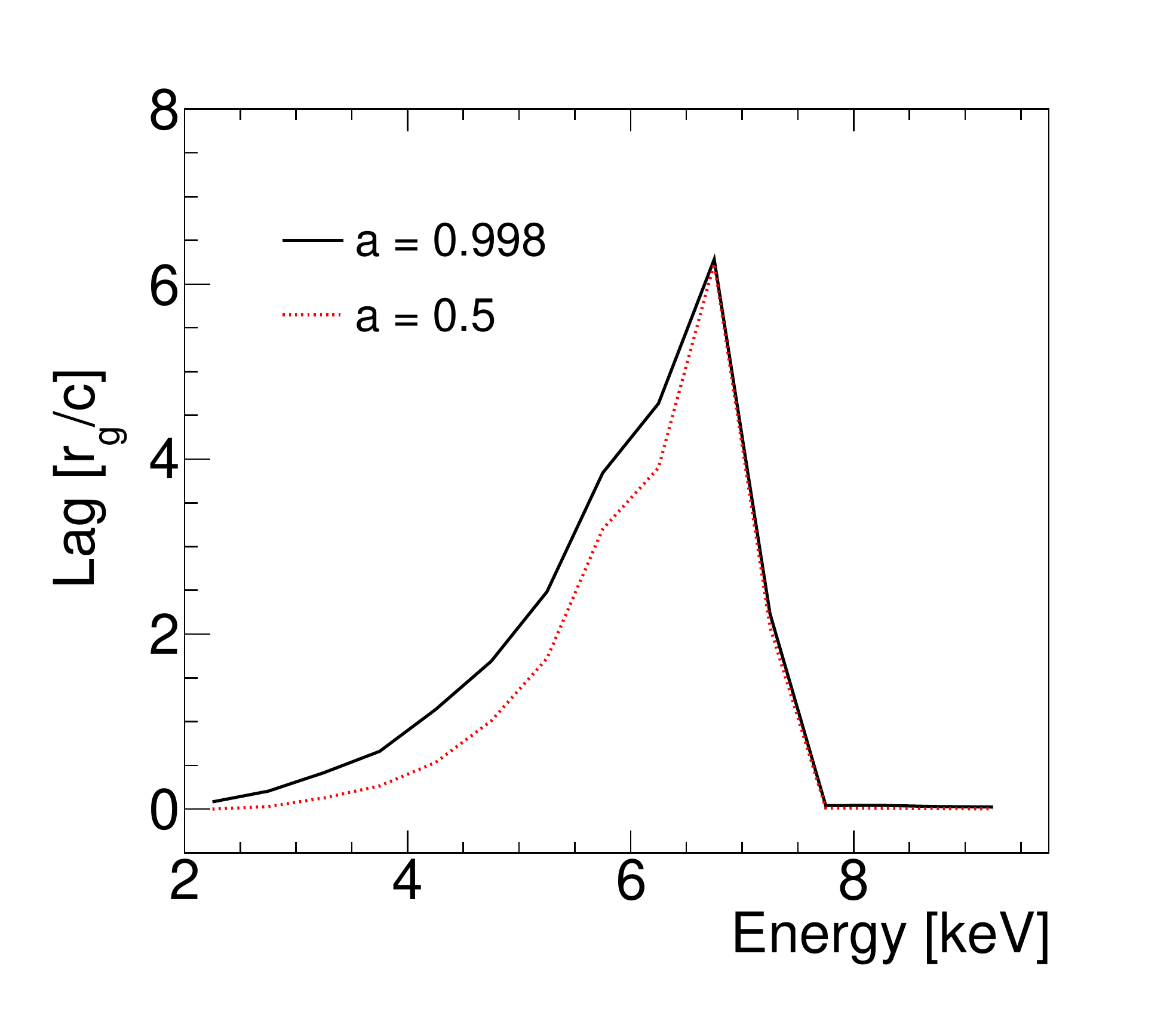}
   			 \caption{\label{fig:ab}}
   		\end{center}
	\end{subfigure}
\end{center}
	\caption{\label{fig:spin} Lag-frequency spectrum (Figure \ref{fig:aa}) and lag-energy spectrum (Figure \ref{fig:ab}) varying spin.}
\end{figure}

After running the simulations, the Fe-K reverberation signatures were calculated assuming various ionization schemes. Unless otherwise specified the values used in the analysis are $a = 0.998$, $h = 5 r_g$, $i = 45^{o}$, $\Gamma = 2$, $a0 = 1$, $r0 = 1$, $\gamma_1 = 1$, $\gamma_2 = 1$.  The lag-energy plots are calculated in the frequency range of (1-2)$\times 10^{-3}$ c/$r_g$. The figures which follow show the effects changing one of these variables has on the lag-frequency and lag-energy spectrum while holding the other values constant.  The parameters describing the relationship between scattering and absorption are seen to have a significant effect on the reverberation signatures.  This is seen in Figure \ref{fig:index1} where the lag can vary by $\sim 1.25 r_g/c$ at low frequencies when varying $\gamma_1$ and in Figure \ref{fig:a0} where differences in lags up to $\sim 2 r_g/c$ are seen at low frequencies when changing $a0$.  When varying these two parameters simultaneously it is possible to get similar lags that vary within $\sim 1 r_g/c$ (Figure \ref{fig:v22}) as well as lags which vary significantly up to $\sim 5 r_g/c$ (Figure \ref{fig:v2}).  However, neither the parameters describing the probability emitting an Fe-K$\alpha$ photon (Figures \ref{fig:index2} and \ref{fig:r0}) nor the photon index (Figure \ref{fig:gamma}) have a strong contribution to the reverberation signatures.  The resulting changes from varying $h$, $i$, and $a$ are shown in Figures \ref{fig:height}, \ref{fig:incl}, and \ref{fig:spin} respectively with $h$ and $i$ having the largest effect on the overall signatures.  While the earlier results had not been previously seen, the effect of varying these three parameters have been studied previously \citep{Cackett2014} with these figures showing consistent results.  Future studies would include modeling the recently observed Compton hump lag as well as considering the angular distribution of the emission from the disk.

%%% Local Variables: 
%%% mode: latex
%%% TeX-master: "thesis-main"
%%% End: 

\chapter{X-Calibur}
\label{X-Calbiur}

\section{Introduction}
X-Calibur is a balloon-borne hard X-ray polarimetry experiment which is scheduled to have a 1 day duration flight in September 2016 from Fort Sumner, New Mexico, and for a $\sim$ 30 day long duration balloon flight from McMurdo (Ross Island) in December 2018 - January 2019.  My contribution to this experiment included work on the alignment of the mirror-polarimeter system.  I performed thermal deformation calculations for the truss design and worked systems to measure the alignment during flight.

\section{Description of X-Calibur} 
\citet{Beilicke2012,Guo2013,Beilicke2014} provide a detailed description of this experiment and a summary of these papers will be presented in this section.  X-Calibur is a scattering polarimeter where the X-rays are focused by a grazing incident X-ray mirror to scatter in a scintillator surrounded by detectors to record the final location of the scattered photons.  The scattering element in this experiment is a low-Z plastic scintillator 1.3 cm in diameter and 14 cm long.  In the energy range of 20-60 keV, for which X-Calibur is sensitive, Compton scattering is the dominant process and photoelectric absorption which dominates at lower energies becomes negligible. The method of scattering polarimetry takes advantage of the fact that photons scatter preferentially in a direction perpendicular to the electric field. The Klein-Nishina cross-section describing Compton scattering is given by
	\begin{equation}
	\frac{d \sigma}{d \Omega} = \frac{r_0^2}{2} \frac{k_1^2}{k_0^2}\left[\frac{k_0}{k_1}+\frac{k_1}{k_0}-2\sin^2\theta \cos^2\eta \right].
	\end{equation}
		\begin{figure}
			\begin{center}
        		\vspace{0pt}
   				 \includegraphics[width=1\textwidth]{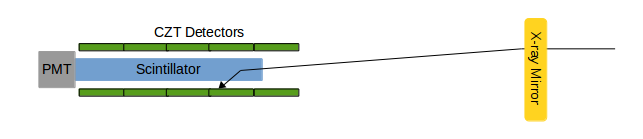}
   			\vspace{10pt}
        		\end{center}
        		\caption{\label{fig:PolDiagram} Schematic of the scattering polarimeter.}
		\end{figure}
Here the angle between the electric field vector of incident photon and the scattering plane is given by $\eta$; $r_0$ is the classical electron radius; the wave vectors before and after scattering are represented by $\mathbf{k_0}$ and $\mathbf{k_1}$ respectively; and $\theta$ is the scattering angle (see e.g. \citet{Rybicki1979}).  From this it is seen that the detectors surrounding the scintillator will observe an azimuthal distribution of events yielding a sinusoidal modulation with a maximum at $\pm 90^o$ to the preferred electric field direction with a $180^o$ periodicity.  For the upcoming flight the scintillator will be surrounded by five high-Z Cadmium Zinc Telluride (CZT) detectors on each of its four sides.  A schematic illustrating the experiment is shown in Figure \ref{fig:PolDiagram}. The detectors have dimensions of 2x2cm, are 2mm thick, and are contacted with a 64-pixel anode grid.  In the required energy range, 2mm thick detectors are sufficient to absorb more than 99\% of the X-rays.  However, the previous configuration of the polarimeter included 8 detectors on each side, 3 detectors 2mm thick and 5 which were 5mm thick.  5mm thick detectors can be calibrated with higher accuracy owing to excellent detection efficiency and energy resolution at high ($>$ 100 keV) energies.  In order to block both the charged and neutral particle background the polarimeter and front-end readout electronics are surrounded by an active CsI(Na) anti-coincidence shield 2.7cm thick at the sides and 5cm thick at the rear end.  A tungsten collimator surrounds the entrance to the scintillator to block X-ray and other particles which where not focused by the X-ray mirror. While the flight of X-Calibur in Fall 2014 was not able to obtain science data due to a fault in the pointing system, \citet{Amini2016} was able to model the background data taken to predict the expected background in the Fall 2016 flight and the long duration flight in Winter 2018.  In order to reduce the systematic uncertainties, the polarimeter and shield is rotated around the optical axis.  The observed X-rays are focused onto the polarimeter through the use of a 225 shell, 40 cm diameter X-ray mirror with a focal length of 8 meters and a field of view (full width half max) of 10 arcminutes.  In order to point at a source, X-Calibur is integrated with the Wallops Arc Second Pointer (WASP) which was developed by the Wallops Flight Facility.  The minimum detectable polarization (MDP), the minimum polarization fraction that can be detected with 99 \% confidence, can be used to characterize the performance of X-Calibur.  Integrating the scattering probability distribution results in the MDP being given by
	\begin{equation}
		MDP \simeq \frac{4.29}{\mu R_{src}} \sqrt{\frac{R_{src}+R_{bg}}{T}}
	\end{equation}
where $R_{src}$ and $R_{bg}$ are the source and background count rates respectively and T is the observation time.  The modulation amplitude, $\mu$, is the modulation amplitude of a 100\% polarized beam given by
	\begin{equation}
		\mu = \frac{C_{max}-C_{min}}{C_{max}+C_{min}}.
	\end{equation}
$C_{min}$ and $C_{max}$ are the minimum and maximum number of counts in the azimuthal Compton scattering distribution.  For X-Calibur, $\mu \approx 0.5$ for all energies. When the scintillator is required to trigger, the MDP is $<$ 5\% for a 4 hour observation of source with a flux equal to that of the Crab Nebula and Pulsar \citep{Guo2013}. 

\section{Thermal Deformations} \label{sec:Thermal}
\begin{sidewaysfigure}
        \centering
        \begin{subfigure}[b]{0.45\textwidth}
   				 \includegraphics[width=\textwidth]{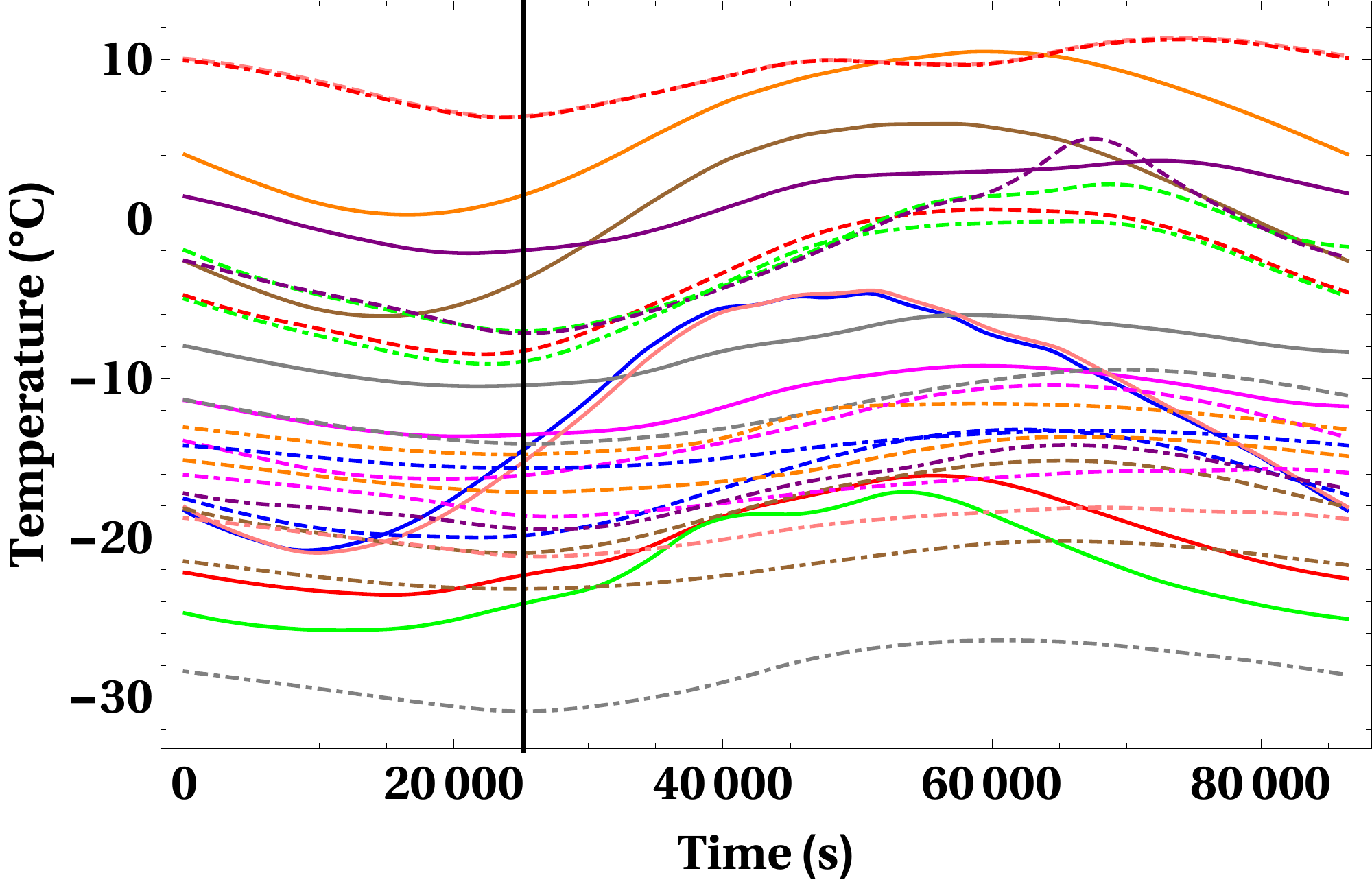}
        \end{subfigure}
       \begin{subfigure}[b]{0.45\textwidth}
   				 \includegraphics[width=\textwidth]{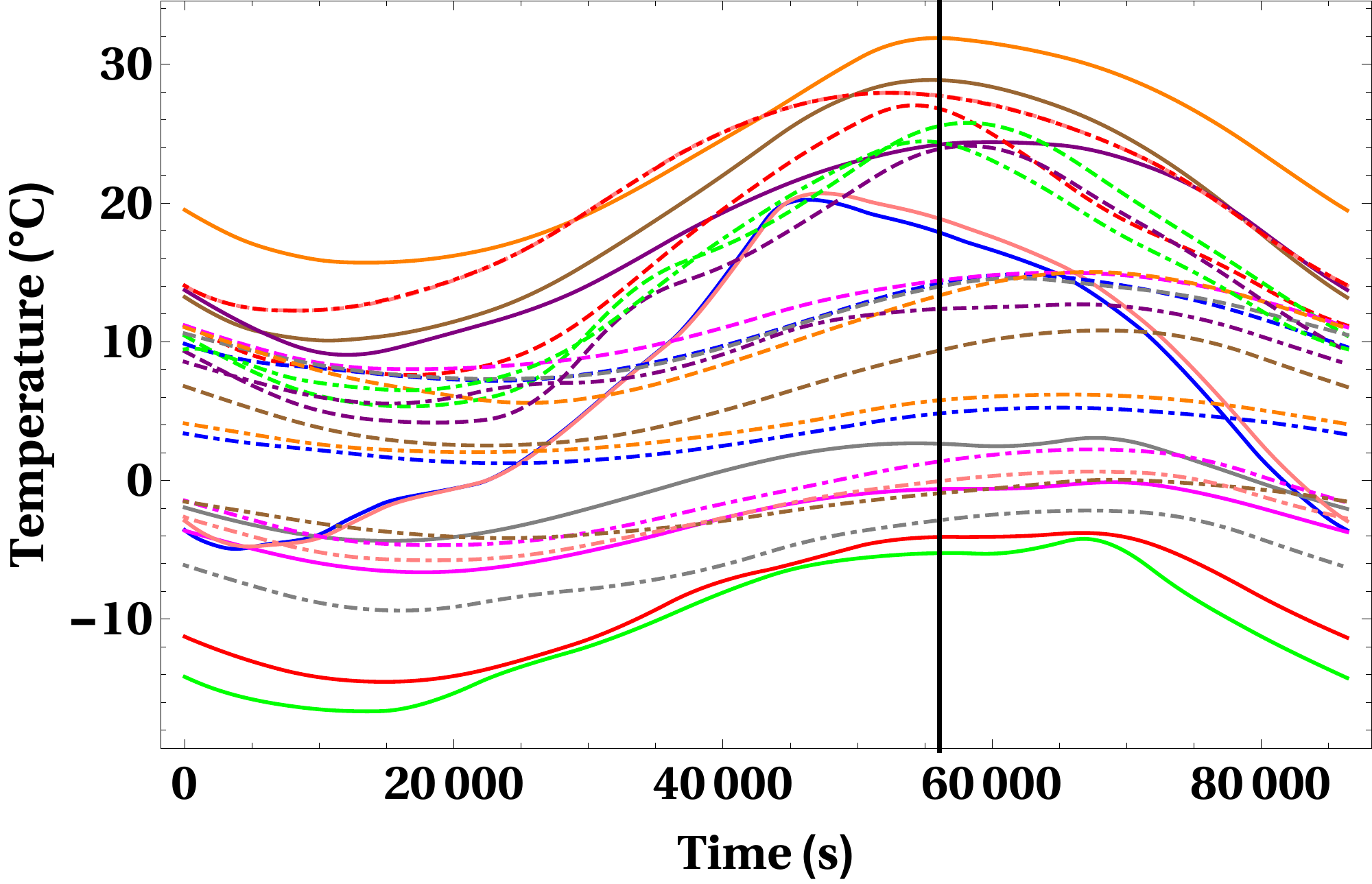}
        \end{subfigure}
        \begin{subfigure}[b]{0.45\textwidth}
   				 \includegraphics[width=\textwidth]{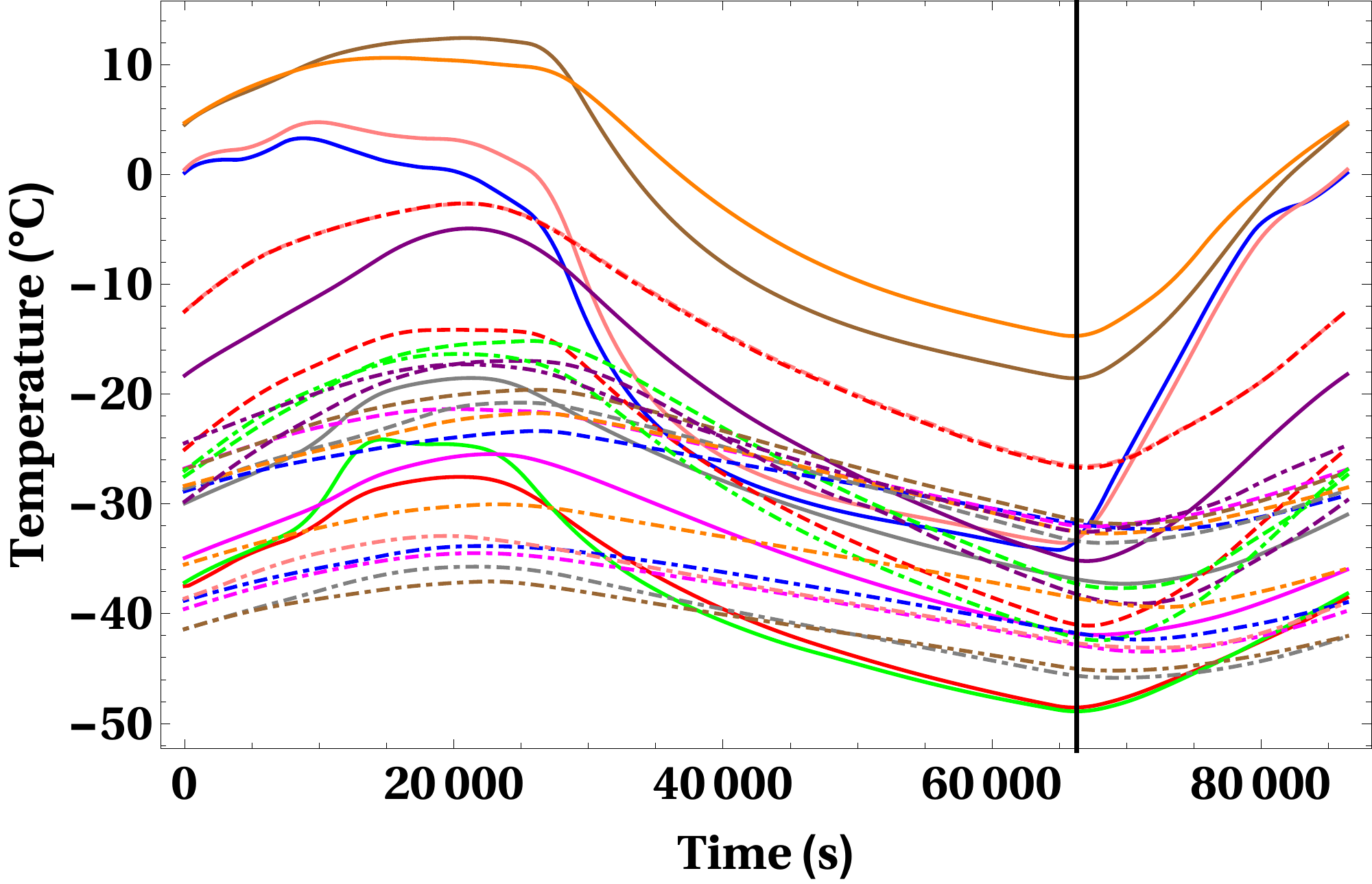}
        \end{subfigure}
       \begin{subfigure}[b]{0.45\textwidth}
   				 \includegraphics[width=\textwidth]{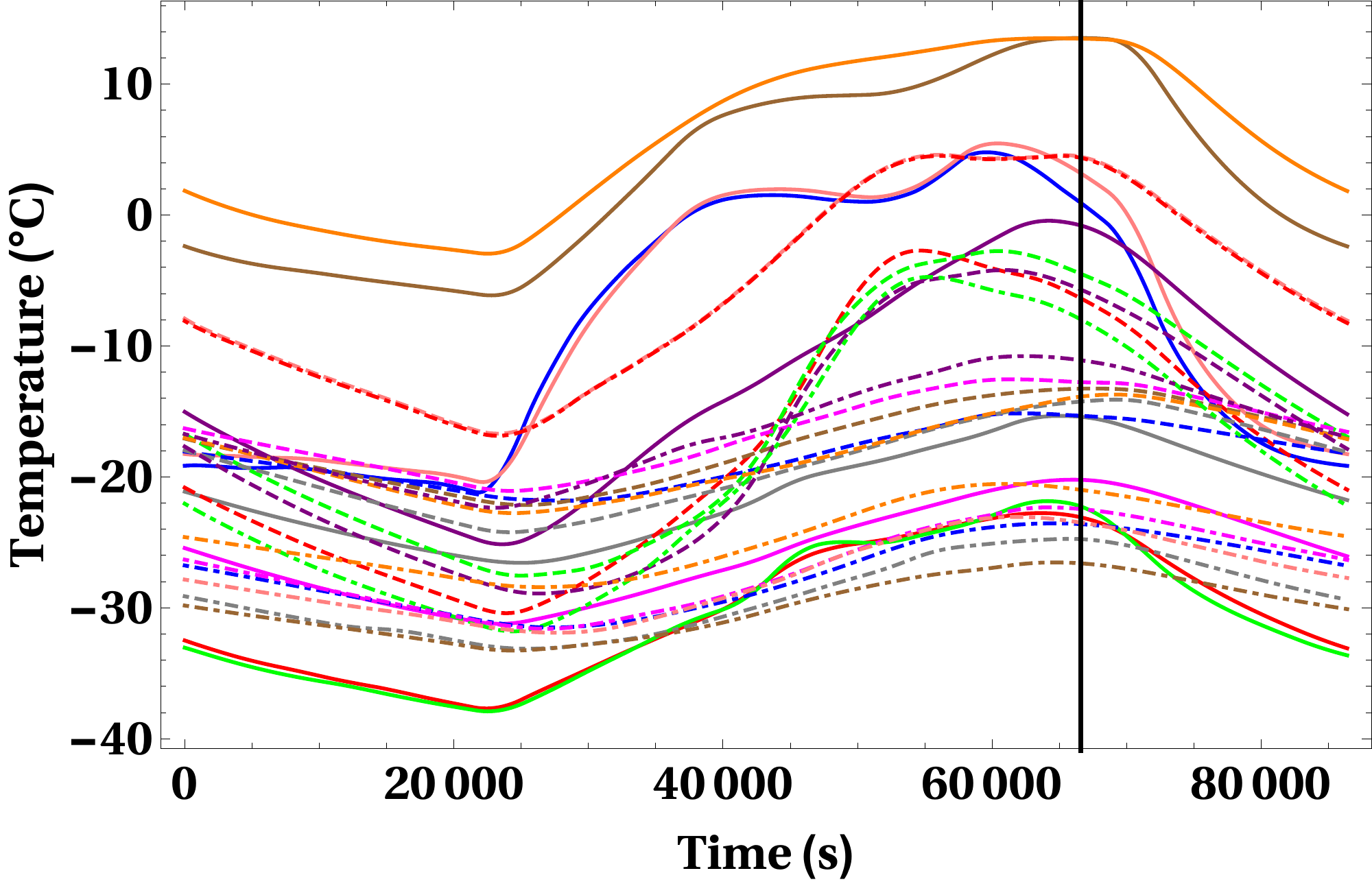}
        \end{subfigure}
        \vspace{12pt}
        \caption{\label{fig:Temperature} Results for the thermal simulations modeling the expected joint temperatures for the Antarctica cold case (top left), Antarctica hot case (top right), Ft. Sumner cold case (bottom left), and the Ft. Sumner hot case (bottom right).  Each curve represents the temperature of a given joint (see Figure \ref{fig:Truss}) over the course of the flight.  The vertical black line indicates the time used in the thermal deformation simulations. Thermal simulation data courtesy of Scott Cannon at the New Mexico State University Physical Science Laboratory.}
\end{sidewaysfigure}
The polarimeter and mirror are mounted to opposite ends of a truss which is attached in the center to the WASP system.  The truss is made of carbon fiber tubes held together with aluminium joints with each end attached to an aluminium honeycomb sheet which will hold the necessary experimental equipment.  Because the diameter of the scintillator is so small, particularly when compared to the 8 m focal length of the mirror, significant attention was paid to making sure the truss does not suffer from deformation due to thermal effects.  To this end simulations were run to determine the most severe temperatures (both hot and cold) that the truss joints would see during both the Ft. Sumner (observing Sco X1) and Antarctica flights (observing Vela X1).  Using these data I performed a thermal deformation analysis on the truss design.  The results from the thermal simulations are shown in Figure \ref{fig:Temperature}, which was provided by Scott Cannon at the New Mexico State University Physical Science Laboratory.  Each colored line corresponds to the temperatures seen by a given joint and the vertical black line shows the time where the most extreme temperatures are seen, which is the time used in my analysis.
The thermal deformation at each joint was calculated in each of the four cases (Antarctica hot, Antarctica cold, Ft. Sumner hot, Ft. Sumner cold) and the total deformation of each chord was calculated. The thermal deformation was calculated using the equation
\begin{equation}
\frac{\Delta L}{L} = \alpha \Delta T
\end{equation}
where the coefficient of thermal expansion of aluminium is 
\begin{equation}
\alpha_{Al} =  2 \times 10^{-5} \frac{m}{m K}
\end{equation}
and the coefficient of thermal expansion of carbon fiber (CF) is
\begin{equation}
\alpha_{CF} = 1.8 \times 10^{-7} \frac{m}{m K}.
\end{equation}
Because $\alpha_{CF}$ is so much smaller than $\alpha_{AL}$ the deformation of the CF tubes is neglected.  $\Delta$T is calculated using the results of the thermal simulations when compared to the room temperature when the truss was assembled, $25^{o}$C. The specific temperatures used in the calculations are given in Table \ref{fig:tempData}.
\begin{table}
\begin{center}
\caption{\label{fig:tempData} Temperatures for each joint in each of the four simulated observations used in the thermal deformation calculations. Thermal simulation data courtesy of Scott Cannon at the New Mexico State University Physical Science Laboratory.}
\begin{tabular}{|c|cccc|c|}
\hline
 & \multicolumn{4}{|c|} {\rule{0pt}{3ex} Temperature ($^o$ C)}   &
   \multicolumn{1}{|c|}{ } \\
       \cline{2-5}
\textrm{\rule{0pt}{3ex}  Node}&
\textrm{Antarctica Cold}&
\textrm{Antarctica Hot}&
\textrm{Ft. Sumner Cold}&
\textrm{Ft. Sumner Hot}&
\textrm{Joint Size (in)}\\
\hline
\rule{0pt}{3ex} 58 & -22.34 & -4.08 & -48.54 & -23.10 & 4 \\
76 & -24.12 & -5.25 & -48.90 & -22.27 & 4 \\ 
128 & -14.47 & 17.87 & -33.26 & 0.92 &  4\\ 
188 & -13.54 & -0.63 & -41.84 & -20.33 & 4  \\ 
240 & -10.44 & 2.65 & -36.87 & -15.42 & 4 \\ 
276 & -3.81 & 28.85 & -18.56 & 13.52 & 4 \\ 
302 & 1.50 & 31.89 & -14.72 & 13.47 & 4 \\ 
330 & -15.22 & 18.86 & -33.18 & 3.18 & 4 \\ 
380 & -1.98 & 24.21 & -35.15 & -0.80 & 4 \\ 
384 & -8.28 & 26.79 & -41.01 & -6.37 & 4 \\ 
388 & -7.05 & 25.55 & -37.28 & -4.49 & 4 \\ 
476 & -19.86 & 14.20 & -31.81 & -15.34 & 7, 14 \\ 
516 & -16.08 & 14.40 & -32.01 & -12.77 & 7, 14 \\ 
751 & -14.11 & 13.97 & -33.39 & -14.25 & 7 \\ 
790 & -20.95 & 9.36 & -31.47 & -13.27 & 7 \\ 
830 & -17.14 & 13.33 & -32.54 & -13.87 & 7  \\ 
886 & 6.44 & 27.73 & -26.55 & 4.44 & 4 \\ 
960 & -7.18 & 23.89 & -38.24 & -5.70 & 4 \\ 
999 & 6.39 & 27.71 & -26.70 & 4.36 & 4 \\ 
1019 & -8.94 & 24.36 & -42.19 & -7.98 & 4 \\ 
1043 & -15.63 & 4.83 & -41.77 & -23.62 & 7 \\ 
1081 & -18.63 & 1.36 & -42.85 & -22.46 & 7 \\ 
1163 & -30.89 & -2.88 & -45.65 & -24.76 & 7 \\ 
1202 & -23.21 & -0.93 & -45.02 & -26.61 & 7\\ 
1296 & -14.77 & 5.78 & -38.60 & -21.00 & 7, 14\\ 
1392 & -21.16 & -0.06 & -42.68 & -23.49 & 7, 14 \\ 
1471 & -19.44 & 12.36 & -32.52 & -11.10 & 4 \\  
\hline
\end{tabular}
\vspace{10pt} 
\end{center}
\end{table}
\begin{sidewaysfigure}
   \begin{center}
         \begin{subfigure}[b]{1\textwidth}
   			\includegraphics[width=\textwidth]{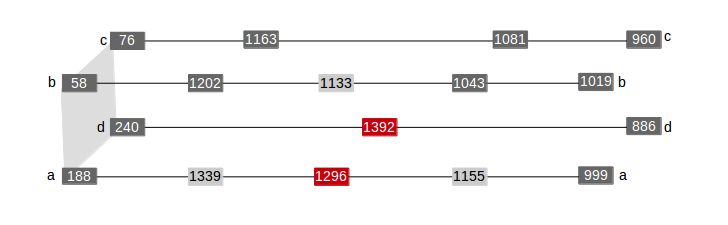}
         \end{subfigure}
         
         \begin{subfigure}[b]{0.2\textwidth}
            \includegraphics[width=\textwidth]{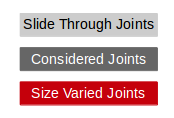}
         \end{subfigure}
         
         \begin{subfigure}[b]{1\textwidth}
            \includegraphics[width=\textwidth]{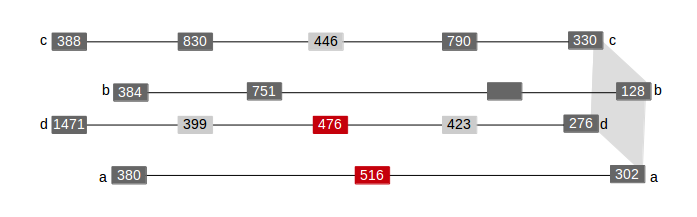}
         \end{subfigure}
   \end{center}
             \caption{\label{fig:Truss} Schematic of the truss design where the top shows the mirror end of the truss and the bottom shows the polarimeter end of the truss.}
\end{sidewaysfigure}
A diagram of both ends of the truss is shown in Figure \ref{fig:Truss}. The light grey joints are designed such that the CF tubes slide through the entire joint and therefore do not effect the total deformation of the chord.  The dark grey joints represent the male/female joints glued to the CF tubes and are used in the deformation calculations. 
 When the first results were calculated it was noticed that there was a significant difference in thermal contraction between the top and bottom chords, which was due to the fact that the top chords contained more aluminium than the bottom chords and therefore had more material to contract.  The calculations were then repeated doubling the length of the joints shown in red in Figure \ref{fig:Truss}.  The results from this analysis are given in Table \ref{fig:thermDef}.  As seen in this table, when the joint is 14 inches long, as opposed to 7 inches, the difference in the contraction of the top and bottom chords is significantly reduced.  From these results it was decided to increase the length of the red joints.  It was also decided to put heaters on select joints so, if the deformation measured during flight is larger than expected, the joints can be heated to straighten out the truss to avoid any issues in detecting the observed sources.
\begin{table}
\begin{center}
\caption{\label{fig:thermDef} Thermal deformation results.}
\begin{tabular}{|c|cccc|cccc|}
\hline 
 & \multicolumn{8}{|c|} {\rule{0pt}{3ex} Chord Deformation (in)} \\
    \cline{2-9}
&   \multicolumn{4}{|c|}{\rule{0pt}{3ex}7 inch joint} &
    \multicolumn{4}{|c|}{14 inch joint} \\
    \cline{2-9}
\textrm{Observation}&
\textrm{\rule{0pt}{3ex}A}&
\textrm{B}&
\textrm{C}&
\textrm{D}&
\textrm{A}&
\textrm{B}&
\textrm{C}&
\textrm{D}\\
\hline
\rule{0pt}{3ex} Antarctica Cold & -0.0247 & -0.0432 & -0.0468 & -0.0285 & -0.0377 & -0.0432 & -0.0468 & -0.0431\\
Antarctica Hot & -0.0069 & -0.0148 & -0.0172 & -0.0092 & -0.0116 & -0.0148 & -0.0172 & -0.0150\\
Ft. Sumner Cold & -0.0425 & -0.0683 & -0.0675 & -0.0429 & -0.0618 & -0.0683 & -0.0675 & -0.0628\\
Ft. Sumner Hot & -0.0248 & -0.0423 & -0.0447 & -0.0264 & -0.0382 & -0.0423 & -0.0447 & -0.0406\\
\hline 
\end{tabular}
\end{center}
\end{table}
 
\section{Alignment Systems} 
Because the alignment of the mirror-polarimeter system is so important there are multiple procedures in place to ensure and monitor this alignment.

	\subsection{Scintillator Alignment}
\begin{figure}
		\begin{center}
        \vspace{0pt}
   				 \includegraphics[width=\textwidth]{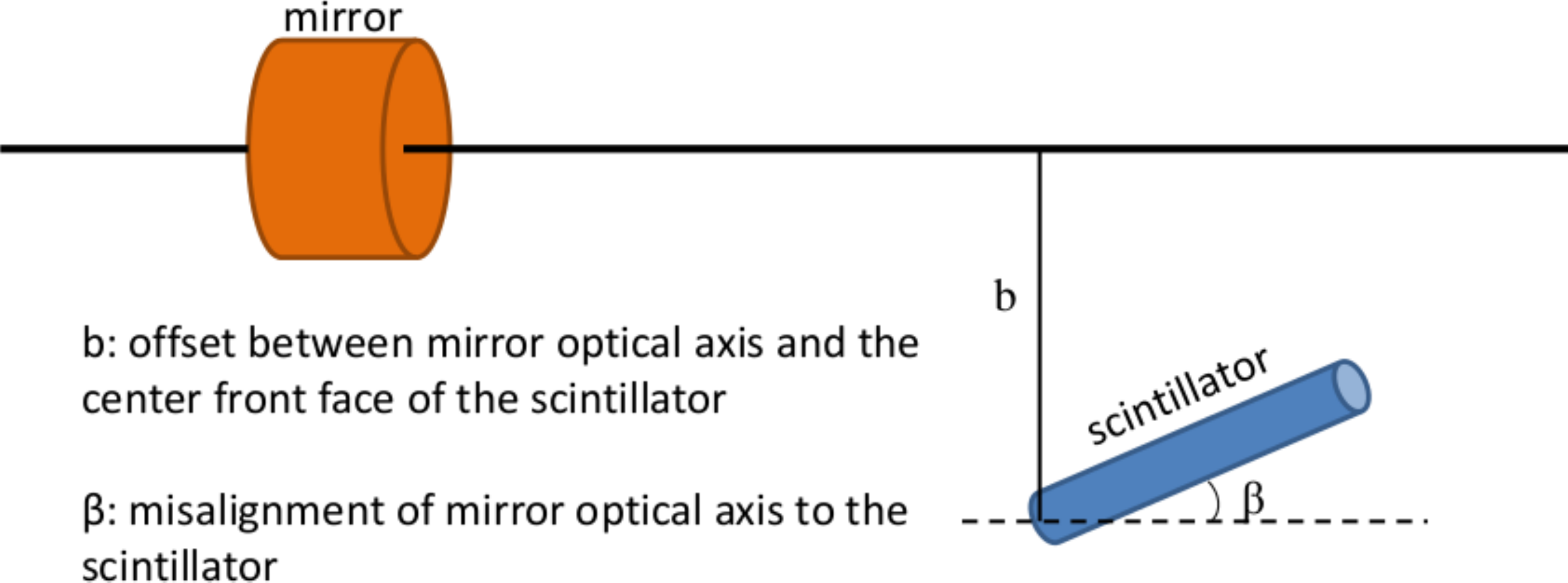}
   		\vspace{10pt}
        	\end{center}
        \caption{\label{fig:ScintParam} Schematic of scintillator alignment.}
\end{figure}

\begin{figure} 
        \centering
        \begin{subfigure}[b]{0.7\textwidth}
   				 \includegraphics[width=\textwidth]{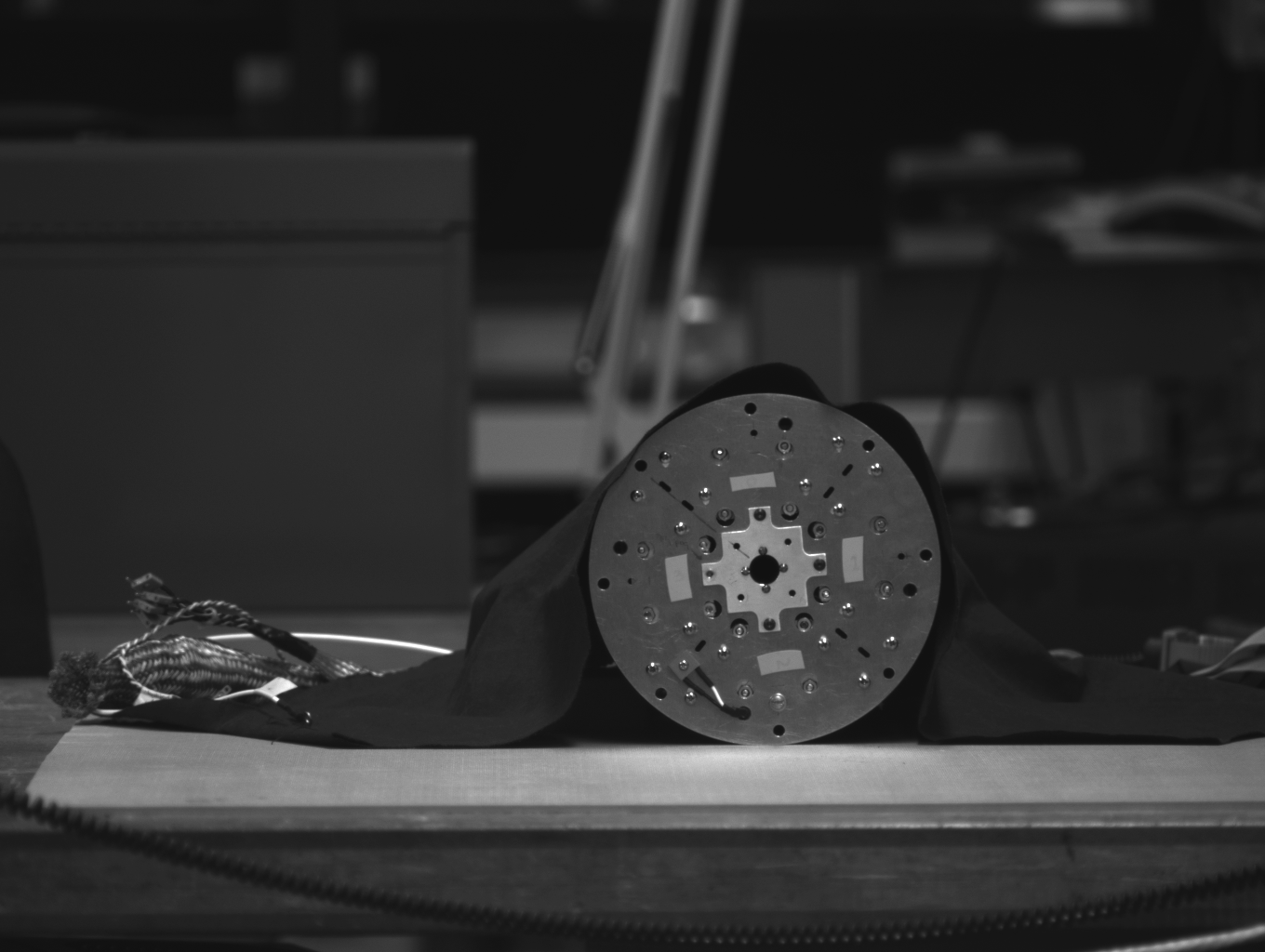}
                \caption{\label{fig:ScintFront} Front of the Polarimeter} 
        \end{subfigure}
        	\quad
         \begin{subfigure}[b]{0.7\textwidth}
              \includegraphics[width=\textwidth]{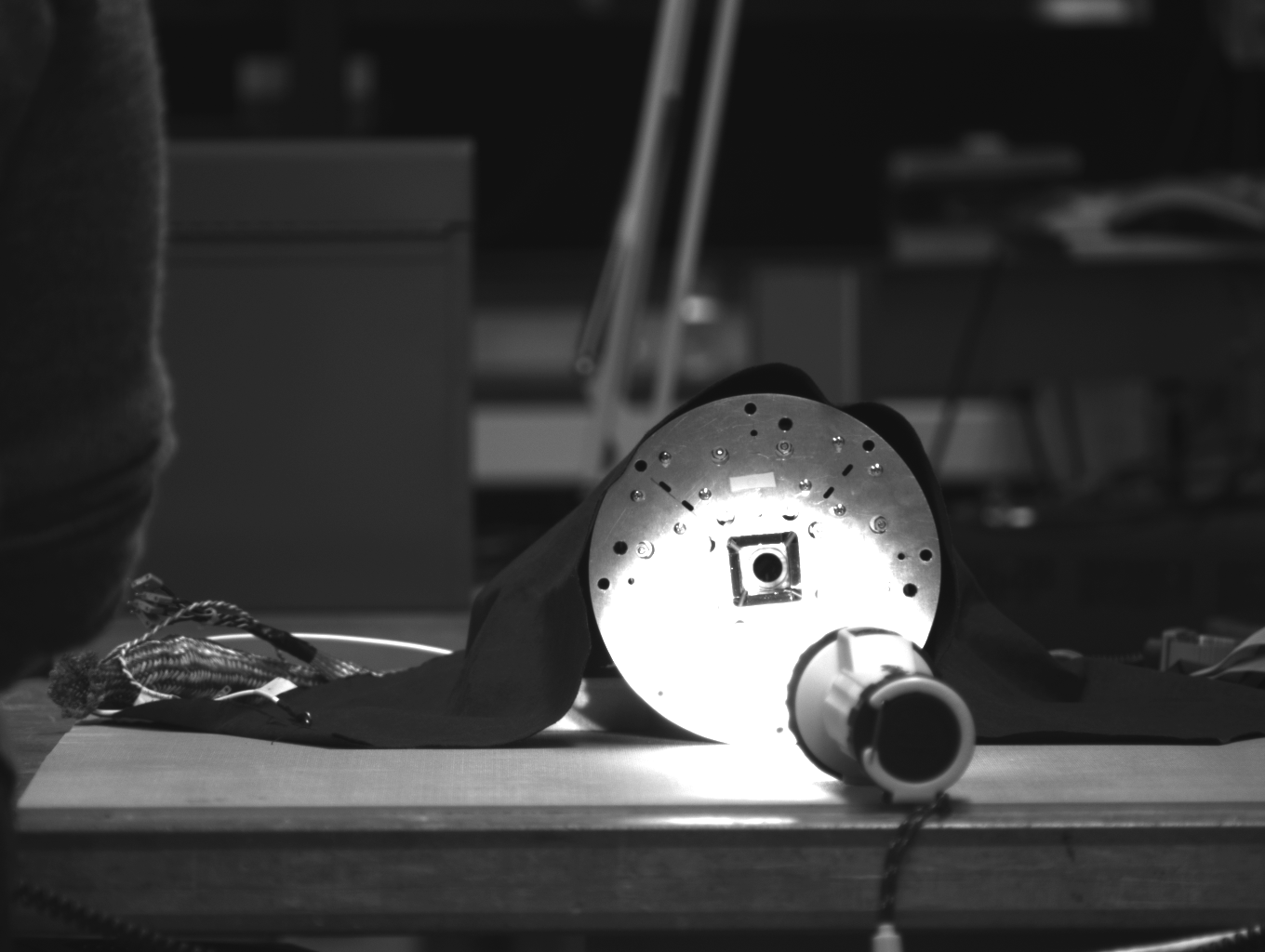}
               \caption{\label{fig:ScintBack} Back of the Polarimeter}
        \end{subfigure}
        \caption{\label{fig:Scint} Images taken for the scintillator alignment.}
\end{figure}

	Before flight, it is important to verify the alignment of the scintillator with the mirror optical axis.  Therefore it is necessary to measure the offset of the scintillator from the optical axis, $b$, and the orientation of the scintillator, $\beta$.  These parameters are shown in Figure \ref{fig:ScintParam}.
To measure these parameters pictures are taken of both the front and back of the polarimeter where the scintillator will be when it is inserted. These images are shown in Figure \ref{fig:Scint}, where the distance between the polarimeter and the camera is $\sim$ 6.4 m.
A program was written to isolate the scintillator region, and the center of each image is calculated using
\begin{equation} \label{eq:unweightedCenter}
\langle x \rangle = \frac{\sum_{j} x_j}{N}
\end{equation}
where N is the number of points and the corresponding uncertainty is
\begin{equation}
\sigma_x = \left( \frac{\langle x^2 \rangle - \langle x \rangle ^2}{N} \right) ^{1/2}.
\end{equation}
Using the fitted center points of the front and back of the scintillator the values for both $b$ and $\beta$ can be calculated.  These results are shown in Table \ref{fig:ScintReq}, and the uncertainties could potentially be reduced by using a combined analysis of multiple images.
\begin{table}
\begin{center}
\caption{\label{fig:ScintReq} Alignment requirements.}
\begin{tabular}{|c|ccc|}
\hline 
\textrm{\rule{0pt}{3ex}measurement}&
\textrm{1-sigma calculated}&
\textrm{3-sigma calculated}&
\textrm{3-sigma requirement}\\
\hline
\rule{0pt}{3ex} $b$ & 3.6 arcseconds & 10.8 arcseconds & 11 arcseconds \\ 
$\beta$ & 0.174 mm & 0.522 mm & $<$1.2 mm\\
\hline 
\end{tabular}
\end{center}
\end{table}

\subsection{Laser-CCD System}
\begin{figure}
        \centering
        \begin{subfigure}[b]{0.49\textwidth}
   				 \includegraphics[width=\textwidth]{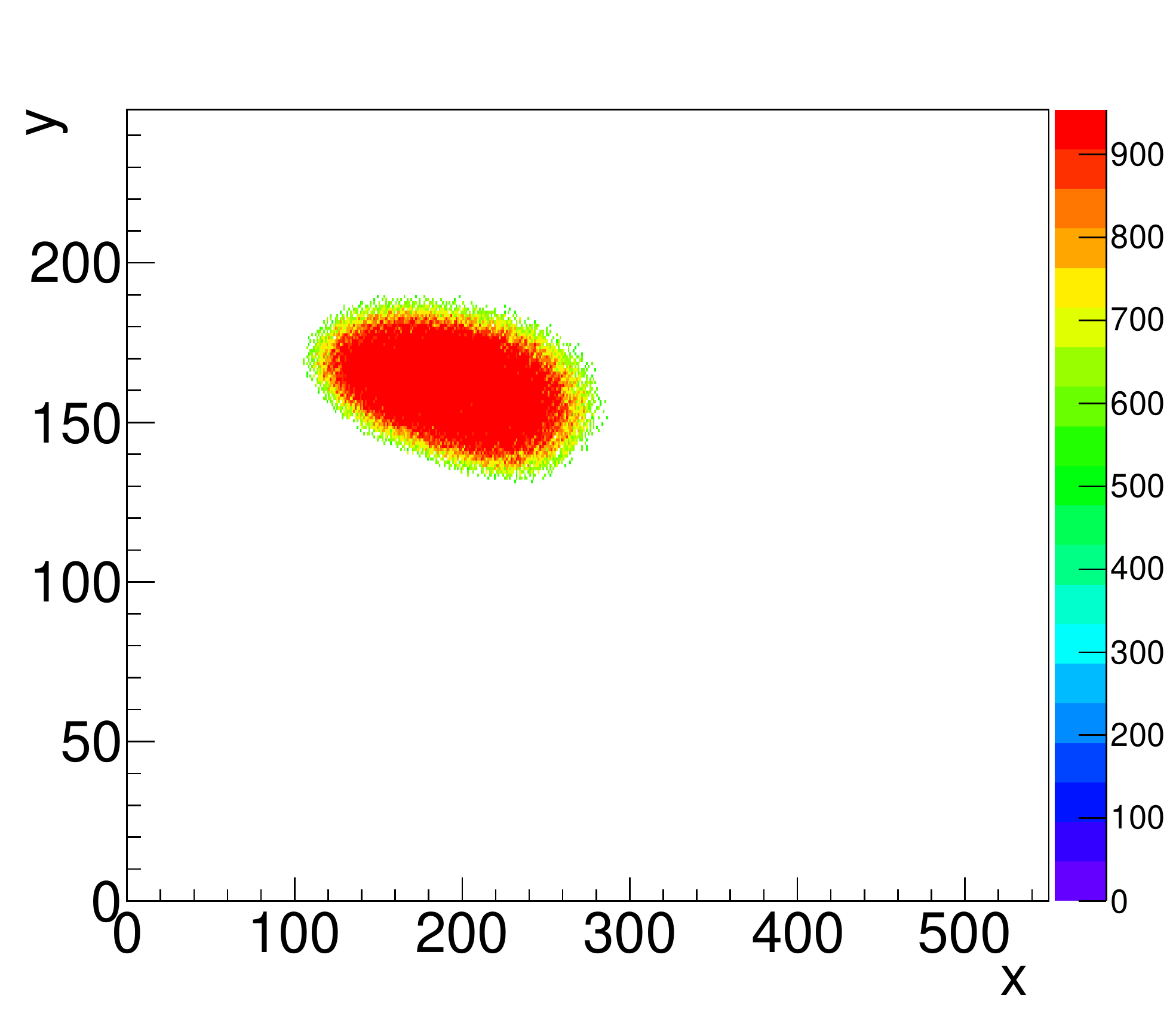}
                \caption{\label{fig:Blob55} Laser spot with a threshold of 55.} 
        \end{subfigure}
       \begin{subfigure}[b]{0.49\textwidth}
   				 \includegraphics[width=\textwidth]{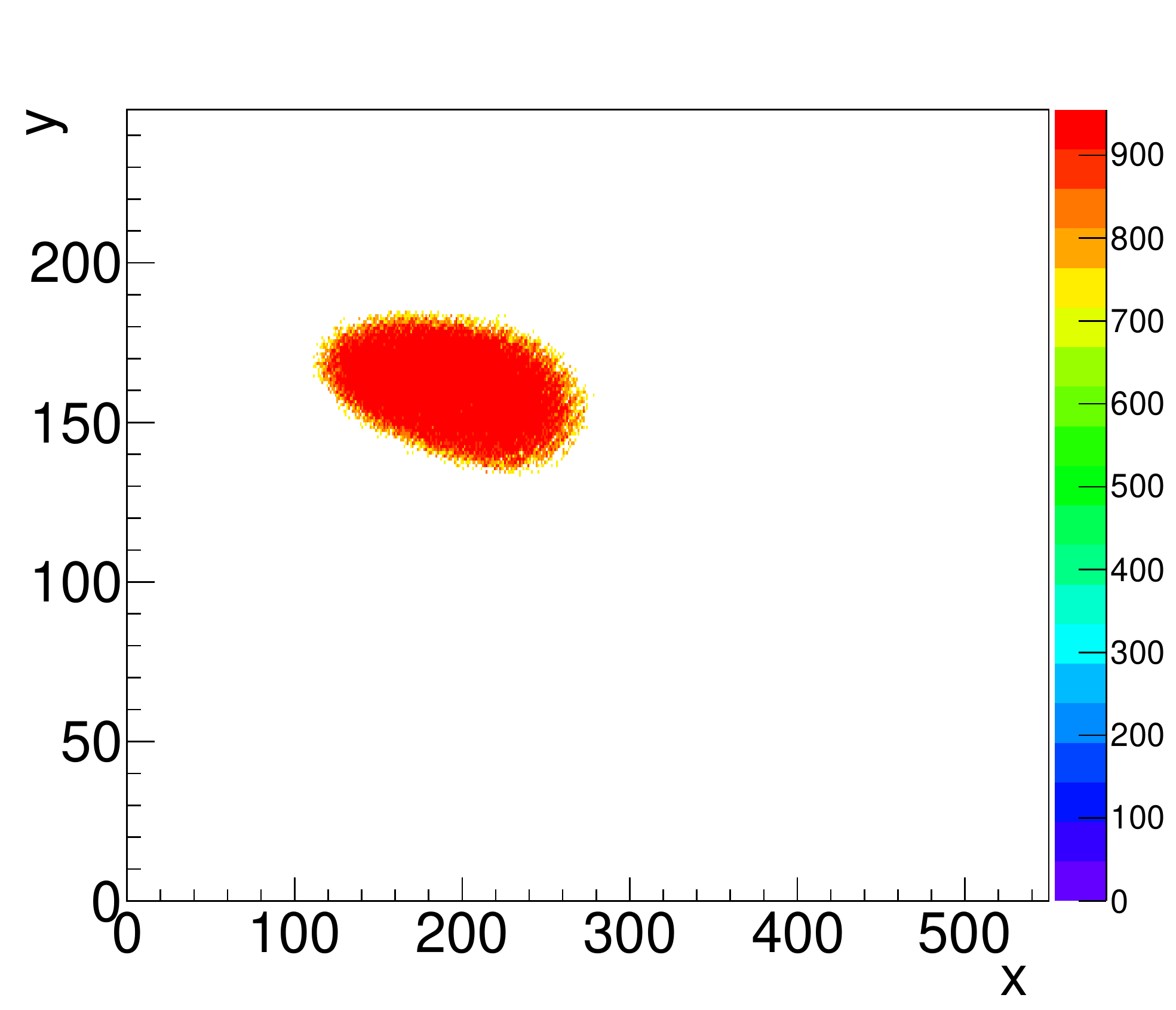}
                \caption{\label{fig:Blob65} Laser spot with a threshold of 65.} 
        \end{subfigure}
        \begin{subfigure}[b]{0.49\textwidth}
   				 \includegraphics[width=\textwidth]{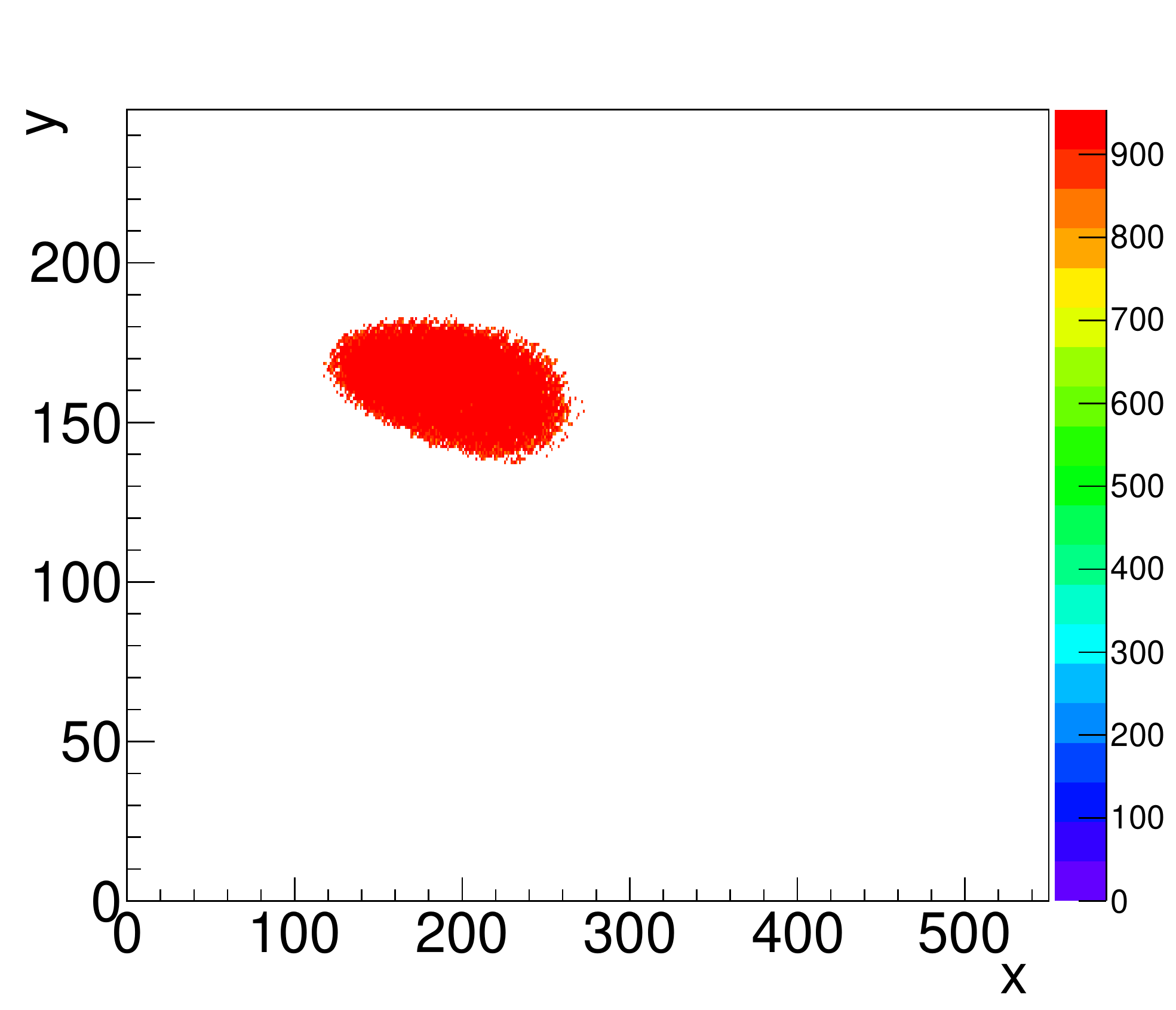}
                \caption{\label{fig:Blob75} Laser spot with a threshold of 75.} 
        \end{subfigure}
        \caption{\label{fig:Blob} Images of the laser diode for various detection threshold values.}
\end{figure}

One method to monitor the alignment of the truss during flight is to use a Laser-CCD system where a laser diode and CCD camera are mounted on opposite ends of the truss.  The location of the spot can be monitored and any movement indicates a deformation in the truss structure.  As mentioned in Section \ref{sec:Thermal}, there will be heaters on the joints which can be used to straighten the truss if it is found to deform.  When the image is taken the user can specify a threshold value above which the intensity of the pixel will be recorded. The images of the laser spot seen with the CCD camera are shown in Figure \ref{fig:Blob} for three different threshold values.
I then went on to examine how well the centroid of the laser spot could be determined using different threshold and fitting methods.  For each of the three thresholds I looked at four different fitting methods.  Method 1 used the centroid fitting method given in Equation \ref{eq:unweightedCenter}.  Method 2 used this same centroid fitting algorithm but with an additional discriminator used to remove any potential stray light.  Method 3 used a weighted centroid fitting algorithm given by 
\begin{equation}
\langle x \rangle = \frac{\sum_{j} x_j I_j}{I_j}
\end{equation}
where $I_j$ is the intensity of each pixel, while Method 4 repeats this with an added discriminator.  The results of this analysis are shown in Table \ref{fig:centFit}.  While overall the location of the center of the laser spot is fairly consistent, there is no difference in results from any method when the higher initial threshold is chosen.  This indicates that the center of the laser spot can be reliably determined regardless of the method used to calculate it.

\begin{table}
\begin{center}
\caption{\label{fig:centFit} Centroid fitting results.}
\begin{tabular}{|c|cc|cc|cc|}
\hline 
 & \multicolumn{6}{|c|} {\rule{0pt}{3ex} Center Point (pixel number)} \\
    \cline{2-7}
&   \multicolumn{2}{|c|}{\rule{0pt}{3ex} Threshold 55} &
\multicolumn{2}{|c|}{Threshold 65} &
\multicolumn{2}{|c|}{Threshold 75} \\
    \cline{2-7}
\textrm{Method}&
\textrm{x}&
\textrm{y} &
\textrm{x}&
\textrm{y} &
\textrm{x}&
\textrm{y} \\
\hline
1 & 196 & 161 & 195 & 160 & 194 & 160 \\
2 & 193 & 161 & 194 & 160 & 194 & 160 \\
3 & 195 & 161 & 195 & 160 & 194 & 160 \\
4 & 193 & 161 & 194 & 160 & 194 & 160 \\
\hline 
\end{tabular}
\end{center}
\end{table}

%%% Local Variables: 
%%% mode: latex
%%% TeX-master: "thesis-main"
%%% End: 

\chapter{\textit{PolSTAR}}
\label{PolSTAR}
This chapter summarizes my contributions to the paper \citet{Krawczynski2015} which describes the proposed X-ray polarimetry mission \textit{PolSTAR}.  All figures in this chapter were published in \citet{Krawczynski2015}.  I contributed Figures \ref{fig:Cyg-X1}, \ref{fig:GRS1915tm}, \ref{fig:MCG}, and \ref{fig:GRS1915pl}, performing \textit{PolSTAR} simulations with a code developed by Henric Krawczynski and Fabian Kislat, as well as contributing Figure \ref{fig:Chi-Square}. 

	\section{The \textit{PolSTAR} Experiment}
\textit{PolSTAR}, the Polarization Spectroscopic Telescope Array, is the satellite version of X-Calibur which was proposed during the 2014 call for NASA small explorer (SMEX) missions.  The details of the proposed mission are presented in \citet{Krawczynski2015} and will be summarized in this chapter.  \textit{PolSTAR} combines the X-Calibur polarimeter with the \textit{NuSTAR} \citep{Harrison2013} detectors and telescope assembly to achieve X-ray polarization measurements in both the hard and soft regimes.  In order to detect the low energy photons that X-Calibur is not sensitive to, the plastic scintillator would be replaced with one made of lithium hydride (LiH).  The LiH scintillator is 6 cm long and 1 cm in diameter and would be surrounded on each side by four CZT detectors with an extra detector behind the stick to image the photons which were not scattered.  The CZT hybrid detectors that would be used are 32x32 pixels and the same as those used in the \textit{NuSTAR} mission. This would not only allow for the mission to achieve the required energy range of 3-50 keV, but also the possibility of the extension of observing the range from 2.5-70 keV. \textit{PolSTAR} would also use the same metrology system, extendible mast structure, and similar optics that \textit{NuSTAR} used.  However, \textit{PolSTAR} would only operate one telescope and would also rotate every 10 minutes to minimize systematic errors. \textit{PolSTAR} would fly in a near circular low earth orbit at 530 km and an inclination of $6^{o}$.
	
	\section{Observing Key Black Hole Parameters}
			\begin{figure}
			\begin{center}
        		\vspace{0pt}
   			 \includegraphics[width=.8\textwidth]{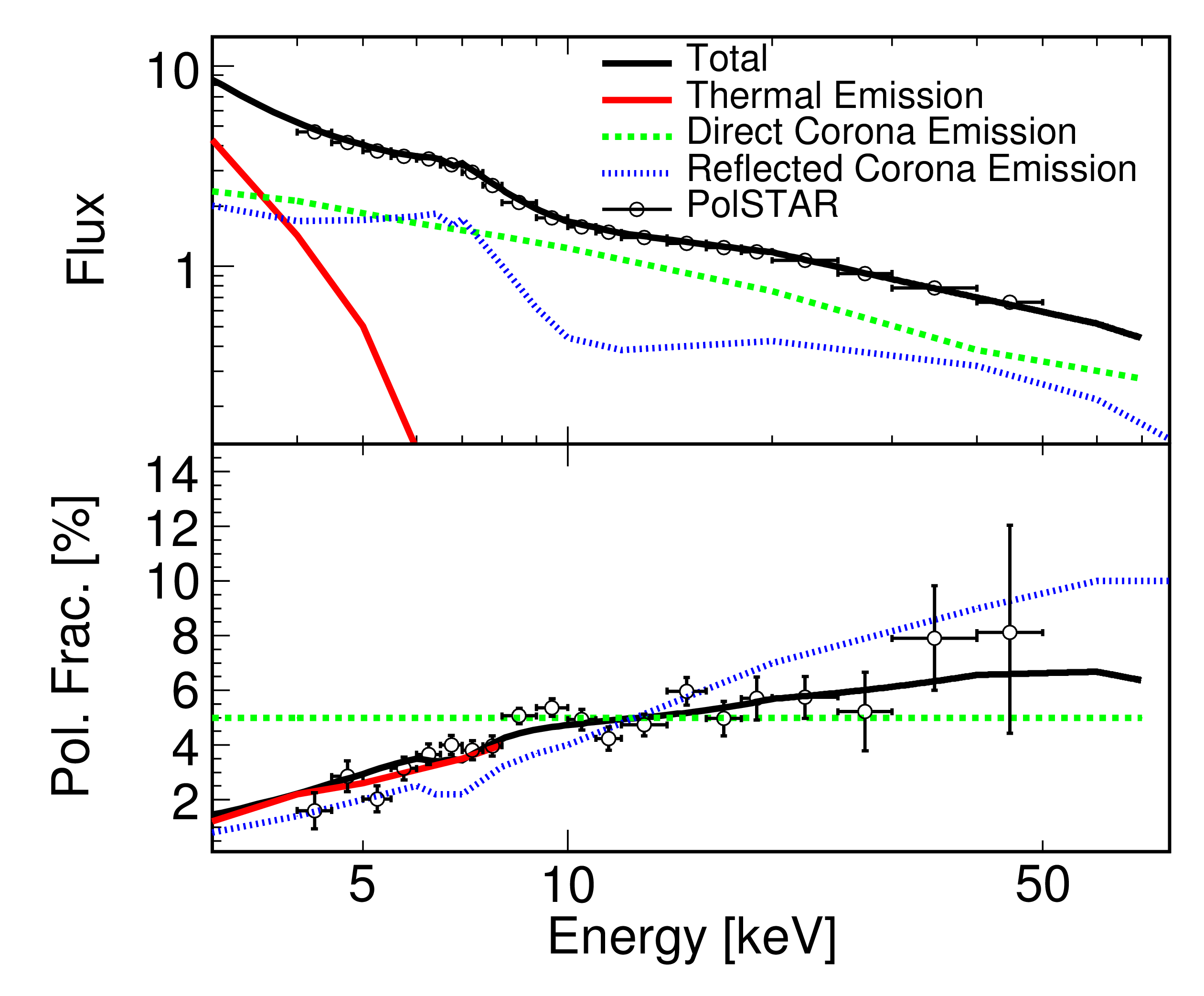}
   			\vspace{10pt}
        		\caption{\label{fig:Cyg-X1} Simulated 3 day \textit{PolSTAR} observation of Cyg X-1. Flux given in units of $10^{-9}$ erg cm$^{-2}$ s$^{-1}$. }
        		\end{center}
		\end{figure}
Because of its 2.5-70 keV energy range, \textit{PolSTAR} would be in a position to detect multiple states of BH accretion systems from the thermal emission at lower energies to the power-law component at higher energies.  Each energy regime provides valuable insight into these extreme environments.  For example, Figure \ref{fig:Cyg-X1} shows the energy spectrum and polarization fraction (also known as polarization degree) for a 10 $M_{\odot}$ BH with $a$ = 0.9, $i$ = $75^{o}$ accreting at 10\% of the Eddington rate.  Simulations for energy spectra and polarization taken from \citet{Tomsick2014} and \citet{Schnittman2009, Schnittman2010} illustrate \textit{PolSTAR}'s capabilities over this large energy range.  The proposed baseline mission included 5 BHBs and 3 SMBHs.
	
\subsection{Measuring Spin and Inclination}
\begin{figure}
			\begin{center}
        		\vspace{0pt}
   			 \includegraphics[width=.8\textwidth]{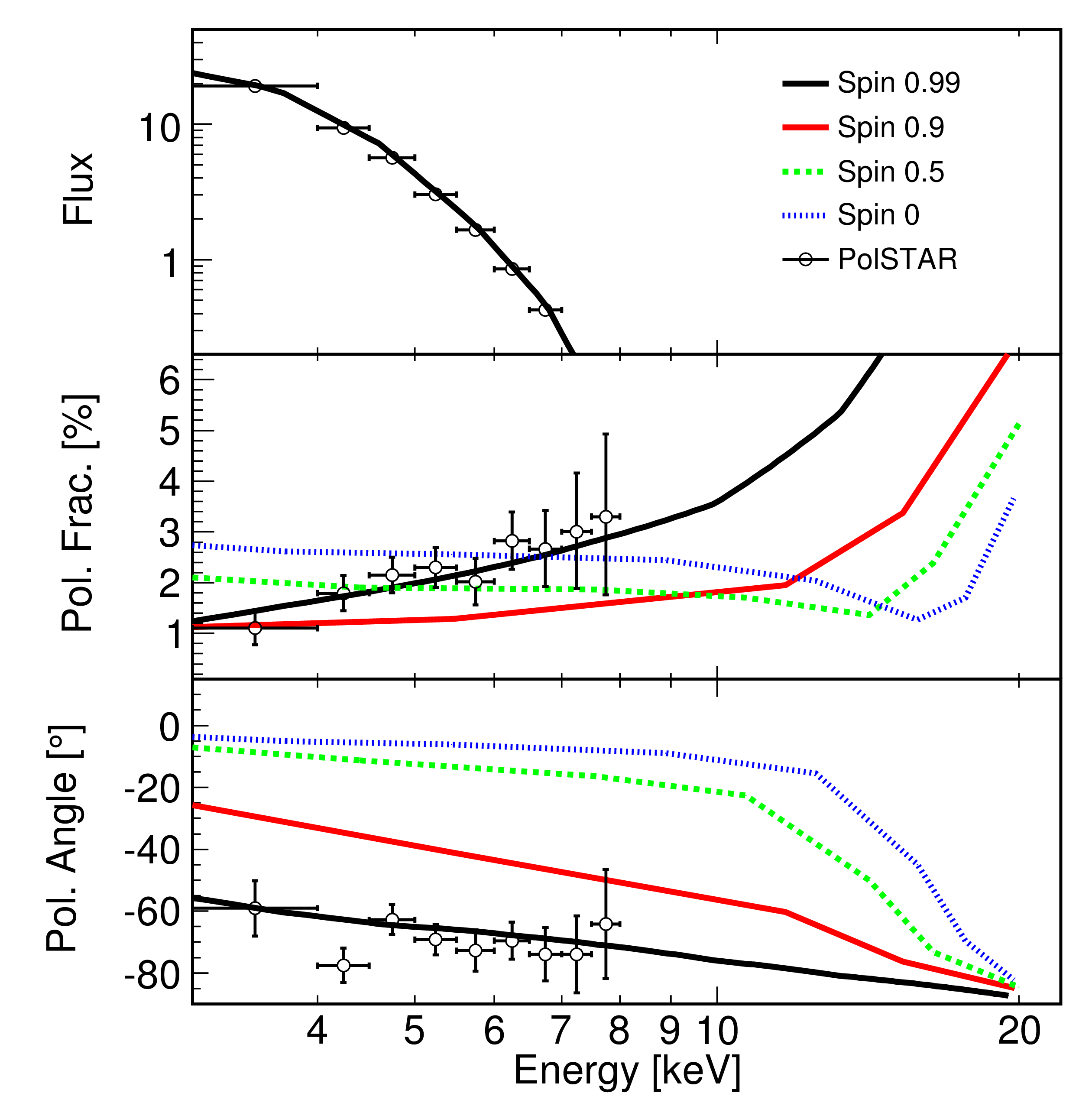}
   			\vspace{10pt}
        		\caption{\label{fig:GRS1915tm} Simulated 7 day \textit{PolSTAR} observation of GRS 1915+105 with $a$ = 0.99 showing the effect of spin on the polarization of the thermal emission. Flux given in units of $10^{-9}$ erg cm$^{-2}$ s$^{-1}$.}
        		\end{center}
		\end{figure}

As mentioned in Chapter 1, it has been shown that X-ray polarization of the thermal disk emission can be used to determine both the spin of the BH \citep{Schnittman2009} and the inclination \citep{Li2009}.  Because of its sensitivity to lower energy X-ray, \textit{PolSTAR} would be able to measure these important values.  For example, Figure \ref{fig:GRS1915tm} shows a simulated 7 day \textit{PolSTAR} observation of GRS 1915+105.  The thermal spectrum is taken from \citet{Ueda2010}, and the polarization results are taken from simulations of a 10 $M_\odot$ BH at an inclination of $75^{o}$ accreting at 10\% of the Eddington luminosity shown in Figure 8 of \citet{Schnittman2009}.  The \textit{PolSTAR} simulated data points are for a BH with $a$ = 0.99.

\begin{figure}
			\begin{center}
        		\vspace{0pt}
   			 \includegraphics[width=.8\textwidth]{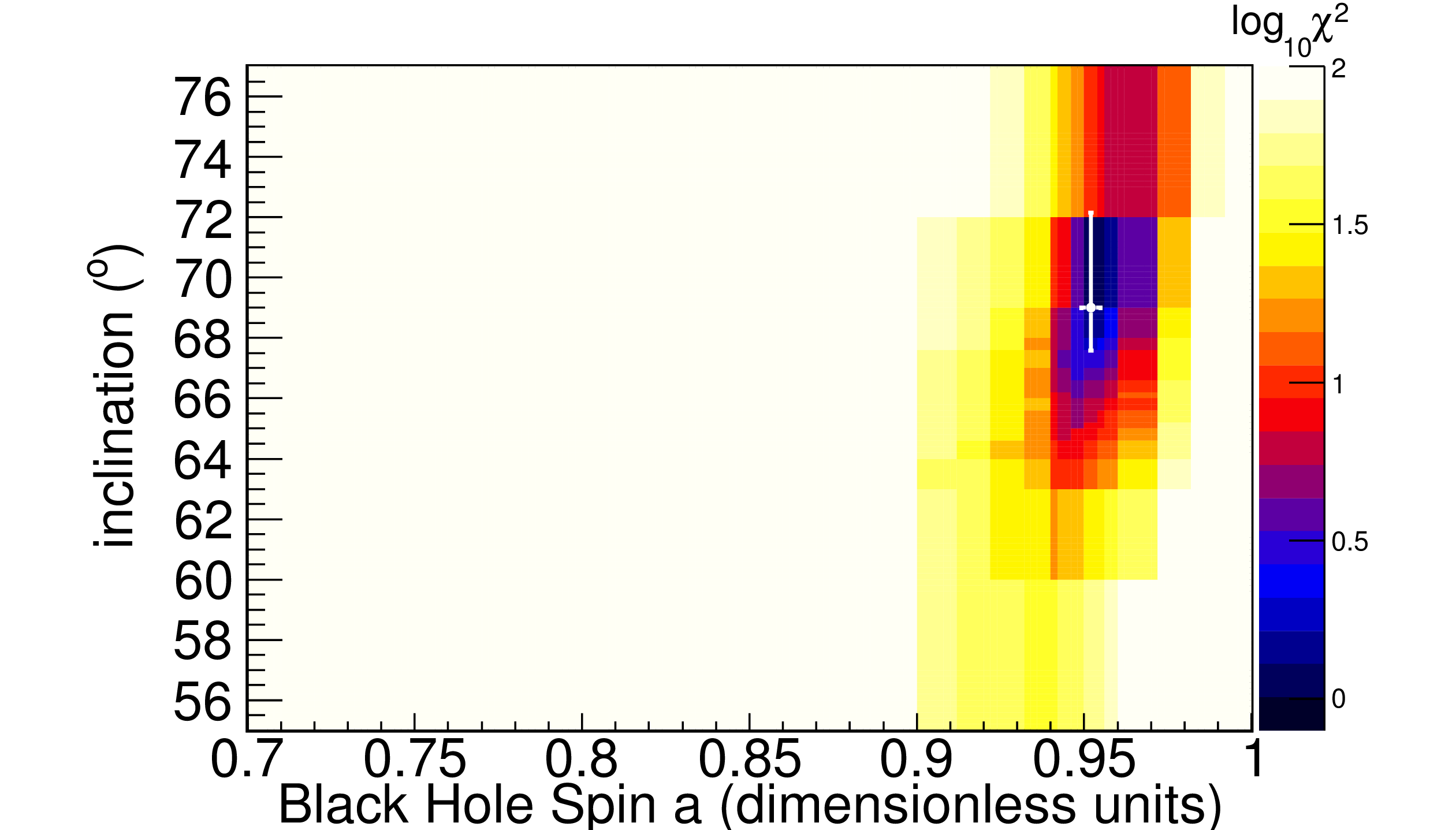}
   			\vspace{10pt}
        		\caption{\label{fig:Chi-Square} Chi-square fit of spin, inclination, and orientation of GRS 1915+105 for a 7-day \textit{PolSTAR} observation.}
        		\end{center}
		\end{figure}

In regards to the inclination, it is often assumed that the accretion disk aligned with the binary plane.  However, it is possible for the inner accretion disk to be misaligned with the binary plane, which would provide observational evidence for the Bardeen-Peterson effect where relativistic frame-dragging around spinning black holes causes the disk to warp (see e.g. \citet{Nealon2015}).  Using templates for the Stokes parameters generated with the code described in Chapter 2, I performed a chi-squared fit to determine the ability of \textit{PolSTAR} to simultaneously measure the spin, inclination, and orientation of the spin axis.  These results are shown in Figure \ref{fig:Chi-Square} for a 7-day \textit{PolSTAR} observation of GRS 1915+105.  The white dot shows the best fit value at $a$ = 0.952, $i$ = $69^{o}$ (assuming true values of $a$ = 0.95 \citep{McClintock2014} and $i$ = $66^{o}$ \citep{McClintock2006b}).  The orientation of the spin axis is not shown in this plot.  The white cross shows the 1-$\sigma$ combined statistical and systematic errors.

\subsection{Constraining Corona Geometry}
\begin{figure}
			\begin{center}
        		\vspace{0pt}
   			 \includegraphics[width=.8\textwidth]{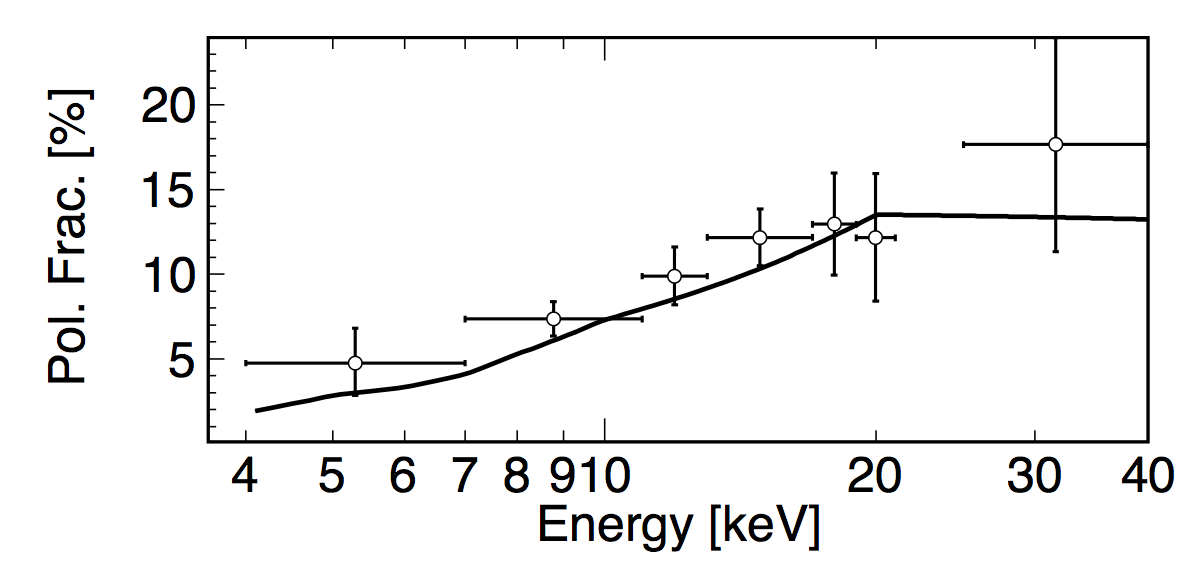}
   			\vspace{10pt}
        		\caption{\label{fig:MCG} Simulated 544 ks \textit{PolSTAR} observation of MCG-5-23-16.}
        		\end{center}
		\end{figure}

\begin{figure}
			\begin{center}
        		\vspace{0pt}
   			 \includegraphics[width=.8\textwidth]{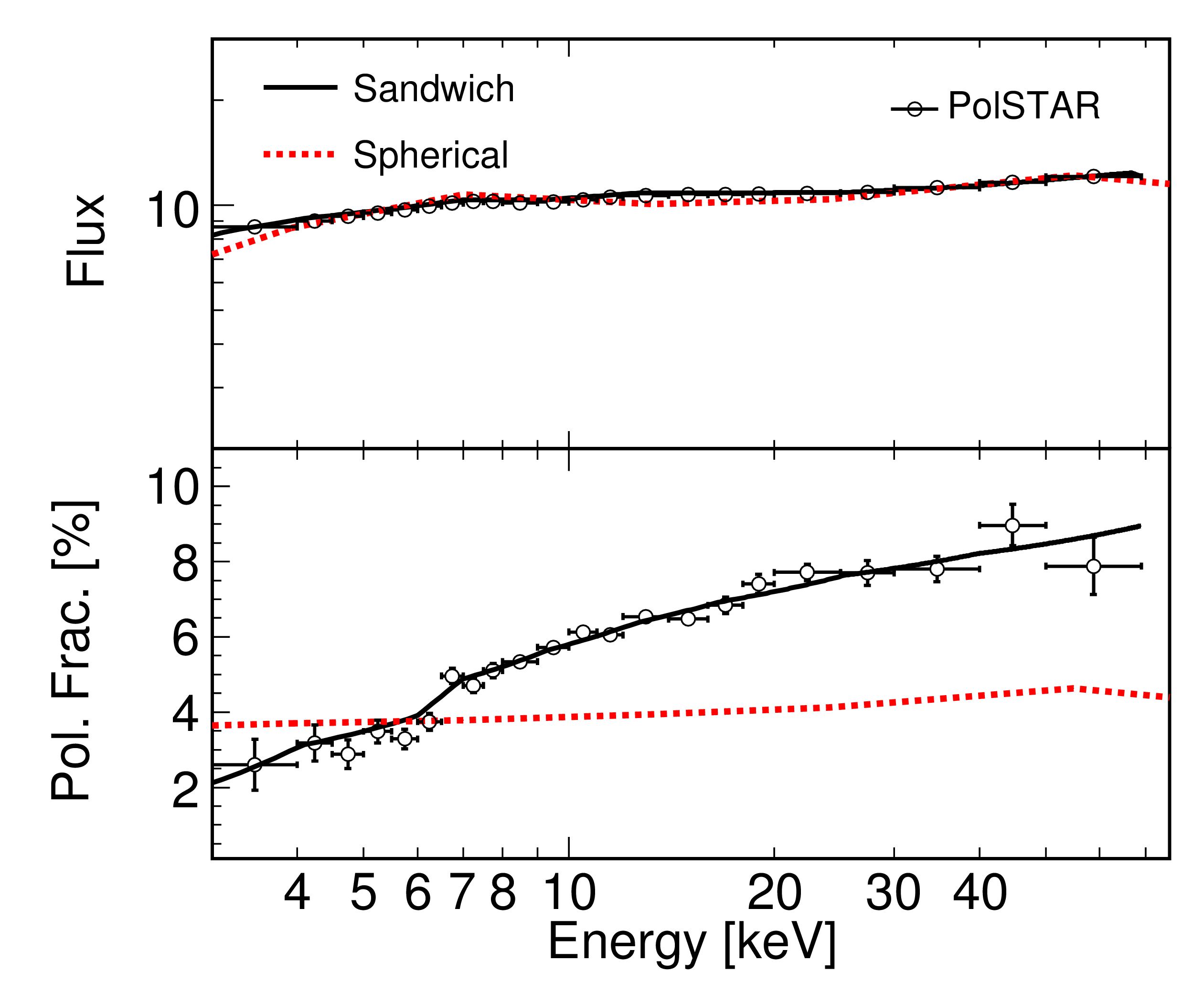}
   			\vspace{10pt}
        		\caption{\label{fig:GRS1915pl} Simulated 3-day \textit{PolSTAR} observation of GRS 1915+105 for a sandwich corona as compared to the predicted polarization for a spherical corona. Flux given in units of $10^{-9}$ erg cm$^{-2}$ s$^{-1}$.}
        		\end{center}
		\end{figure}

Polarization also proves a valuable way to place constraints on the geometry of the corona.  Because of \textit{PolSTAR}'s sensitivity at high energies it is well placed to provide such measurements.  This is illustrated in Figure \ref{fig:MCG} which shows the simulated \textit{PolSTAR} measurement of the polarization fraction for a 544 ks observation MCG-5-23-16, a rapidly rotating SMBH \citep{Brenneman2013}, using simulations for a lamp-post corona \citep{Dovciak2011,Marin2012}. 

In order to further determine the ability of \textit{PolSTAR} to provide insight into the corona geometry, simulations were done to compare the polarization of two different corona geometries: the spherical corona (Figure \ref{fig:CoronaDiagramb}) and the sandwich corona (Figure \ref{fig:CoronaDiagramc}).  These simulations were
run for a 10 $M_\odot$ BH with $a$ = 0.9, $i$ = $75^{o}$, accreting at 10 \% of the Eddington rate as shown in Figure 3 of \citet{Schnittman2010}.  From this 3-day simulated observation of GRS 1915+105 it is seen that
\textit{PolSTAR} would be able to distinguish between these two corona geometries as the sandwich corona results in a higher polarization fraction at higher energies.
		
\section{Other Scientific Motivation}
While my studies focused on the abilities of \textit{PolSTAR} to measure the spin, inclination, and orientation of BHB
and the coronal properties of BHB and SMBHs, there are many other areas into which \textit{PolSTAR} would be able to provide valuable insight.  For example, a prediction of GR is that spacetime can become twisted causing a precession of the material in the accretion disk. This Lense-Thirring precession is a potential explanation for low frequency QPOs such as those observed in GRS 1915+105.  If Lense-Thirring precession is causing this		   timing signal there would be a corresponding polarization signature that would be detectable with \textit{PolSTAR}.  Another valuable source to observe are blazars, SMBHs with relativistic jets along the line of sight to Earth.  These jets are important in understanding galaxies and galaxy clusters as they are responsible for heating their surroundings \citep{Fabian2012}.  It is thought that these jets are strongly affected by magnetic fields with a popular model describing a helical magnetic field threading through the jet \citep{DeVilliers2005, McKinney2006, McKinney2009}.  This would result in regular swings in the observed polarization angle.
The various emission mechanisms proposed for magnetars, highly magnetized neutron stars, result in significantly different polarization signatures.  Because the magnetized plasma near the surface is birefringent, the quiescent thermal emission is almost entirely polarized  \citep{Ozel2001, Lai2003}.  However, a moderate polarization fraction would be expected for the non-thermal which results from  thermal photons undergoing resonant cyclotron and Compton upscattering.  Furthermore, the fallback model which states that thermal and bulk Compton scattering results in the observed non-thermal emission would correspond to a lower polarization fraction \citep{Trumper2010,Trumper2013}.  Because magnetars and accreting pulsars include magnetic fields significantly stronger than those produced on Earth, polarization measurements of these sources can provide a way to test Quantum Electrodynamics (QED) in regimes not achievable elsewhere \citep{Fernandez2011, Kii1987}.  While \textit{PolSTAR} was not selected for the next phase of study, two X-ray polarimeters, \textit{IXPE} \citep{Weisskopf2014} and \textit{PRAXyS} \citep{Jahoda2015}, were chosen for the next phase and would be able to study these systems in the lower energy range along with \textit{XIPE}, which is an X-ray imaging polarimetry mission under consideration by ESA.

%%% Local Variables: 
%%% mode: latex
%%% TeX-master: "thesis-main"
%%% End: 

\chapter{Summary and Outlook}
\label{SummaryOutlook}
As a way to study GR in the strong gravity regime and the properties of inner accretion flow, I used and added to a ray-tracing code that models the X-ray emission from BH accretion disks.  The code originally modeled the polarization and spectra of the thermal emission from BHBs \citep{Krawczynski2012} and I added in the capability to simulate the reflected power-law emission from both BHBs and SMBHs, which was described in detail in Chapter \ref{RayTracingCode}.  This includes modeling the Fe-K$\alpha$ line and Compton hump, as well as their corresponding polarization and reverberation signatures.  These simulations can be performed assuming any axially symmetric metric. I included four non-GR metrics: 2 phenomenological in nature \citep{Johannsen2011a,Glampedakis2006} and 2 motivated by alternative theories of gravity \citep{Aliev2005,Pani2011}.  In Chapter \ref{AlternativeMetrics}, summarizing the paper \citet{Hoormann2016}, I examined how joint timing, spectral, and polarization observations of BHBs and SMBHs could be used to constrain any potential deviations from GR as a way to test the No-Hair theorem.  This study focused on comparing versions of these metrics which all yield the same value for the $r_{ISCO}$.  While it was found that varying the metric only results in small differences in the spectral, timing, and polarization signatures for solutions with the same $r_{ISCO}$, this finding allows us to use measurements of the $r_{ISCO}$ to exclude large portions of the parameter space of alternative theories of gravity.  Observations of rapidly spinning BHs turn out to be particularly useful to constrain deviations from GR.  I have used observations of the stellar mass BH Cyg X-1 \citep{Gou2011} to derive sensitive constraints on the alternative metrics of \citet{Johannsen2011a} and \citet{Aliev2005}. In addition to constraining deviations from GR, this code can also be used to model the Fe-K reverberation signatures as a way to model the ionization of the accretion disk.  From these results it is seen that the ionization scheme characterizing absorption versus scattering of the photons interacting with the disk has the greatest effect on the corresponding Fe-K reverberation signal.

The simulations used in this thesis are phenomenological GR ray-tracing simulations in which the geometries of both the disk and the corona are not derived self-consistently.  This code specifically assumes a thin disk described by \citet{Novikov1973} and a lamp-post corona geometry. In recent years, general relativistic magnetohydrodynamic (GRMHD) simulations have been introduced which solve for the structure of the disk and corona from first principals (see Chapter \ref{GRRMHD}).  In order to assess the validity of the NT model, GRMHD simulations were performed and found that the NT model does hold for luminosities  $0.01 L_{Edd} < L < 0.3 L_{Edd}$ \citep{Kulkarni2011,Penna2012,Sadowski2016}.  On the other hand, recent UV/optical observations suggest accretion disks may be larger than expected for AGN \citep{Morgan2010, Fausnaugh2016}, indicating the presence of additional accretion flow components.

Current and upcoming missions will continue to test the theoretical models. \textit{XMM-Newton} \citep{Cottam2001}, \textit{Swift} \citep{Hurley2003}, and \textit{NuSTAR} \citep{Harrison2013} provide valuable data to perform reverberation analyses.  The Neutron star Interior Composition ExploreR (\textit{NICER}) is an X-ray timing and spectroscopy mission sensitive from 0.2-12 keV which is intended to launch in 2016 as a payload on the International Space Station \citep{Gendreau2012}.  There are also three X-ray polarimetry missions currently under study (\textit{IXPE, PRAXyS}, and \textit{XIPE}) which would allow a way to obtain geometrical information about sources that are too small to image directly.  ESA plans to launch \textit{Athena} in 2028, consisting of a calorimeter spectrometer and wide field imager.  \textit{Athena} which will be able to further study the extreme environments surrounding SMBHs \citep{Barret2013}.  The first two detections of gravitational waves from the merger of binary BHs were announced this year \citep{Abbott2016,Abbott2016b}.  The detection was made by LIGO, the Laser Interferometer Gravitational-Wave Observatory, which has the potential to detect many more mergers of systems containing BHs and/or neutron stars, providing further insight into the extreme gravitational fields surrounding these objects.  The Event Horizon Telescope (EHT), made up of a global array of radio telescopes, will be able to study the inner regions of the accretion flow of Sgr $A^\ast$ with an angular resolution comparable to its event horizon \citep{Doeleman2010}.  Therefore, the combination of new telescopes and theoretical modeling with numerical codes indicates that significant progress will be made in the understanding of BHs in upcoming years.

%%% Local Variables: 
%%% mode: latex
%%% TeX-master: "thesis-main"
%%% End: 

%\appendix                        % now we start appendicies

% A good place for the bibliography.
%
\bibliographystyle{plainnat}
\addcontentsline{toc}{chapter}{References}

\begin{spacing}{1.0}
\bibliography{./thesis-References}
\end{spacing}
\nocite{*}

%
% The Vita should be the last thing
%
%\begin{thesisauthorvita}
%
% This is just a sample of what to do in a vita
%
%\input{vita}
%\end{thesisauthorvita}
%\iffalse
%{\flushleft
%\textit{Note:} Use month and year in which your degree will be conferred.}
%\fi

\begin{thesisshorttitlepage}
%\iffalse
%{\small 

%\textbf{NOTE:}	This is a sample of a ``short title'' page.  Please change the
%line above to use an appropriate ``short title'' for your thesis, insert your
%last name, and include your degree and year in which the degree will be earned.
%Separate elements using commas, as illustrated in the sample above.
%\uline{Your ``short title'' cannot exceed 35 characters, counting spaces}.  It
%does not matter if there is a page number at the bottom of the page.  

%\textbf{IMPORTANT:} This page should be printed and taped securely to each of
%the three manila envelopes used to submit your final hard copies.  Remove this
%page before submitting your final copies (i.e., this page should not be
%included in either your electronic submission or your hard copy submissions).
%See Appendix~\ref{app:procedures} for further details if necessary.
%}
%\fi
\end{thesisshorttitlepage}

%%% Local Variables: 
%%% mode: latex
%%% TeX-master: "thesis-main"
%%% End: 

\end{document}